\documentclass[%
 aip,
 amsmath,amssymb,
 reprint,%
]{revtex4-1}

\usepackage{graphicx}% Include figure files
\usepackage{dcolumn}% Align table columns on decimal point
\usepackage{bm}% bold math
\usepackage[utf8]{inputenc}
\usepackage[T1]{fontenc}
\usepackage{mathptmx}

\usepackage{basis}

\newcommand{\blue}[1]{{#1}}

\begin{document}

\preprint{AIP/123-QED}

\title{Efficient Sampling of Thermal Averages of Interacting Quantum Particle Systems with Random Batches}
% Force line breaks with \\

\author{Xuda Ye}
	\affiliation{School of Mathematical Sciences, Peking University}
	\email{abneryepku@pku.edu.cn}
\author{Zhennan Zhou}% 
	\affiliation{Beijing International Center for Mathematical Research, Peking University}
	\email{zhennan@bicmr.pku.edu.cn}

\date{\today}

\begin{abstract}
An efficient sampling method, the pmmLang+RBM, is proposed to compute the quantum thermal average in the interacting quantum particle system. Benefiting from the random batch method (RBM), the pmmLang+RBM reduces the complexity due to the interaction forces per timestep from $O(NP^2)$ to $O(NP)$, where $N$ is the number of beads and $P$ is the number of particles. Although the RBM introduces a random perturbation of the interaction forces at each timestep, the long time effects of the random perturbations along the sampling process only result in a small bias in the empirical measure of the pmmLang+RBM from the target distribution, which also implies a small error in the thermal average calculation. We numerically study the convergence of the pmmLang+RBM, and quantitatively investigate the dependence of the error in computing the thermal average on the parameters including the batch size, the timestep, etc. We also propose an extension of the pmmLang+RBM, which is based on the splitting Monte Carlo method and is applicable when the interacting potential contains a singular part. 
\end{abstract}

\maketitle

\section{Introduction}
Simulating complex chemical systems with quantum effects has been an appealing subject in computational physics and chemistry.
In quantum systems,
thermal properties are fully described by the canonical ensemble, and a considerable number of methods of calculating thermal averages are based on the path integral representation \cite{feymann,pathintegral1,pathintegral2,pathintegral3}, which reformulates the quantum system as a classical ring polymer system. In the past decades, the path integral Monte Carlo (PIMC) \cite{pimc1,pimc2,pimc3} and the path integral molecular dynamics (PIMD) \cite{pimd1,pimd2,pimd3} techniques have been developed and successfully applied to the calculation of quantum properties including reaction rates \cite{rate1,rate2,rate3}, correlation functions \cite{correlation1,nonprecondition0} and quantum tunneling \cite{tunnel1,tunnel2}.\par
In this paper we focus on long-range interacting quantum particle systems. The complexity to calculate the interaction forces of a $P$-particle system is $O(P^2)$, and efficient computational methods are thus needed to simulate such big systems. Compared to short-range interactions which can be easily treated with cutoff \cite{cutoff1} or data structures like the cell list \cite{celllist1,celllist2}, the long-range nature of the interaction potential makes it more difficult to reduce the complexity due to the interaction forces.\par
So far a large variety of methods have been proposed to calculate the interaction forces for a particle system with electrostatic interactions. The most representative methods in this class are Ewald summation \cite{ewald1,ewald2} for problems with periodic boundaries and fast multipole method (FMM) \cite{fmm1} for open systems, and the complexity of the interaction forces can be reduced to $O(P\log P)$ or even $O(P)$. However, an obvious drawback of these methods is that they rely on the specific expression of the interaction potential, i.e., the Coulomb potential, and can hardly be applied to more general interacting systems.\par
Besides the methods which aim to calculate the interaction forces directly, there are also methods focusing on modifying the dynamics while maintaining the physical properties of interest. 
For example, the generalized Verlet algorithm \cite{modify1} reduces the number of times of calculating interaction forces by introducing distance classes, and the random batch method (RBM) \cite{rbm} simplifies the dynamics by random sampling at each iteration.\par
%% ---- benefites of RBM ---- %%
In this paper, the RBM is employed to reduce the complexity due to the interaction forces.
The RBM is an efficient sampling method recently introduced for interacting particle systems, and has been applied to reduce the computational cost of certain examples in real time dynamics and ensemble average calculation \cite{rbm1,rbm2}.
In each timestep,
the RBM randomly groups all particles into small batches, and the evolution of the whole system is replaced by the evolution in those batches respectively.
Due to the random nature of the grouping in the RBM, the average force felt by each particle in the RBM is statistically the same as the force in the original dynamics.
Since interaction forces are only calculated in the small batches, the complexity in a timestep is reduced from $O(P^2)$ to $O(P)$.
Compared to other efficient sampling methods like the Ewald summation and the FMM, the RBM is easier to implement and more flexible to apply in complex interacting particle systems.\par
After including quantum effects, simulation of interacting particle systems meets additional challenges.
In the context of the PIMD, \blue{suppose there are $N$ beads in the reformulated ring polymer system. }The complexity due to interaction forces in \blue{this} reformulated system per timestep is $O(NP^2)$, which is $N$ times of the complexity in the classical case. In addition, the ring polymer system suffers from the stiffness of the ring polymer potential when the number of beads $N$ is large. To resolve the stiffness, one can either precondition the dynamics by modifying the ring polymer mass matrix \cite{precondition1,precondition} or using the staging coordinates \cite{staging,precondition2}, or employ other non-preconditioned methods \cite{nonprecondition0,nonprecondition1,nonprecondition2,nonprecondition3}.\par
In large interacting quantum systems, the compatibility between the RBM and the technique for resolving the stiffness must be taken into account. We choose to employ the preconditioned mass-modified Langevin dynamics (pmmLang) \cite{precondition} rather than the staging coordinates to precondition the dynamics, since the pmmLang sticks to the use of the physical coordinates.\par 
In this paper, we will show that the combination of the pmmLang and the RBM (\blue{denoted by} the pmmLang+RBM) is a suitable sampling method in the large interacting quantum particle systems.
The pmmLang+RBM costs only $O(NP)$ complexity in a timestep to compute the interaction forces, and is able to obtain the accurate thermal average up to a small bias. Also, the pmmLang+RBM can be naturally combined with the splitting Monte Carlo method \cite{rbm1} to apply to systems \blue{with} singular interacting potentials.\par
Now we present our basic setup of this paper. In the $P$-particle quantum system in $\mathbb R^{3P}$ with long-range interactions, we aim to calculate the thermal average of some observable operator $\hat A$. Suppose the potential of the system is $\hat V = V(\hat q)$, with $V(q)$ given by
\begin{equation}
	V(q) = \sum_{i=1}^P V^{(o)}(q^i)
	+\sum_{1\Le i<j \Le P}
	V^{(c)}(q^i-q^j)
	\label{eq:V}
\end{equation}
where
$q = (q^1,\cdots,q^P)\in\mathbb R^{3P}$ is the position and momentum of the $P$ particles, $V^{(o)}(q^i)$ is the external potential of $q^i$ in $\mathbb R^3$, and $V^{(c)}(q^i-q^j)$ is the interacting potential between $q^i,q^j$. Assume each particle has mass $m$, then
the Hamiltonian operator of the system is
\begin{equation}
	\hat H = \frac{\hat p^2}{2m} + V(\hat q),~~~~
	q,p\in\mathbb R^{3P}
\end{equation}
and the thermal average of $\hat A$ is given by
\begin{equation}
	\avg{\hat A} = \frac1Z
	\Tr[e^{-\beta\hat H}\hat A]
	,~~~~
	Z = \Tr[e^{-\beta\hat H}]
	\label{eq:<A>}
\end{equation}
where $\beta>0$ is the inverse temperature.\par
In the path integral representation, the quantum system (\ref{eq:V}) is reformulated as a classical ring polymer system with the potential function
\begin{equation}
U_N(\q) =  \frac{m}{2\beta_N^2}\sum_{k=1}^N |q_k-q_{k+1}|^2 +
\sum_{k=1}^N V(q_k)
\label{eq:U_N}
\end{equation}
where $\beta_N = \beta/N$ and $\q = (q_1,\cdots,q_N)^\T\in\mathbb R^{N\times3P}$ is the position of the ring polymer. In this way,
the thermal average $\avg{\hat A}$ can be approximated by the ensemble average in the Boltzmann distribution $\pi(\q)\propto e^{-\beta_N U_N(\q)}$. The pmmLang+RBM we propose in this paper, which aims to sample the distribution $\pi(\q)$, can be implemented in the following steps:\par
\begin{enumerate}
	\setlength{\itemsep}{0pt}
	\item Derive the pmmLang \blue{for} the ring polymer system \blue{$U_N(\q)$} to sample the Boltzmann distribution $\pi(\q)$.
	\item Numerically integrate the sampling path of the pmmLang, where the interaction forces are \blue{efficiently} computed by the RBM. Time averages of the weight functions are used to approximate the thermal average $\avg{\hat A}$.
	\item (optional) If the interacting potential $V^{(c)}(q)$ contains a singular part, combine the pmmLang+RBM with the splitting Monte Carlo method.
\end{enumerate}
The pmmLang+RBM is simple and efficient to calculate the thermal average of the interacting quantum particle systems, and benefits from both the RBM and the pmmLang.
Due to the RBM, the complexity of interaction forces in a timestep is reduced to $O(NP)$, and the total complexity in a timestep is $O(N\log NP)$, which has satisfactory scaling properties for both $N$ and $P$.
When the interaction potential contains a singular part (e.g., the Lennard-Jones and the Morse potential), the pmmLang+RBM method can be extended by the use of the splitting Monte Carlo method, which lifts the constraint of using extremely small time steps.\par
The paper is organized as follows. In Section \ref{sec:PIMD}, we introduce the 
the PIMD for interacting particle systems and the difficulties in numerical simulation, then derive the pmmLang.
In Section \ref{sec:RBM}, we introduce the random batch method and discuss of the error analysis of the pmmLang+RBM.
In Section \ref{sec:split}, we introduce the splitting Monte Carlo method and its combination with the pmmLang+RBM, the pmmLang+RBM+split. In Section \ref{sec:tests}, we present the numerical results of the pmmLang+RBM and the pmmLang+RBM+split and \blue{report} the error in the the calculation of thermal averages.
\section{PIMD for interacting particle systems}
\label{sec:PIMD}
\subsection{Ring polymer representation and Langevin sampling}
In the path integral representation, the thermal average
(\ref{eq:<A>}) is approximated as
\begin{equation}
	\avg{\hat A}\approx
	\frac1{Z_N}
	\int\d\q\times e^{-\beta_N U_N(\q)}\times W_N(\q),
	\label{eq:<A> approx}
\end{equation}
where $\beta_N=\beta/N$, and
$$
\q = \begin{bmatrix}
	q_1 \\ \vdots \\ q_N
\end{bmatrix} = 
\begin{bmatrix}
	q_1^1 & \cdots & q_1^P \\
	\vdots & \ddots & \vdots \\
	q_N^1 & \cdots & q_N^P
\end{bmatrix}
\in
\mathbb R^{N\times 3P}
$$
is the position of the ring polymer. In the coordinate notation $q_k^i$, the subscript $k$ indicates the bead index in the ring polymer representation, and the superscript $i$ indicates the particle index in the physical system.
Besides, $U_N(\q)$ is the total potential of the ring polymer system defined in (\ref{eq:U_N}),
\begin{equation}
Z_N = \int \d\q\times e^{-\beta_N U_N(\q)}
\end{equation}
is the partition function, and $W_N(\q)$ is the weight \blue{corresponding to} the observable operator $\hat A$. \blue{Standard theory of the path integral\cite{pimc1} shows that} the approximation (\ref{eq:<A> approx}) is exact as the number of beads $N\rightarrow\infty$.
Define the Boltzmann distribution as
\begin{equation}
	\pi(\q) = \frac1{Z_N} e^{-\beta_N U_N(\q)},~~~
	\q\in\mathbb R^{N\times3P}
	\label{eq:pi(q)}
\end{equation}
then (\ref{eq:<A> approx}) can be equivalently written as
\begin{equation}
	\avg{\hat A} \approx \avg{W_N(\q)}_\pi :=
	\int W_N(\q) \pi(\q) \d\q,
\end{equation}
i.e., $\avg{\hat A}$ is approximated as the ensemble average of the weight function $W_N(\q)$ in the distribution $\pi(\q)$.\par
In this paper, the observable operators of interest are the position-dependent operator $\hat A = A(\hat q)$ and the kinetic energy operator $\hat A = \hat p^2/(2m)$, where $A(q)$ is an analytic function in $\mathbb R^{3P}$. 
For the position-dependent operator, the weight $W_N(\q)$ is \blue{simply}
\begin{equation}
	W_N(\q) = \frac1N
	\sum_{k=1}^N A(q_k).
	\label{eq:W_N position}
\end{equation}
For the kinetic energy operator, the weight $W_N(\q)$ \blue{is chosen as} the virial estimator \cite{virial}
\begin{equation}
	W_N(\q) = 
	\frac{3P}{2\beta}
	+
	\frac1{2N}
	\sum_{k=1}^N
	(q_k-\bar q)^\T \nabla V(q_k),
	\label{eq:W_N virial}
\end{equation}
where
\begin{equation}
\bar q = \frac1N\sum_{k=1}^N q_k\in\mathbb R^{3P}
\end{equation}
is the center of the ring polymer.\par
To compute the ensemble average $\avg{W_N(\q)}_\pi$, we employ the path integral molecular dynamics (PIMD) technique to sample the distribution $\pi(\q)$.
By introducing an auxiliary momentum
variable $\p\in\mathbb R^{N\times 3P}$,	
the PIMD couples the 
the Hamiltonian dynamics and a thermostat scheme to preserve the invariant distribution.
In practice, the widely used thermostats include the Andersen thermostat \cite{Andersen}, the Nos\'e-Hoover thermostat \cite{Nose}, the Langevin thermostat, etc. In this paper we focus on the last one to develop an efficient sampling algorithm with random batches, and \blue{our} underlying dynamics is \blue{thus} the second-order Langevin dynamics.\par
For heuristic purposes, we present in the following the PIMD with the Langevin thermostat. Introduce the positive definite mass matrix $M\in\mathbb R^{N\times N}$ (maybe different from the physical mass $m$) and \blue{extend $U_N(\q)$ to the Hamiltonian}
\begin{equation}
	H_N(\q,\p) = 
	\frac12\avg{\p,M^{-1}\p}_F + U_N(\q),
	\label{eq:H(q,p)}
\end{equation}
where $\avg{\cdot,\cdot}_F$ is the Frobenius inner product in $\mathbb R^{N\times3P}$. By adding damping and diffusion terms in the Hamiltonian dynamics of (\ref{eq:H(q,p)}), we obtain the second-order Langevin dynamics \blue{in the PIMD},
\begin{equation}
	\begin{aligned}
		\d\q & = M^{-1}\p\d t, \\
		\d\p & = - \nabla U_N(\q) \d t - \gamma \p\d t
		+
		\sqrt{\frac{2\gamma M}{\beta_N}}\d\B.
	\end{aligned}
	\label{dyn:Lang momentum}
\end{equation}
where $\gamma>0$ is the friction constant and $\B$ is standard Brownian motion in $\mathbb R^{N\times 3P}$.\par
The invariant distribution of (\ref{dyn:Lang momentum}) is
\begin{equation}
	\pi(\q,\p) \propto
	\exp\bigg(\hspace{-2pt}-\beta_N
	\Big(
	\frac12\avg{\p,M^{-1}\p}_F + U_N(\q)
	\Big)
	\bigg),
	\label{eq:pi(q,p)}
\end{equation}
whose marginal distribution in $\q$ is exactly $\pi(\q)$ as in (\ref{eq:pi(q)}). From the ergodicity of the Langevin dynamics, $\avg{W_N(\q)}_\pi$ can be computed by the infinite time average of \blue{$W_N(\q(t))$ with the classical trajectory $(\q(t),\p(t))$ propogated by} (\ref{dyn:Lang momentum}), i.e.,
\begin{equation}
	\avg{W_N(\q)}_{\pi} = 
	\lim_{T\rightarrow\infty}
	\frac1T\int_{0}^T W_N(\q(t))\d t.
	\label{eq:<W_N> time average}
\end{equation}
\blue{In practice}, one employs the discrete time trajectory $\q(j\Delta t)$ \blue{numerically solved by (\ref{dyn:Lang momentum})} to approximate \blue{the time integral in (\ref{eq:<W_N> time average})}, i.e.,
\begin{equation}
	\avg{W_N(\q)}_{\pi} = 
	\lim_{J\rightarrow\infty}
	\frac1J
	\sum_{j=1}^J W_N(\q(j\Delta t)),
\end{equation}
where $\Delta t$ is the timestep.\par
In the large interacting particle system, the most costly part of the Langevin dynamics (\ref{dyn:Lang momentum}) is computing the gradient $\nabla U_N(\q)$, which involves heavy calculation of the interaction forces. As we shall elaborate in the next part, the complexity due to interaction forces is $O(NP^2)$.
\subsection{Simulation bottleneck in Langevin sampling}
In this paper we are mainly concerned with the efficiency of calculating the thermal average (\ref{eq:<A>}) for a large quantum interacting particle system, where the numerical challenges mainly originate from two aspects of reasons. On the one hand, given a quantum interacting particle system, the number of beads $N$ in the path integral representation needs to be sufficiently large to approximate the thermal average $\avg{\hat A}$. On the other hand,
in order to accurately calculate a wide class of physical quantities, it is often desirable to simulate in large size systems, for example, the liquid water, since the number of particles $P$ should be large enough to incorporate the correct scientific phenomenon.\par
When $N$ and $P$ are large, one of the major difficulties in the numerical simulation of (\ref{dyn:Lang momentum}) is that the computational cost per timestep is extremely heavy. To perform a complexity analysis, we write the total $\nabla U_N(\q)$ as
\begin{equation}
	U_N(\q) = \frac12\avg{\q,L\q}_F + \sum_{k=1}^N
	V(q_k),
\end{equation}
where $L\in\mathbb R^{N\times N}$ is the second order difference matrix
\begin{equation}
L = \frac m{\beta_N^2}
\begin{bmatrix}
	2 & -1 & & \cdots & & -1  \\
	-1 & 2 & -1& \cdots \\
	& -1 & 2 & \cdots \\
	\vdots & \vdots  &\vdots & \ddots & -1\\
	& & & -1 & 2 & -1\\
	-1 &&&&-1&2
\end{bmatrix},
\label{eq:L}
\end{equation}
The gradient $\nabla U_N(\q)$ is thus given by
\begin{equation}
	\nabla U_N(\q) = L\q + 
	\begin{bmatrix}
		\nabla V(q_1) \\
		\vdots \\
		\nabla V(q_N)
	\end{bmatrix} \in \mathbb R^{N\times 3P},
\end{equation}
where each $\nabla V(q_k)$ involves the full interaction forces in the system (\ref{eq:V}). \blue{In fact,}
\begin{equation}
	\nabla V(q_k) = 
	\begin{bmatrix}
		\dfrac{\partial V(q_k)}{\partial q_k^1} & \cdots & 
		\dfrac{\partial V(q_k)}{\partial q_k^P}
	\end{bmatrix}
	\in\mathbb R^{3P}.
\end{equation}
and the interaction force felt by the $i$-th particle \blue{$q_k^i$} is
\begin{equation}
	\frac{\partial V(q_k)}{\partial q_k^i} =  \sum_{i=1}^P
	\nabla V^{(o)}(q_k^i) + 
	\sum_{j\neq i}
	\nabla V^{(c)}(q_k^i-q_k^j)
\end{equation}
Therefore, the complexity of calculating each $\nabla V(q_k)$ is $O(P^2)$, and the complexity of calculating the full gradient $\nabla U_N(\q)$ is $O(NP^2)$.\par
\blue{It is worth noting that} the $O(P^2)$ computational cost per timestep is the common difficulty in the simulation of interacting particle systems, regardless of the thermostat scheme. As long as one  uses molecular dynamics approaches, heavy calculation of the interaction forces is inevitable. In the PIMD where an $N$-bead ring polymer is used, there are $N$ duplicates of the original \blue{interacting} system, and the computational cost becomes $O(NP^2)$.\par
Except for the interaction forces $\nabla V^{(c)}(q_k^i-q_k^j)$, calculation of the weight $W_N(\q)$ can also have $O(NP^2)$ time complexity in a single timestep.
For the position-dependent operator $\hat A = A(\hat q)$, if \blue{the analytic function $A(q)$ is given by}
\begin{equation}
	A(q) = \frac1P 
	\sum_{1\Le i<j\Le P}
	V^{(c)}(q^i-q^j),
\end{equation}
then the complexity to compute the weight function (\ref{eq:W_N position}) is $O(NP^2)$. \blue{Also,} the virial estimator (\ref{eq:W_N virial}) for the kinetic energy involves \blue{the full gradient} $\{\nabla V(q_k)\}_{k=1}^N$, thus the complexity is $O(NP^2)$.\par
The $O(NP^2)$ complexity in the calculation of the interaction forces and the weight is a huge impediment on efficient sampling, especially for large $P$. Hence we seek effective means to accelerate the simulation of the PIMD Langevin dynamics (\ref{dyn:Lang momentum}) by reducing the computational cost per timestep to $O(NP)$.
In the classical case, the Ewald summation \cite{ewald1} and FMM \cite{fmm1} have been successfully applied to reduce the complexity of interaction forces, but they are complicated to implement, cumbersome in the ring polymer representation, and only efficacious for specific interacting potentials, for example, the Coulomb potential.
Therefore, we aim to develop a method to reduce the $O(NP^2)$ complexity, which is easy-to-use, consistent with the PIMD framework, and applicable for any interaction potential $V^{(c)}(q)$. For this reason, we propose a sampling algorithm motivated by the recently proposed random batch method (RBM) \cite{rbm}.\par
In the RBM, the group of $P$ particles is randomly divided into small batches, where the division is chosen independently in different timesteps. During one timestep, the $P$ particles are restricted to interact within their own batches, and in this way the total complexity of interaction forces is reduced from $O(NP^2)$ to $O(NP)$. A more detailed description of incorporating the RBM in Langevin sampling is presented in Section \ref{sec:RBM}.
\subsection{Choice of the preconditioning method}
Before continuing the discussion on the RBM, we take a short detour to address another ubiquitous numerical issue in simulating the PIMD when $N$ is large. The ring polymer potential $U_N(\q)$ leads to a stiff term in the Langevin dynamics (\ref{dyn:Lang momentum}), which \blue{prevents} the use of large time steps in integrating the sampling path. To get a glimpse of the stiffness, note that the condition number of the matrix $L\in\mathbb R^{N\times N}$ defined in (\ref{eq:L}) is
\begin{equation}
	\mathrm{cond}(L) := 
	\frac{\lambda_{\max}(L)}{\lambda_{\min}(L)} = 
	\Big(\sin\frac\pi N\Big)^{-1},
\end{equation}
hence if the mass matrix is simply chosen as
\begin{equation}
	M = mI,
\end{equation}
the timestep $\Delta t$ needs to be small as $O(1/N)$ to integrate the highest oscillation mode of (\ref{dyn:Lang momentum}), which means the Langevin dynamics (\ref{dyn:Lang momentum}) shows stiffness when $N$ is large.\par
The numerical stiffness can be resolved by certain preconditioning methods, which is often realized by introducing a proper change of coordinates or choosing a proper mass matrix. For example, the staging coordinates transformation \cite{staging} given by
\begin{equation}
	\tilde q_1 = q_1,~~
	\tilde q_k = 
	q_k - \frac{(k-1)q_{k+1}+q_1}k,~~
	k=2,\cdots,N.
\end{equation}
has shown to be a powerful technique in relaxing the use of time steps.
However, the staging coordinates is not compatible with the RBM unless one repeatedly uses the staging coordinates transformation and its inverse \blue{(see appendix \ref{sec:precondition} for a detailed discussion)}, which are inconvenient for large systems. In fact, a preconditioning method which can be directly implemented in the physical coordinates is preferable to be combined with the RBM.\par
For this reason, we adopt the preconditioned mass-modified Langevin dynamics (pmmLang) \cite{precondition} as the preconditioning method, which we present in the following. \blue{In the Langevin dynamics (\ref{dyn:Lang momentum}),} we choose the mass matrix in such a specific form
\begin{equation}
	M = L^\alpha := L + \alpha I
\end{equation}
where the regularization parameter $\alpha>0$ is \blue{introduced to make sure $L^\alpha$ is positive definite}. Let
\begin{equation}
	\v = M^{-1}\p = (L^\alpha)^{-1}\p \in\mathbb R^{N\times 3P}
\end{equation}
be the velocity, and define the modified potential
\begin{equation}
	U^\alpha(\q) = \sum_{k=1}^N V(q_k) - \frac{\alpha}2 |\q|^2,
\end{equation}
\blue{then (13) becomes}
\begin{equation}
	\begin{aligned}
		\d\q & = \v\d t, \\
		\d\v & = - \q\d t - (L^\alpha)^{-1} \nabla U^\alpha(\q)\d t \\
		& \hspace{1.5cm} - \gamma\v\d t + \sqrt{\frac{2\gamma(L^\alpha)^{-1}}{\beta_N}}\d\B.
	\end{aligned}
	\label{dyn:pmmLang velocity}
\end{equation}
\blue{The obtained Langevin dynamics (\ref{dyn:pmmLang velocity}) is exactly the pmmLang we aim to derive.} Compared to the staging coordinates, the pmmLang can be conveniently implemented in the physical coordinates and is naturally compatible with the RBM. \blue{A detailed derivation of the pmmLang and the discussion on its implementation is given in Appendix \ref{sec:precondition}}.\par
\blue{Given the regularity parameter $\alpha>0$, }
the invariant distribution of the pmmLang (\ref{dyn:pmmLang velocity}) is
\begin{equation}
	\pi(\q,\v) \propto 
	\exp\bigg(-\beta_N
	\Big(
	\frac12\avg{\v,L^\alpha\v}_F + 
	\blue{\frac12\avg{\q,L^\alpha\q}_F + 
	U^\alpha(\q)}
	\Big)
	\bigg).
	\label{eq:pi(q,v)}
\end{equation}
\blue{Since the right hand sides of (\ref{eq:pi(q,p)}) and (\ref{eq:pi(q,v)}) are exactly the same except for the velocity transformation $\v = (L^\alpha)^{-1}\p$}, the pmmLang (\ref{dyn:pmmLang velocity}) produces the correct invariant distribution $(L^\alpha)^{-1}\pi(\q)$ in the marginal of $\q$. If we \blue{could} neglect the $\nabla U^\alpha(\q)$ term in (\ref{dyn:pmmLang velocity}), the Hamiltonian part of the pmmLang is simply
\begin{equation}
	\begin{aligned}
	\d\q & = \v\d t, \\
	\d\v & = -\q\d t,
	\end{aligned}
\end{equation}
thus there is no stiffness in (\ref{dyn:pmmLang velocity}). \blue{With further analysis, the pmmLang (\ref{dyn:pmmLang velocity}) can be shown to be a successful preconditioning method\cite{precondition} in the PIMD.\par
It is worth pointing out that the pmmLang has also been studied in some recent work \cite{hmc1,hmc2}, where the authors obtained a quantitative analysis of the convergence rate with its dependence on the regularization constant $\alpha$. 
A qualitative corollary of their result is that $\alpha$ being close to zero or too large both slowdowns the simulation efficiency of the pmmLang. To our point of view, if $\alpha$ is close to zero, the entries of the matrix $(L^\alpha)^{-1}$ in (\ref{dyn:pmmLang velocity}) will be large as $O(1/\alpha)$, hence the magnitude of $(L^\alpha)^{-1}\nabla U_N(\q)$ restricts the timestep for numerical stability.} For mathematical well-posedness, $\alpha$ should be chosen \blue{to ensure} that the modified potential $U^\alpha(\q)$ is confined, i.e.,
\begin{equation}
	\lim_{\q\rightarrow\infty} U^\alpha(\q) = +\infty.
\end{equation}
\blue{While it is not a trivial task to choose the optimal parameter $\alpha$ in the pmmLang, there are other non-preconditioned numerical integrators for the PIMD Langevin dynamics which are parameter free \cite{nonprecondition0,nonprecondition3}. Nevertheless, how to choose $\alpha$ is not the focus of this work, and we employ the pmmLang as a preconditioning method in the PIMD under certain simplified conditions given below.}\par
We assume the external potential $V^{(o)}(q)$ in the quantum system (\ref{eq:V}) is harmonic, i.e., blue{for some constant $\alpha_0>0$ we have}
\begin{equation}
	V^{(o)}(q) = \frac{\alpha_0}2|q|^2,~~~~q\in\mathbb R^3
	\label{eq:V^o}
\end{equation}
We choose regularization parameter $\alpha>0$ to coincide with the constant $\alpha_0$ in (\ref{eq:V^o}), thus \blue{the modified potential} $U^\alpha(\q)$ is simply the sum of all \blue{interacting potentials},
\begin{equation}
	U^\alpha(\q) = \sum_{k=1}^N
	\sum_{1\Le i<j\Le N}
	V^{(c)}(q_k^i-q_k^j),
	\label{eq:U^alpha}
\end{equation}
and the pmmLang (\ref{dyn:pmmLang velocity}) can be equivalently written as

\begin{widetext}
\begin{equation}
	\begin{aligned}
		\d\q^i & = \v^i\d t, \\
		\d\v^i & = -\q^i\d t - (L^\alpha)^{-1}\sum_{j\neq i} \nabla V^{(c)}(\q^i-\q^j)  \d t -\gamma\v^i \d t + \sqrt{\frac{2\gamma(L^\alpha)^{-1}}{\beta_N}}\d\B^i.
	\end{aligned}
	~~~(i=1,\cdots,P)
	\label{dyn:pmmLang}
\end{equation}
\end{widetext}
\mbox{}\indent
For simplicity, the incorporation of the RBM will be discussed only for (\ref{dyn:pmmLang}), \blue{under the assumption of (\ref{eq:V^o})}. Clearly, the RBM can be applied without (\ref{eq:V^o}), and there will be additional terms in $U^\alpha(\q)$ \blue{except the interacting potentials in (\ref{eq:U^alpha})}. These additional terms contributes only $O(NP)$ complexity in a single timestep and does not change the major difficulties, since the $O(NP^2)$ complexity is only due to the interaction forces $\nabla V^{(c)}(q_k^i-q_k^j)$.
\section{Random Batch Method in Langevin Sampling}
\label{sec:RBM}
\subsection{RBM for the pmmLang}
In the simulation of large interacting particle systems, high computational cost per timestep has always been an impediment to efficient sampling. With the setup in the previous sections, to integrate the sampling trajectory for a single timestep, it requires a cost of $O(NP^2)$ complexity to compute all the interaction forces $V^{(c)}(\q^i-\q^j)$. Since the calculation of the interaction forces is rather expensive when $P$ is large, reducing the complexity due to interaction forces is crucial to attain high efficiency in the PIMD simulation of interacting particle systems.
\par
The recently proposed random batch method (RBM) provides a simple approach to fulfill such a reduction of cost in force evaluation. The RBM avoids calculation of the full interaction by randomly dividing the group of $P$ particles into small-size batches, and only allowing interactions within each small batch with adjusted interaction strength. Since for each timestep, a new random division is conducted, after sufficiently many times, each particle will not only has an equal probability to interact with any other particles in the group, but also will have plenty of random interactions with the rest of the particles.
This is why we expect the RBM to produce the correct statistical properties of the interacting system in the long time simulation. One does not necessarily need to trace the interactions between a typical particle with every other particle to have an accurate prediction of the statistical properties, but rather, the interactions with other particles may vastly cancel each other and result in an average field to the typical particle. For heuristic purpose, we present the RBM for the pmmLang (\ref{dyn:pmmLang}) in the following.\par
During each timestep, we first randomly divide the group of $P$ particles into $n$ batches $\mathcal C_1,\cdots,\mathcal C_q$, where each batch is of size $p$ and $n = P/p$. The batch size $p$ should be far less than $P$ to avoid massive calculation of interaction forces, and be at least two to capture the pairwise interactions $\nabla V^{(c)}(\q^i-\q^j)$.
Then we approximate the full interaction of the $P$-particle system with the interaction forces within each small batch $\mathcal C_l$, and adjust the interaction strength to ensure statistical consistency.
To be specific, the interaction force felt by the $i$-th particle $\q^i\in\mathbb R^{N\times3P}$ in the original pmmLang (\ref{dyn:pmmLang}) is
\begin{equation}
	\sum_{j\neq i}
	\nabla V^{(c)}(\q^i-\q^j)
	\in
	\mathbb R^{N\times3},
	\label{eq:full force}
\end{equation}
With the use of RBM, each $\q^i$ is assigned to a batch $\mathcal C_l$ for some $l=1,\cdots,q$, and hence the approximation of the interaction force (\ref{eq:full force}) within $\mathcal C_l$ is
\begin{equation}
	\frac{P-1}{p-1}
	\sum_{j\in\mathcal C_l,j\neq i}
	\nabla V^{(c)}(\q^i-\q^j).
	\label{eq:batch force}
\end{equation}
Recall that the division $\mathcal C_1,\cdots,\mathcal C_q$ is randomly generated for each timestep, thus for fixed $\q^i$, the batch $\mathcal C_l$ which contains $\q^i$ is also random. Therefore, (\ref{eq:batch force}) can be interpreted as a $(p-1)$-term sample approximation of a $(P-1)$-term summation (\ref{eq:full force}). Also, the approximation (\ref{eq:batch force}) is unbiased, i.e.,
\begin{equation}
	\mathbb E\bigg(
	\frac{P-1}{p-1}
	\sum_{j\in\mathcal C_l,j\neq i}
	\nabla V^{(c)}(\q^i-\q^j)
	\bigg) = 
	\sum_{j\neq i}
	\nabla V^{(c)}(\q^i-\q^j).
\end{equation}
In this way, the random-batch approximated pmmLang within the batch $\mathcal C_l$ is given by

\begin{widetext}
\begin{equation}
	\begin{aligned}
		\d\q^i & = \v^i\d t, \\
		\d\v^i & = -\q^i\d t -
		\frac{P-1}{p-1}
		(L^\alpha)^{-1}\sum_{j\in\mathcal C_l,j\neq i} \nabla V^{(c)}(\q^i-\q^j)  \d t -\gamma\v^i \d t + \sqrt{\frac{2\gamma(L^\alpha)^{-1}}{\beta_N}}\d\B^i,
	\end{aligned}
	~~~(i\in\mathcal C_l)
	\label{dyn:pmmLang batch}
\end{equation}
\end{widetext}
\noindent
and the RBM for the pmmLang (\blue{denoted by} the pmmLang+RBM) in a single timestep $\Delta t$ just evolves (\ref{dyn:pmmLang batch}) for all batches $\mathcal C_l$, $l=1,\cdots,n$, as shown in Algorithm \ref{dyn:pmmLang RBM}. In consecutive timesteps, the previous divisions are discarded and new sets of divisions are randomly chosen. Hence a fixed particle $\q^i\in\mathbb R^{N\times3}$ can be assigned to different batches $\mathcal C_l$ in different timesteps, and only interacts with the particles in the current batch.\par
\begin{algorithm*}
	\caption{RBM for the pmmLang (\ref{dyn:pmmLang}) in a timestep $\Delta t$}
	Randomly divide the $P$ particles into $n$ batches $\mathcal C_1,\cdots,\mathcal C_n$ of size $p$, where $n=P/p$. \\
	\For{$l=1,\cdots,n$}{
		\mbox{}\\
		Evolve the pmmLang within the batch $\mathcal C_l$ in a timestep $\Delta t$:
		$$
		\begin{aligned}
			\d\q^i & = \v^i\d t, \\
			\d\v^i & = -\q^i\d t - \frac{P-1}{p-1}(L^\alpha)^{-1}\sum_{j\in\mathcal{C}_l,j\neq i}\nabla V^{(c)}(\q^i-\q^j)\d t
			-\gamma\v^i \d t + \sqrt{\frac{2\gamma(L^\alpha)^{-1}}{\beta_N}}\d\B.
		\end{aligned}~~~~
		(i\in\mathcal{C}_l)
		$$
	}
	\label{dyn:pmmLang RBM}
\end{algorithm*}
In Algorithm \ref{dyn:pmmLang RBM}, to generate a random division $\mathcal C_1,\cdots,\mathcal C_q$, one may use the Fisher–Yates shuffle \cite{shuffle} with $O(P)$ complexity to obtain a random permutation of $1,\cdots,P$ and then divide it into $n$ batches in order.
Note that there are $\frac{p(p-1)}2$ pairs of interactions within each batch $\mathcal C_l$, the complexity of Algorithm \ref{dyn:pmmLang} due to interaction forces is
\begin{equation}
	n \times
	N\times \frac{p(p-1)}2
	 = 
	 N\times \frac{P(p-1)}2
	  = O(NPp),
\end{equation}
hence the batch size $p$ should be small to attain high efficiency in the pmmLang+RBM. In particular, if $p$ is chosen as small integers such as $2$ and $4$, the complexity due to interaction forces is only $O(NP)$. In this way, we can employ the pmmLang+RBM (Algorithm \ref{dyn:pmmLang RBM}), which is much more efficient than the original pmmLang (\ref{dyn:pmmLang}), to compute the ensemble average $\avg{W_N(\q)}_\pi$.
\par
Finally, we point out that the RBM has an analogy to the stochastic gradient descent (SGD) \cite{sgd,sgd2,sgd3,mbgd}, the stochastic optimization method widely used in machine learning. 
In machine learning problems, the size of the data set can be extremely large, and the SGD approximates the full gradient with the average calculated from samples in a randomly chosen small batch.
Although the RBM and the SGD shares the idea of full gradient approximation, the high dimensional natures of their underlying problems are different. In an optimization problem where the SGD applies, the dimension of the parameter space does not increases with respect to the size of the data set. However, in an interacting particle system where the RBM applies, the dimension of the \blue{coordinate} space growth linearly with the number of particles \blue{$P$}.
\subsection{Error analysis in PIMD with the RBM}
In this section, we aim to investigate the error in calculating the ensemble average $\avg{W_N(\q)}_\pi$ due to the use of the RBM.
By the ergodicity of the pmmLang (\ref{dyn:pmmLang}), $\avg{W_N(\q)}_\pi$ can be accurately computed from the time average
\begin{equation}
	\frac1J \sum_{j=1}^J 
	W_N(\q(j\Delta t))
	\label{avg:pmmLang}
\end{equation}
as the number of samples $J\rightarrow\infty$,
where $\{\q(t)\}_{t\Ge0}$ is the pmmLang trajectory evolved by (\ref{dyn:pmmLang}) and $\Delta t$ is the timestep specified in Algorithm \ref{dyn:pmmLang RBM}.
Correspondingly, the pmmLang+RBM uses the following time average
\begin{equation}
	\frac1J \sum_{j=1}^J 
	W_N(\tilde\q(j\Delta t)))
	\label{avg:pmmLang RBM}
\end{equation}
to estimate $\avg{W_N(\q)}_\pi$,
where $\{\tilde\q(t)\}_{t\Ge0}$ is the pmmLang+RBM trajectory evolved by Algorithm \ref{dyn:pmmLang RBM}. Therefore, the error analysis in the RBM is actually asking whether the time average (\ref{avg:pmmLang RBM}) is a good estimator of $\avg{W_N(\q)}_\pi$ for sufficiently large $J$.\par
In general, exploring the effects of the consecutive uses of random-batch approximations (\ref{eq:batch force}) along the sampling process is difficult. To our knowledge, the strong error in a contracting dynamics \cite{rbm} and the weak error in a short period \cite{rbm3} have been rigorously justified, whereas none of those theoretical results apply to the RBM in the long time sampling process.\par
If the adoption of the RBM results in a small strong error in the long time simulation, i.e., the pmmLang+RBM trajectory $\tilde\q(t)$ almost coincides with the pmmLang trajectory $\q(t)$ all the time, we can certainly deduce that the RBM also leads to a small error in the ensemble average calculation. However, this is not true. As we shall show in the following, it is likely that the pmmLang+RBM trajectory $\tilde\q(t)$ drifts significantly apart from $\q(t)$ in a short time, but still $\tilde\q(t)$ provides a faithful and accurate approximation of $\avg{W_N(\q)}_\pi$ in the long time simulation. We shall further rationalize the error induced by the use of the RBM in the PIMD in the following.
\subsubsection{Strong error analysis}
The strong error characterizes the deviation of the numerically computed trajectory from the exact one in form of the mean squared error \cite{stochastic}. The strong error of the pmmLang+RBM is then defined as
\begin{equation}
e(t) = \sqrt{\mathbb E |\tilde\q(t)-\q(t)|^2},~~~~
t\Ge0
\label{err:strong}
\end{equation}
where we assume the pmmLang+RBM trajectory $\tilde\q(t)$ and the pmmLang trajectory $\q(t)$ are driven by the same Brownian motion $\B(t)$ and the same initial state. In addition, the expectation is taken over all possible Brownian motions and choices of random batches in the time interval $[0,t]$. When (\ref{dyn:pmmLang})(\ref{dyn:pmmLang batch}) are integrated exactly, the deviation of $\tilde\q(t)$ from $\q(t)$ is solely due to the use of the random batches.\par
According to Algorithm \ref{dyn:pmmLang RBM}, the pmmLang+RBM trajectory $\tilde\q(t)$ coincides with $\q(t)$ when the batch size $p=P$ or in the limit $\Delta t\rightarrow0$.
In fact, the batch size $p=P$ implies the interaction forces are calculated accurately, while in the limit $\Delta t\rightarrow0$ each particle is supposed to be driven by the average effects. Nevertheless, such parameters are impractical to apply in the pmmLang+RBM, because choosing either $p=P$ or $\Delta t\rightarrow0$ greatly increases the total computation cost and contradicts with our original intention to use the RBM. To be specific, the cost due to interaction forces of the pmmLang+RBM in the time interval $[0,T]$ is
$$
	 O\bigg(
	\frac{p}{\Delta t}
	TP^2
	\bigg),
$$
which is extremely large when $p=P$ or $\Delta t\rightarrow0$.\par
When the parameters $p,\Delta t$ are chosen such that the pmmLang+RBM becomes an efficient sampling method, i.e., the batch size $p$ is small and the timestep $\Delta t$ is relatively large, we cannot expect the pmmLang+RBM to produce the accurate pmmLang trajectory $\q(t)$. In contrast, the strong error $e(t)$ shortly grows large in \blue{a short time period} , as we show in the following example.\par
\begin{figure}
	\centering
	\includegraphics[width=0.48\textwidth]{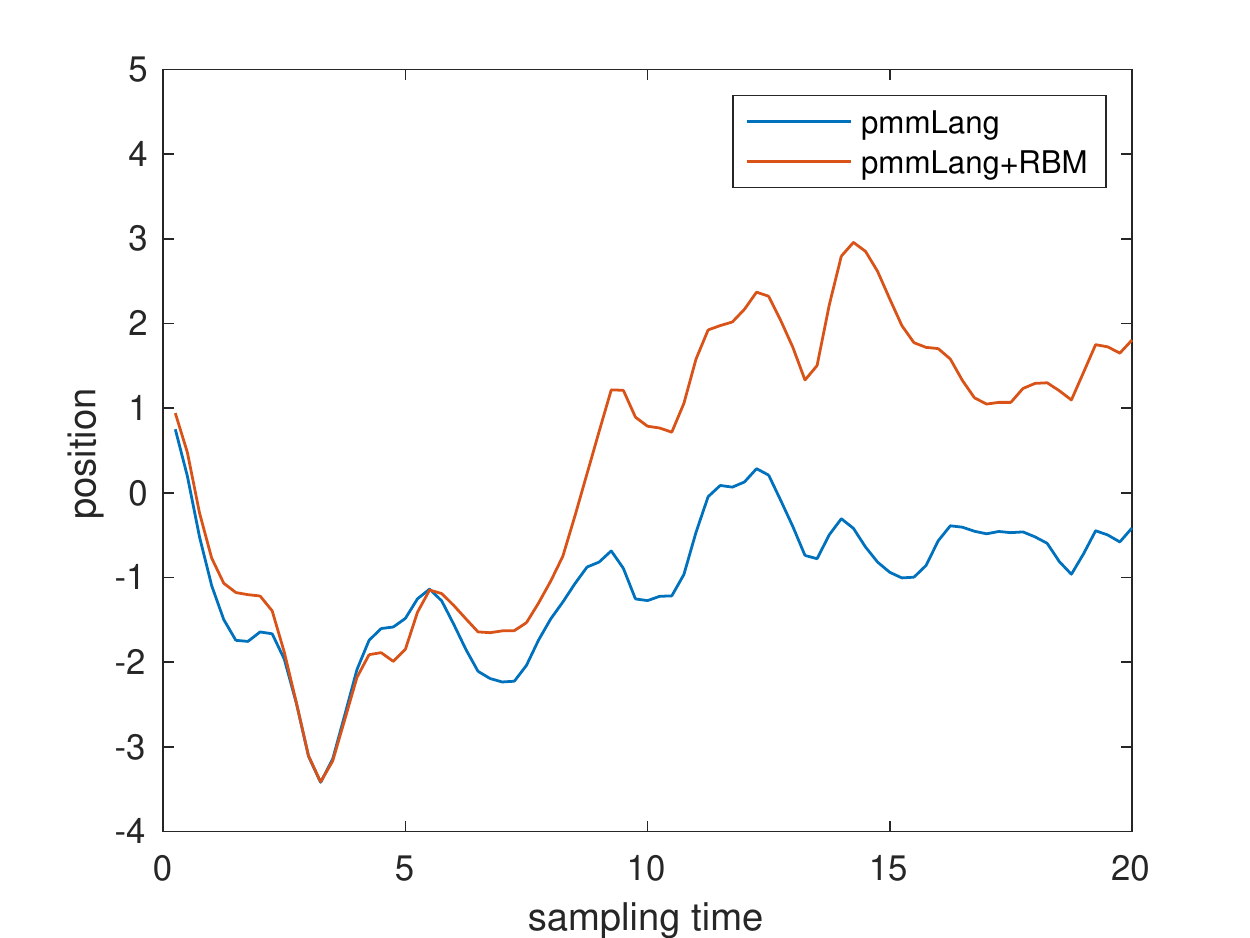}
	\includegraphics[width=0.48\textwidth]{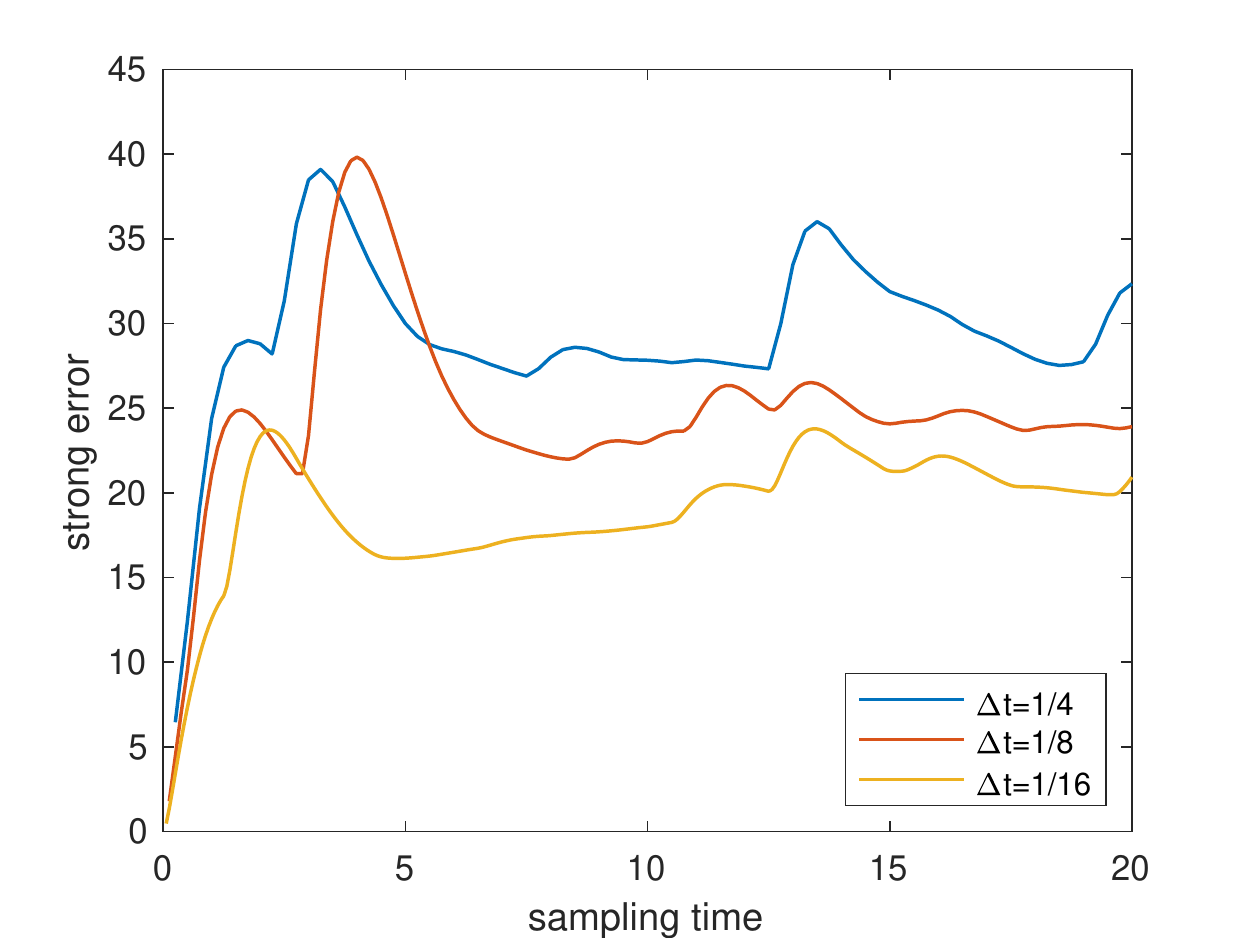}
	\caption{Top: typical samples of the pmmLang trajectory $\q(t)$ and the pmmLang+RBM trajectory $\tilde\q(t)$ with the timestep $\Delta t = 1/4$. Only the first component of $\q\in\mathbb R^{N\times3P}$ is plotted.
	Bottom: the strong error $e(t)$ defined in (\ref{err:strong}) with $\Delta t = 1/4,1/8,1/16$. 10000 independent trajectories are used to compute the expectations.
	In the Coulomb interacting system,
	the mass $m=1$, the inverse temperature $\beta=4$, the number of particles $P=8$, the number of beads $N=32$, the total sampling time $T=20$ and the batch size $p=2$.}
	\label{fig:strong}
\end{figure}

In the Coulomb interacting system (\blue{the potential function is defined in (\ref{Coulomb})} in Section 
\ref{sec:tests}), we show in Figure \ref{fig:strong} how the pmmLang+RBM trajectory $\tilde\q(t)$ deviates from $\q(t)$ and the strong error $e(t)$ grows large along the sampling process.
It can be seen from Figure \ref{fig:strong} that the pmmLang+RBM trajectory $\tilde\q(t)$ drifts apart from $\q(t)$ at about $t\approx5$, and the strong error $e(t)$ grows large \blue{at $t\approx1$}, no matter how small the timestep $\Delta t$ is.
Therefore, there will always be large deviation in the pmmLang+RBM trajectory $\tilde\q(t)$ from the pmmLang trajectory $\q(t)$, unless one chooses $p=P$ or $\Delta t\rightarrow0$. As we shall show in the next, requiring a small path-wise error is totally unnecessary in thermal average calculation.
\subsubsection{Weak error analysis}
\begin{figure}
	\centering
	\includegraphics[width=0.48\textwidth]{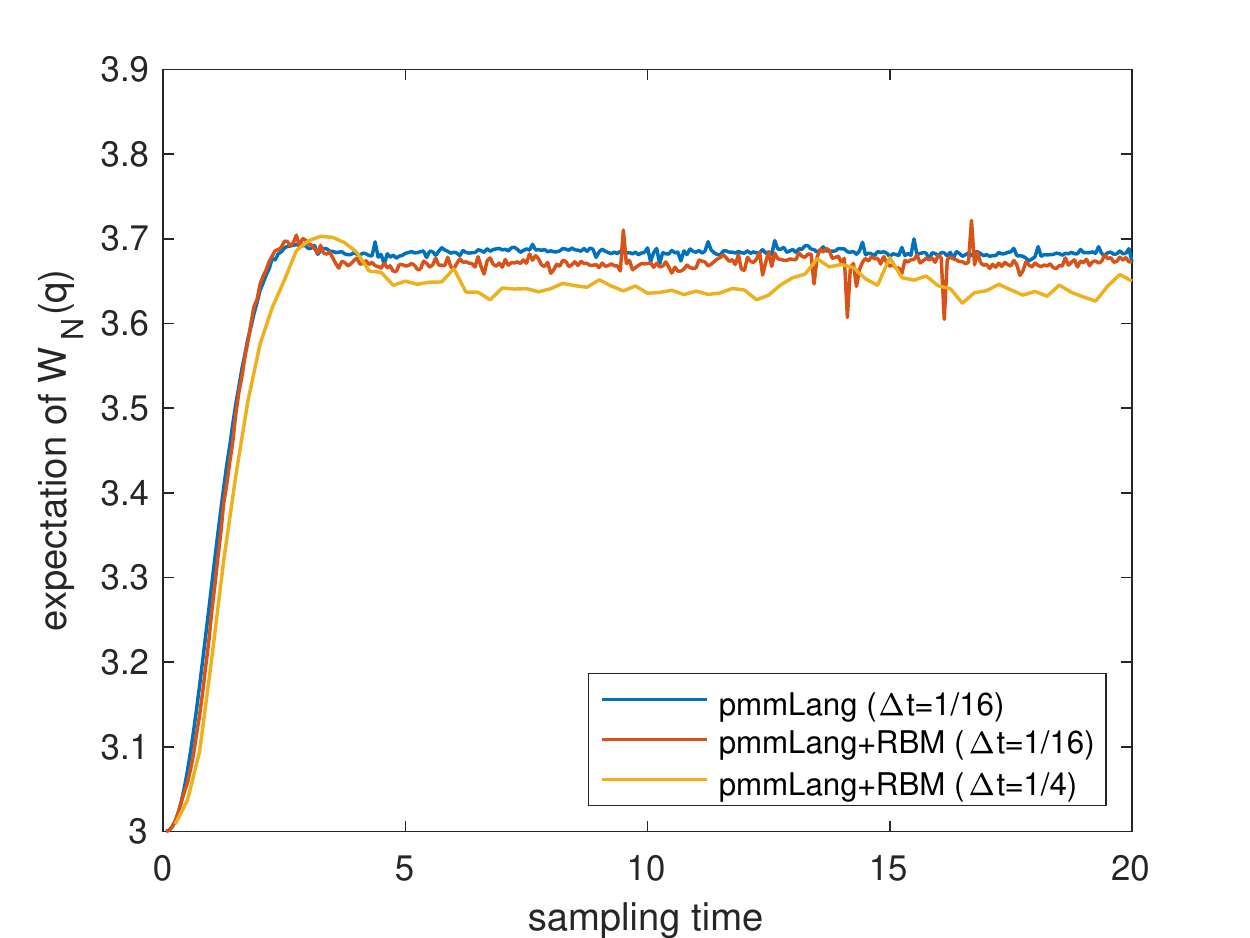}
	\caption{
	Expectations $\mathbb EW_N(\q(t))$ and $\mathbb EW_N(\tilde\q(t))$ computed by the pmmLang and the pmmLang+RBM in the Coulomb interacting system, \blue{where the observable operator is the kinetic energy.}
	The mass $m=1$, the inverse temperature $\beta = 4$, the number of particles $P=8$, the number of beads $N=32$, the timestep $\Delta t = 1/4,1/16$, the total sampling time $T=20$ and the batch size $p=2$. 10000 independent trajectories are used to compute the expectations. The blue curve is associated with the pmmLang with $\Delta t = 1/16$, and the red and yellow curves are associated with the pmmLang+RBM with $\Delta t = 1/16$ and $1/4$ respectively.
	The difference between the curves of the pmmLang and the pmmLang+RBM is exactly the weak error $e(t)$ defined in (\ref{err:weak}).}
	\label{fig:weak}
\end{figure}

The weak error measures the difference of the approximate and the exact values of the ensemble average \cite{stochastic}. In the PIMD, the weight function of interest is $W_N(\q)$, which takes different forms for different observables. To be specific, the weak error of the pmmLang+RBM is defined as
\begin{equation}
	e(t) = \mathbb E W_N(\tilde\q(t)) - \mathbb E W_N(\q(t)),
	\label{err:weak}
\end{equation}
where the expectation is taken over all possible Brownian motions and choices of random batches, \blue{where the observable operator is the kinetic energy, and the corresponding weight function of interest is the virial estimator (\ref{eq:W_N virial}).} To give a first impression of the weak error, we plot in Figure \ref{fig:weak} the expectations $\mathbb EW_N(\q(t))$ and $\mathbb E W_N(\tilde\q(t))$ with various timesteps in the Coulomb interacting system.
Unlike the strong error (\ref{err:strong}), the curve of $\mathbb E W_N(\tilde\q(t))$ associated with the pmmLang+RBM is very close to the curve of $\mathbb EW_N(\q(t))$ associated with the pmmLang, and the weak error (\ref{err:weak}) remains small all along the sampling process.
It can be seen from Figure \ref{fig:weak} that the ring polymer system quickly goes into equilibrium about $t\approx2.5$, for both the pmmLang and the pmmLang+RBM dynamics, hence the introduction of the batch force approximation (\ref{eq:batch force}) does not influence the convergence mechanism of the original pmmLang (\ref{dyn:pmmLang}).
Also, the stochastic error in computing $W_N(\q(t))$ due to the use of the force approximation (\ref{eq:batch force}) cancels over all possible choices of random batches, which finally results in a small bias on the average $\mathbb EW_N(\q(t))$, even if the batch size $p$ is small and the timestep $\Delta t$ is relatively large.\par
However, in the long time simulation it's impossible to use $\mathbb EW_N(\q(t))$ or $\mathbb EW_N(\tilde\q(t))$ to estimate $\avg{W_N(\q)}_\pi$, since the calculation of the expectations in (\ref{err:weak}) requires a large number of sampling trajectories. In practice, it is more feasible to use the time averages (\ref{avg:pmmLang})(\ref{avg:pmmLang RBM}) to compute $\avg{W_N(\q)}_\pi$, which are obtained by numerically integrating one sampling path, respectively.\par
If the numerical error is neglected, we know that due to the ergodicity of the Langevin dynamics, the time average (\ref{avg:pmmLang}) generated by the pmmLang converges to the correct limit as the number of samples $J\rightarrow\infty$, i.e.
\begin{equation}
	\lim_{J\rightarrow\infty}
	\frac1J
	\sum_{j=1}^J 
	W_N(\q(j\Delta t)) = 
	\avg{W_N(\q)}_\pi.
	\label{eq:limit pmmLang}
\end{equation} 
We can actually view such a convergence from the perspective of the relative entropy. Recall that we aim to sample  the Boltzmann distribution $\pi(\q)$ as in (\ref{eq:pi(q)}) and for any distribution $f(\q)$, we can use the relative entropy \cite{re1,re2}
\begin{equation}
D(f||\pi)= \mathbb E^f \left( \log\frac{f}{\pi} \right)
\end{equation}
to measure how much the distribution $f$ deviates from the target distribution $\pi$. The relative entropy is always nonnegative, and it is zero only when the two distributions are identical. Note that we can define a family of the empirical distributions from the samples generated by the pmmLang, which are given by
\begin{equation}
	\mu_J(\q) = 
	\frac1J 
	\sum_{j=1}^J \delta(\q-\q(j\Delta t)).
\end{equation}
Thus, the convergence (\ref{eq:limit pmmLang}) can be interpreted as 
\begin{equation}
\lim_{J \rightarrow \infty} D( \mu_J ||\pi) =0.
\end{equation}
The diminishing of the relative entropy manifests the time irreversibility and the ergodicity of the sampling trajectory. That is, the thermostat effect of the Langevin dynamics brings in dissipation to the relative entropy with respect to the invariant measure, such that the empirical measure $\mu_J(\q)$ converges to the target Boltzmann distribution $\pi(\q)$ in the weak sense.\par
However, when the random batches are used along a sampling path, the empirical distributions are repeated perturbed, mostly likely away from the Boltzmann distribution. In fact, when the random batches are used for each time, although the force can be viewed as an unbiased approximation from the total interaction force, the sampling path deviates from the Langevin dynamics without the use of random batches with a deterministic bias for each specific choice of the random divisions.  Let us denote the empirical measure of the pmmLang+RBM by $\tilde\mu_J(\q)$, which is defined by
\begin{equation}
	\tilde\mu_J(\q) = 
	\frac1J 
	\sum_{j=1}^J \delta(\q-\tilde\q(j\Delta t)).
\end{equation}
Because the random divisions are chosen independently, we cannot expect $\tilde\mu_J(\q)$ to converge as $J \rightarrow \infty$. However, as we shall demonstrate with the following numerical example, the accumulation of the bias for each use of the random batches does not add up to an unbounded error. On the contrary, the perturbations of the interaction forces along the sampling path only results in a small bias in $\tilde\mu_J(\q)$ from the target distribution $\pi(\q)$.\par
\begin{figure}
	\includegraphics[width=0.48\textwidth]{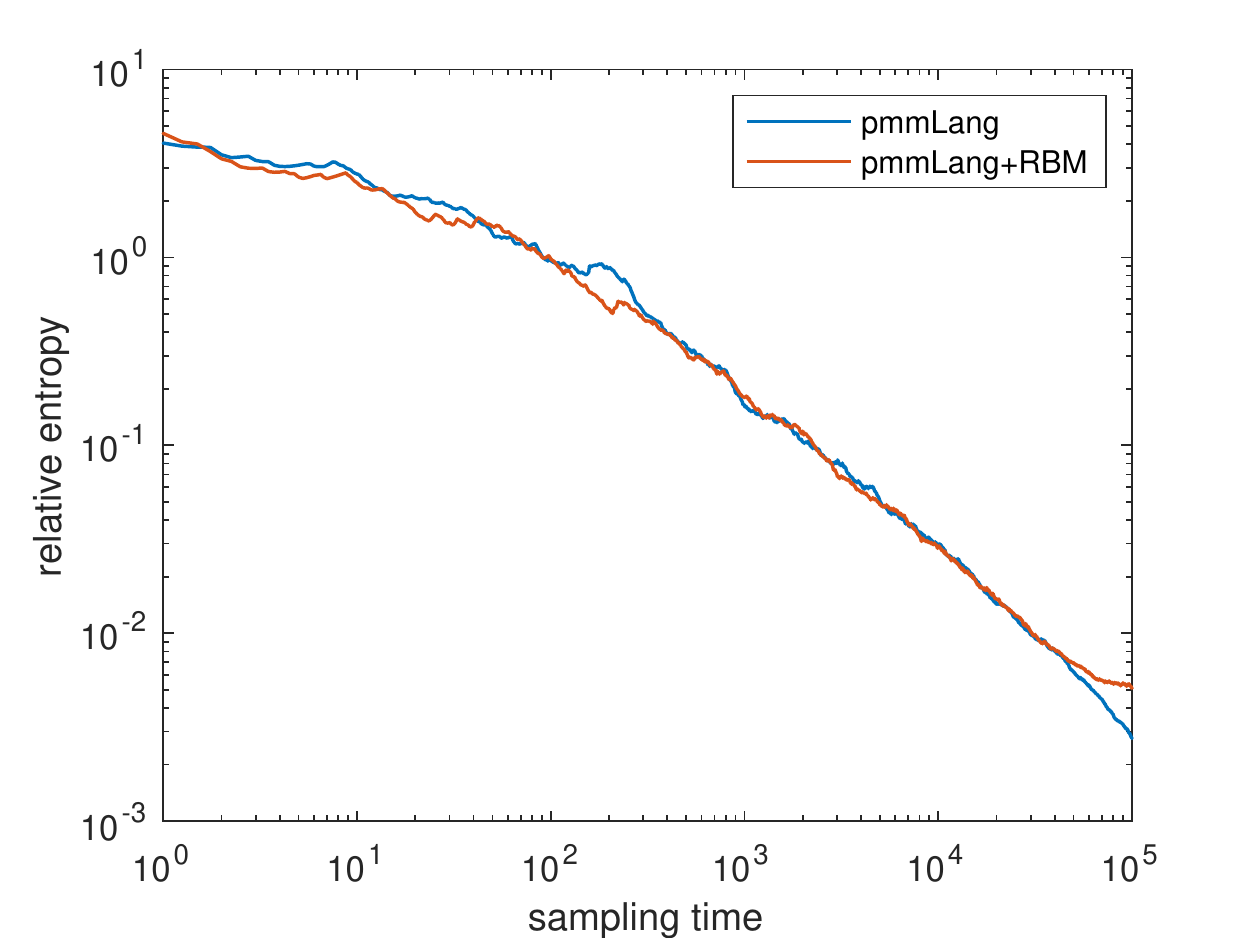}
	\includegraphics[width=0.48\textwidth]{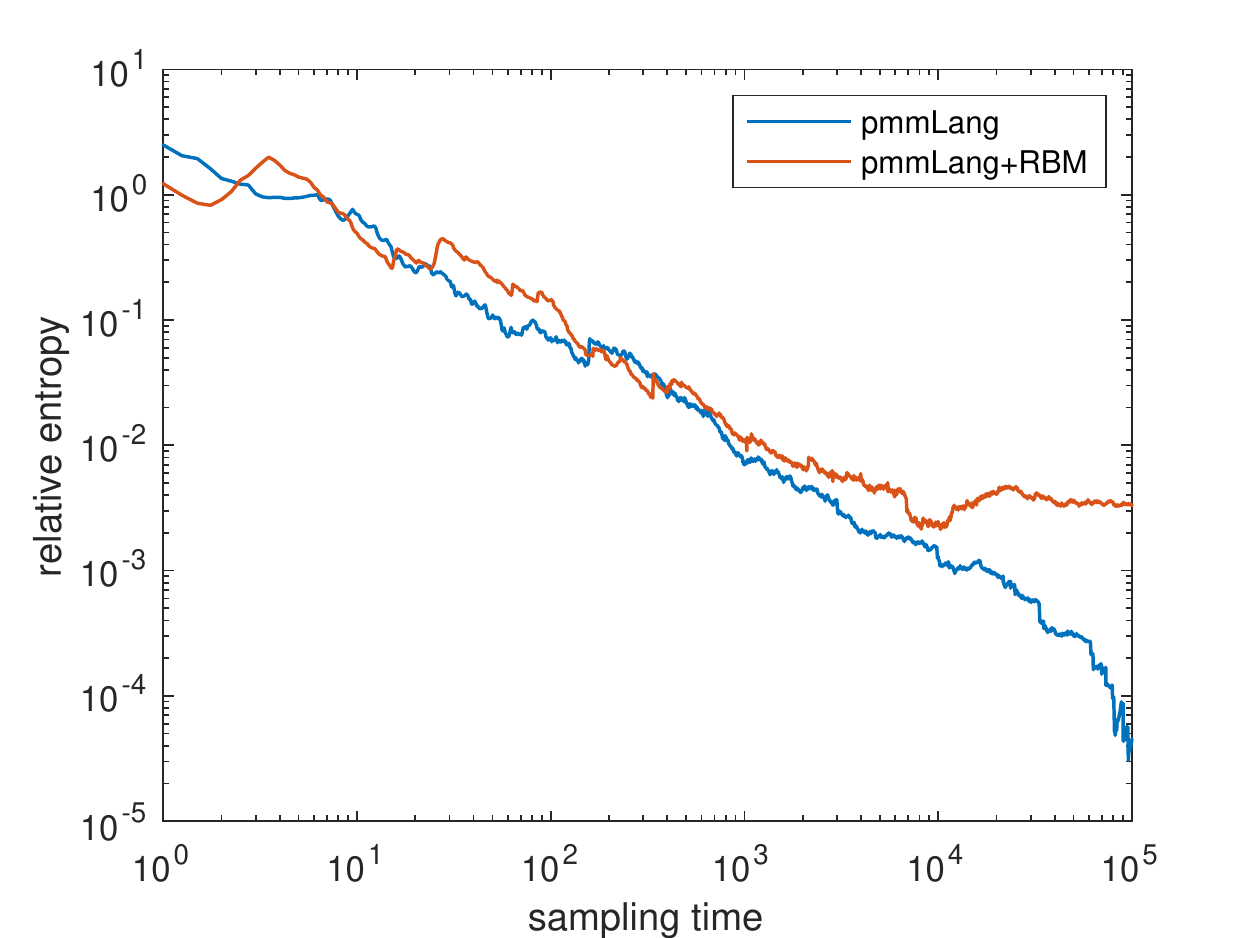}
	\caption{The relative entropy in the numerical simulation of the Coulomb interacting system. The top and bottom figures are associated with the observables in the position $q_1^1\in\mathbb R^3$ and the pairwise distance $|q_1^1-q_1^2|\in\mathbb R$ respectively.
	The mass $m=1$, the inverse temperature $\beta = 4$, the number of particles $P=8$, the number of beads $N=32$, the timestep $\Delta t = 1/4$, the total sampling time \blue{$T=10^5$} and the batch size $p=2$. The target distribution $\pi(\q)$ is computed with \blue{$T=5\times10^5$}.}
	\label{fig:relative entropy}
\end{figure}

We present a numerical experiment to observe the difference between $\tilde\mu_J(\q)$ and $\pi(\q)$ for sufficiently large $J$.
Note that $\pi(\q)$ is defined over the high-dimensional coordinate space $\mathbb R^{N\times3P}$, it's intractable to directly compute $D(\tilde\mu_J||\pi)$, hence we instead numerically simulate the distributions of certain observables associated with $\tilde\mu_J(\q)$ and $\pi(\q)$. 
In this example, we choose the observable operator to be the position $q_1^1\in\mathbb R^3$ and the pairwise distance $|q_1^1-q_1^2|\in\mathbb R$. Recall that the superscript denotes the index of particles, while the subscript denotes the index of beads. We plot the relative entropy between $\tilde\mu_J(\q)$ and $\pi(\q)$ in the observables along the sampling process in Figure \ref{fig:relative entropy}. As a comparison, the relative entropy for $\mu_J(\q)$, the empirical measure of the pmmLang, is also plotted.
\par
We observe from Figure \ref{fig:relative entropy} that the empirical measure of the pmmLang+RBM converges to $\pi(\q)$ as fast as the pmmLang when the sampling time \blue{$T<10^4$}. However, as \blue{$T>10^4$}, there is a significant slowdown in the convergence of the pmmLang+RBM, \blue{especially in the bottom figure where the observable is the pairwise distance $|q_1^1-q_1^2|\in\mathbb R$}. Finally, there is a certain bias of the empirical measure $\tilde\mu_J(\q)$ from the target distribution $\pi(\q)$. \blue{The slowndown in the bottom panel of Figure \ref{fig:relative entropy} is more noticeable than the top one, because the RBM modifies the interaction forces and directly impacts the calculation of the pairwise distance. Meanwhile, the RBM has relatively less influence on the marginal distribution of a single particle.}\par
\begin{figure}
	\centering
	\includegraphics[width=0.48\textwidth]{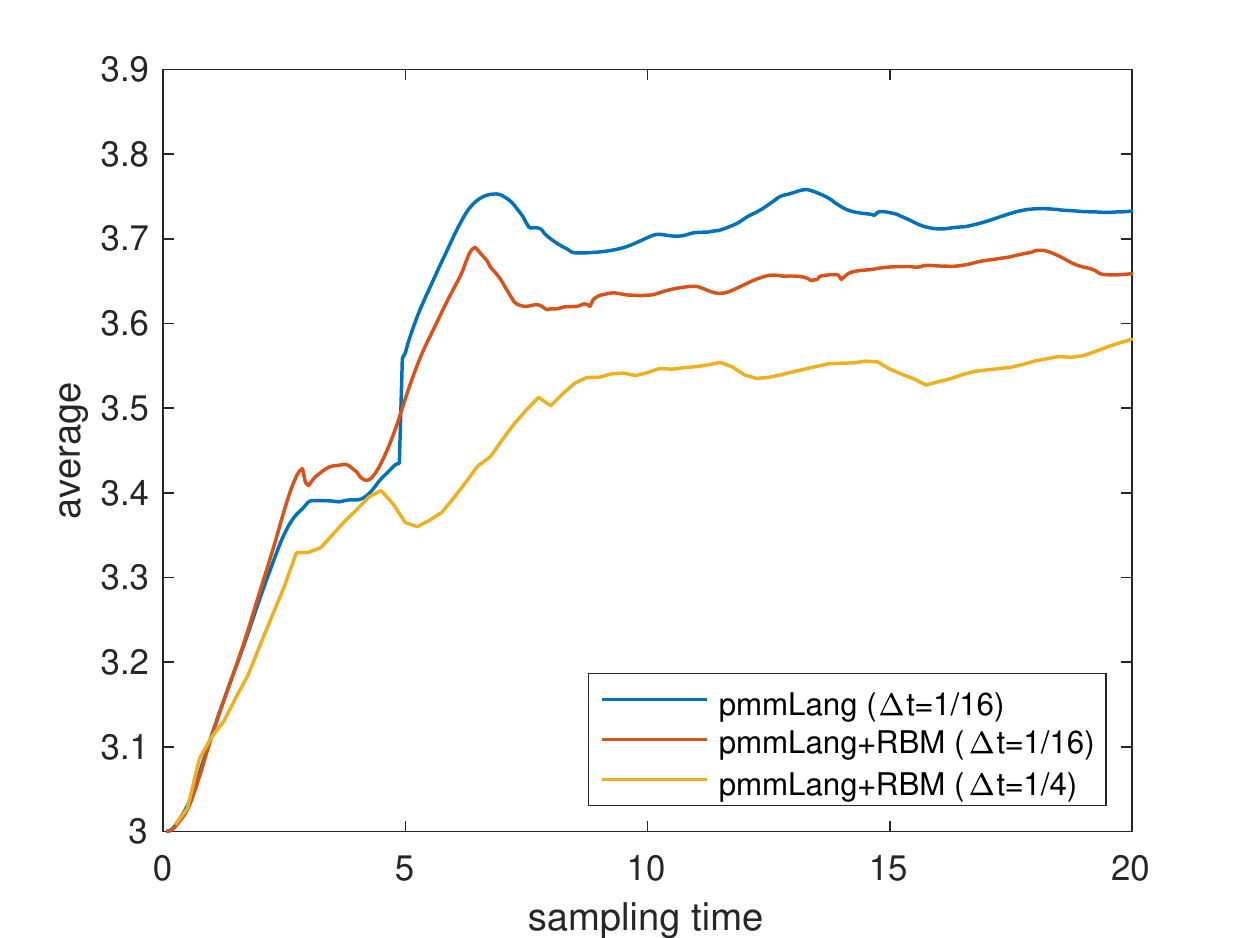}
	\includegraphics[width=0.48\textwidth]{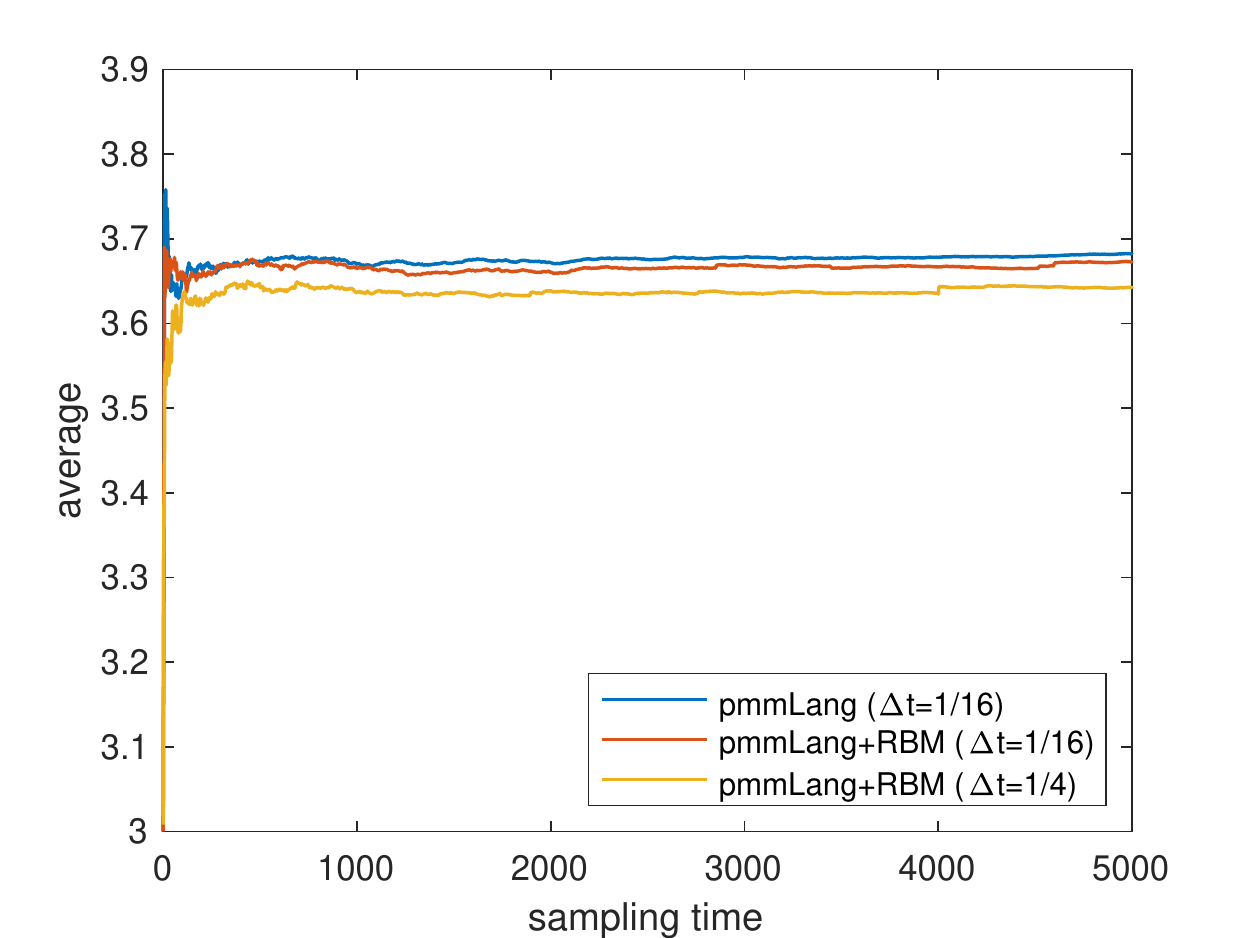}
	\caption{The pmmLang time average (\ref{avg:pmmLang}) and the pmmLang+RBM time average (\ref{avg:pmmLang RBM}) in the Coulomb interacting system, where the observable operator is the kinetic energy.
	The top and bottom figure show the two phases of the Langevin sampling respectively.
	The mass $m=1$, the inverse temperature $\beta = 4$, the number of particles $P=8$, the number of beads $N=32$, the timestep $\Delta t = 1/4,1/16$, the total sampling time $T = 5000$ and the batch size $p=2$.}
	\label{fig:observable}
\end{figure}

The convergence of $\tilde\mu_J(\q)$ to $\pi(\q)$ also implies the convergence of the time average (\ref{avg:pmmLang RBM}) to $\avg{W_N(\q)}_\pi$, which can be verified in the following example.
We plot the time averages (\ref{avg:pmmLang})(\ref{avg:pmmLang RBM}) of the pmmLang and the pmmLang+RBM in Figure \ref{fig:observable}, where the observable operator is the kinetic energy.
It can be seen from Figure \ref{fig:observable} that although (\ref{avg:pmmLang RBM}) deviates from (\ref{avg:pmmLang}) at the first phase of the sampling process, (\ref{avg:pmmLang RBM}) gradually converges to a fixed limit as time evolves. Furthermore, the bias of (\ref{avg:pmmLang}) from $\avg{\hat A}$ is small and diminishes if we shrink the timestep.\par
To understand the the approximation property of the pmmLang+RBM, we present a heuristic explanation on the decay of the relative entropy $D(\tilde\mu_J||\pi)$. Our key observation lies in the fact that the use of the random batches does not affect the thermostat part of the Langevin dynamics. And thus, the pmmLang+RBM can be viewed as an integrated process, where the thermostat part is constantly driving the empirical distribution approaching the target measure and the use of random batches gives rise to a sequence of random perturbations which presents the convergences of the empirical distribution. The numerical results strong suggests that the thermostat effect is dominating the random perturbations, and as a consequence, although the empirical distribution $\tilde\mu_J(\q)$ does not converge, it stays in a relatively small vicinity of the target distribution  $\pi(\q)$ for $J$ sufficiently large. To further analyze the dominance of the thermostat in such an integrated process, we continue the discussion in two distinct phases of sampling.\par
In the first phase of sampling, the samples generated by the Langevin dynamics are statistically dependent on the choice of the initial state, and within a fairly short time, the sampling path becomes uncorrelated with the initial state. This period roughly corresponds to the upper plot in Figure 3. During this phase of sampling, we observe that the time average of pmmLang+RBM is actually not very close to that of pmmLang, but the trends of both running averages are similar: after a short time, both of them start to fluctuate within a same small neighborhood. We remark that in practice, the samples from the first phase are "burned-in" anyway due to their low qualify, but it is crucial that the consecutive random perturbations are suppressed by the thermostatting mechanism of the Langevin dynamics such that the running average of the pmmLang+RBM is able to approach a small vicinity of the true value by the end of the first phase. In probability language, it means the \blue{difference} between the probability distribution function of the stochastic process pmmLang+RBM and the target distribution \blue{significantly reduces within a short time period, after which the pmmLang+RBM is able to produce the correct samples from the target distribution}.\par
In the second phase of sampling, the Langevin dynamics effectively produces a vast amount of samples of the target distribution such that the empirical measure is converging to the target Boltzmann distribution $\pi(\q)$, which is manifested by the decaying of the relative entropy. When the random batches are used, the generated position samples $\tilde\q(j\Delta t)$ from pmmLang+RBM are not, unfortunately, unbiased samples from $\pi(\q)$. Whereas, the thermostatting effect that dissipates the relative entropy with respect to $\pi(\q)$ still cause that the difference between the empirical measure $\tilde\mu_J(\q)$ and the Boltzmann distribution $\pi(\q)$ diminishes in time. This argument is further confirmed by the numerical tests as shown in Figure \ref{fig:relative entropy}, where  we observe that with relative entropy of $D(\tilde \mu_J ||\pi)$ from pmmLang+RBM decays exponentially in time, although the convergence speed may be slightly reduced comparing with the decaying of $D( \mu_J ||\pi)$. This implies the large number of small biases induced by the use of the RBM in long time sampling only add to a small bias in the empirical measure. Furthermore, we conclude that the use of random batches only lead to small random perturbations to the sample values of the observables, and thus by the law of large numbers, we expect a small bias in the thermal average calculations. \par
We summarize the weak error analysis for pmmLang+RBM in the following mathematical statement: there exists a small constant $C_{p,\Delta t}$ depending only on the batch size $p$, and the timestep $\Delta t$ such that for $J\gg1$,
\begin{equation}
	\bigg|\frac1J
	\sum_{j=1}^J 
	W_N(\tilde\q(j\Delta t)) - 
	\avg{W_N(\q)}_\pi\bigg| 
	< C_{p,\Delta t}.
	\label{eq:limit RBM}
\end{equation}
In other words, the fluctuation-dissipation relationship for the pmmLang+RBM can be established in an approximate sense, when the RBM mechanism is incorporated. The rigorous justification of this statement is not yet complete and beyond the scope of this paper.
\subsection{Efficient calculation of the weight function}
In Section \ref{sec:PIMD}, we have mentioned that the calculation of the weight $W_N(\q)$ also has $O(NP^2)$ complexity, if the observable operator is the kinetic energy or in the form of a summation over pairwise contributions. Using again the idea of random batches, we can make an unbiased approximation of $W_N(\q)$ and reduce the complexity of calculating $W_N(\q)$ from $O(NP^2)$ to $O(NP)$.\par
For the kinetic energy $\hat A = \hat p^2/(2m)$, rewrite the virial estimator (\ref{eq:W_N virial}) as
\begin{equation}
	W_N(\q) = \frac{3P}{2\beta} + 
	\frac1{2N}
	\avg{\q-\bar q,\alpha\q + \nabla U^\alpha(\q)}_F
	\label{eq:W_N virial rewritten}
\end{equation}
where the modified potential $U^\alpha(\q)$ \blue{defined} in (\ref{eq:U^alpha}) is the sum of all interacting potentials $V^{(c)}(q_k^i-q_k^j)$. We can directly obtain an unbiased approximation of $\nabla U^\alpha(\q)$ (and thus $W_N(\q)$) from the random-batch approximation (\ref{eq:batch force}). For the position-dependent operator $\hat A = A(\hat q)$ with $A(q)$ given in the form
\begin{equation}
	A(q) = \frac1P 
	\sum_{1\Le i<j\Le P}
	a(q^i-q^j),
\end{equation}
we randomly pick a batch $\mathcal C$ of size $p$ from the group of $P$ particles (a full division is not required \blue{now}) and approximate $A(q)$ as
\begin{equation}
	A(q) \approx \frac{P-1}{p(p-1)}
	\sum_{i,j\in\mathcal C,i<j}
	a(q^i-q^j)
	\label{eq:A RBM}
\end{equation}
which is unbiased.
Hence the weight $W_N(\q)$ can be efficiently computed by
\begin{equation}
	W_N(\q) \approx 
	\frac{P-1}{Np(p-1)}
	\sum_{k=1}^N\sum_{i,j\in\mathcal C,i<j}
	a(q_k^i-q_k^j)
\end{equation}\par
Finally, we emphasize that such approximation of the weight $W_N(\q)$ does not change the original dynamics.
\section{Splitting Monte Carlo method}
\label{sec:split}
\begin{algorithm*}
	\caption{Splitting Monte Carlo method for the pmmLang (\ref{dyn:pmmLang}) in a timestep $\Delta t$}
	Evolve the pmmLang in a timestep $\Delta t$:
	$$
	\begin{aligned}
		\d\q^i & = \v^i\d t, \\
		\d\v^i & = -\q^i\d t - (L^\alpha)^{-1}\sum_{j\neq i} \nabla V_1^{(c)}(\q^i-\q^j)  \d t -\gamma\v^i \d t + \sqrt{\frac{2\gamma(L^\alpha)^{-1}}{\beta_N}}\d\B^i.
	\end{aligned}
	~~~(i=1,\cdots,P)
	$$
	
	Let $(\q^*,\v^*)$ be the proposal calculate above.
	Set $(\q,\v) = (\q^*,\v^*)$ with probability
	$$
	a(\q,\q^*) = \min\{1,e^{-\beta_N(U_2(\q^*)-U_2(\q))}\},
	$$
	
	otherwise set $(\q,\v) = (\q,-\v)$.
	\label{dyn:pmmLang split}
\end{algorithm*}
\begin{algorithm*}
	\caption{RBM with splitting Monte Carlo for the pmmLang (\ref{dyn:pmmLang}) in a timestep $\Delta t$}
	Randomly divide the $P$ particles into $n$ batches $\mathcal C_1,\cdots,\mathcal C_n$ of size $p$, where $n=P/p$. \\
	\For{$l = 1,\cdots,n$}{
		\mbox{}\\
		Evolve the pmmLang within the batch $\mathcal C_l$ in a timestep $\Delta t$:
		$$
		\begin{aligned}
			\d\q^i & = \v^i\d t \\
			\d\v^i & = -\q^i\d t - \frac{P-1}{p-1}(L^\alpha)^{-1}\sum_{j\in\mathcal{C}_l,j\neq i}\nabla V_1^{(c)}(\q^i-\q^j)\d t
			-\gamma\v^i \d t + \sqrt{\frac{2\gamma(L^\alpha)^{-1}}{\beta_N}}\d\B
		\end{aligned}~~~~
		(i\in\mathcal{C}_l)
		$$
	}
	Let $(\q^*,\v^*)$ be the proposal calculated above.
	Set $(\q,\v) = (\q^*,\v^*)$ with probability
	$$
	a(\q,\q^*) = \min\{1,e^{-\beta_N(U_2(\q^*)-U_2(\q))}\},
	$$
	
	otherwise set $(\q,\v) = (\q,-\v)$.
	\label{dyn:pmmLang RBM split}
\end{algorithm*}
In the previous section, we have shown that the pmmLang+RBM is an efficient approximate integrator of the pmmLang (\ref{dyn:pmmLang}). However, when the interaction potential $V^{(c)}(q)$ is singular (e.g., Lennard-Jones or Morse potential), extremely small timesteps $\Delta t$ are needed to integrate the dynamics near the singular kernel, no matter if the RBM is used.\par
In the simulation of the pmmLang, the singular kernel of $V^{(c)}(q)$ limits the timestep when the collision occurs, i.e., there are two particles $\q^i,\q^j$ close enough to each other.
Although the collision is a rare event, the requirement for small timesteps largely increases the computational cost. To overcome the singularity of the interaction potential, the pmmLang (\ref{dyn:pmmLang}) should be modified to avoid making use of gradients of singular potentials,
be compatible with the random batch method, and do not change the invariant distribution (\ref{eq:pi(q,v)}) of the pmmLang (\ref{dyn:pmmLang}).\par
In this section we introduce a modified version of the pmmLang, which is based on the splitting Monte Carlo method \cite{rbm1}. Briefly speaking, we split the potential $U^\alpha(\q)$ into the smooth part $U_1(\q)$ and the singular part $U_2(\q)$.
At each timestep, a proposal point $\q^*$ is generated by evolving the pmmLang driven by $U_1(\q)$, and is accepted or rejected according to the potential difference $U_2(\q^*) - U_2(\q)$. Therefore, the splitting Monte Carlo method can be seen as a variant of the Metropolis-adjusted Langevin algorithm (MALA) \cite{mala}.\par 
To illustrate the splitting Monte Carlo method, we split the interaction potential $V^{(c)}(q)$ into
\begin{equation}
	V^{(c)}(q) = V_1^{(c)}(q) + V_2^{(c)}(q),~~~
	q\in\mathbb R^3
\end{equation}
where $V_2^{(c)}(q)$ is short-ranged and captures the singular part of $V^{(c)}(q)$, and thus $V_1^{(c)}(q)$ is smooth. Define
\begin{align}
	U_1(\q) & = \sum_{k=1}^N
	\sum_{1\Le i<j\Le P}
	V^{(c)}_1(q_k^i-q_k^j), 
	\label{U1} \\
	U_2(\q) & = \sum_{k=1}^N
	\sum_{1\Le i<j\Le P}
	V^{(c)}_2(q_k^i-q_k^j),
	\label{U2}
\end{align}
then $U^\alpha(\q)$ is split into the sum of the smooth potential $U_1(\q)$ and the singular potential $U_2(\q)$.
On the one hand, the pmmLang driven by the smooth $U_1(\q)$ is
\begin{equation}
	\begin{aligned}
		\d\q^i & = \v^i\d t, \\
		\d\v^i & = -\q^i\d t - (L^\alpha)^{-1}\sum_{j\neq i} \nabla V_1^{(c)}(\q^i-\q^j)  \d t \\
		& ~~~~ -\gamma\v^i \d t + \sqrt{\frac{2\gamma(L^\alpha)^{-1}}{\beta_N}}\d\B^i, 	~~~~(i=1,\cdots,P)
	\end{aligned}
	\label{dyn:pmmLang U1}
\end{equation}
whose invariant distribution is
\begin{align}
	\pi_1(\q,\v) \propto \exp\bigg(-\beta_N
	\Big(
	\frac12\avg{\v,L^\alpha\v}_F\, + \hspace{1cm} \notag \\
	\frac12\avg{\q,L^\alpha\q}_F + U_1(\q)
	\Big)\bigg).
\end{align}
On the other hand, when the proposal $\q^*$ is accepted with probability
\begin{equation}
	a(\q,\q^*) =  \min\{1,e^{-\beta_N(U_2(\q^*)-U_2(\q))}\},
	\label{eq:acceptance rate}
\end{equation}
the Metropolis algorithm for the singular potential $U_2(\q)$ samples its distribution
\begin{equation}
	\pi_2(\q) \propto e^{-\beta_N U_2(\q)},
\end{equation}
Therefore, the key to assure that the coupled dynamics of the pmmLang (\ref{dyn:pmmLang U1}) driven by $U_1(\q)$ and the Metropolis algorithm for $U_2(\q)$ is the obvious relation
\begin{equation}
	\pi(\q,\v) \propto \pi_1(\q,\v) \pi_2(\q),
\end{equation}\par
Since the pmmLang is the second-order Langevin dynamics, special care should be taken to preserve the detailed balance. According to \cite{db}, the detailed balance of the second-order Langevin dynamics is given by
\begin{align}
	\pi(\q,\v) T_{\Delta t}((\q,\v),(\q',\v')) = \hspace{2cm} \notag \\ \pi(\q',-\v') T_{\Delta t}((\q',-\v'),(\q,-\v)),
	\label{eq:detailed balance}
\end{align}
where $T_{\Delta t}((\q,\v),(\q',\v'))$ is the transition probability density from $(\q,\v)$ to $(\q',\v')$ in a timestep $\Delta t$. The velocities $\v,\v'$ appearing in the right hand side of (\ref{eq:detailed balance}) are flipped as $-\v,-\v'$, thus in the Metropolis algorithm for $U_2(\q)$, the velocity $\v$ should also be flipped if the proposal $\q^*$ is rejected. To sum up, the splitting Monte Carlo method for the pmmLang (\blue{denoted by} the pmmLang+split) is given in Algorithm \ref{dyn:pmmLang split}.
The correctness of Algorithm \ref{dyn:pmmLang split} is guaranteed by the following theorem.
\begin{theorem}
	Let $T_{\Delta t}((\q,\v),(\q',\v'))$ be the transition probability density of the splitting Monte Carlo method (Algorithm \ref{dyn:pmmLang split}) in a timestep $\Delta t$, which satisfies
	\begin{equation}
		\int T_{\Delta t}((\q,\v),(\q',\v'))
		\d\q'\d\v' = 1,
	\end{equation}
	then the detailed balance holds
	\begin{align}
		 \pi(\q,\v) T_{\Delta t}((\q,\v),(\q',\v')) = \hspace{2cm} \notag \\
		 \pi(\q',-\v') T_{\Delta t}((\q',-\v'),(\q,-\v)).
		\label{eq:detailed balance split}
	\end{align}
	The detailed balance (\ref{eq:detailed balance split}) implies $\pi(\q,\v)$ is the invariant distribution of the splitting Monte Carlo method (Algorithm \ref{dyn:pmmLang split}),
	\begin{equation}
		\int \pi(\q,\v) T_{\Delta t}((\q,\v),(\q',\v'))
		\d\q\d\v
		= \pi(\q',\v').
	\end{equation}
	\label{thm:split MC}
\end{theorem}\par
\begin{figure*}
	\centering
	\includegraphics[width=0.4\textwidth]{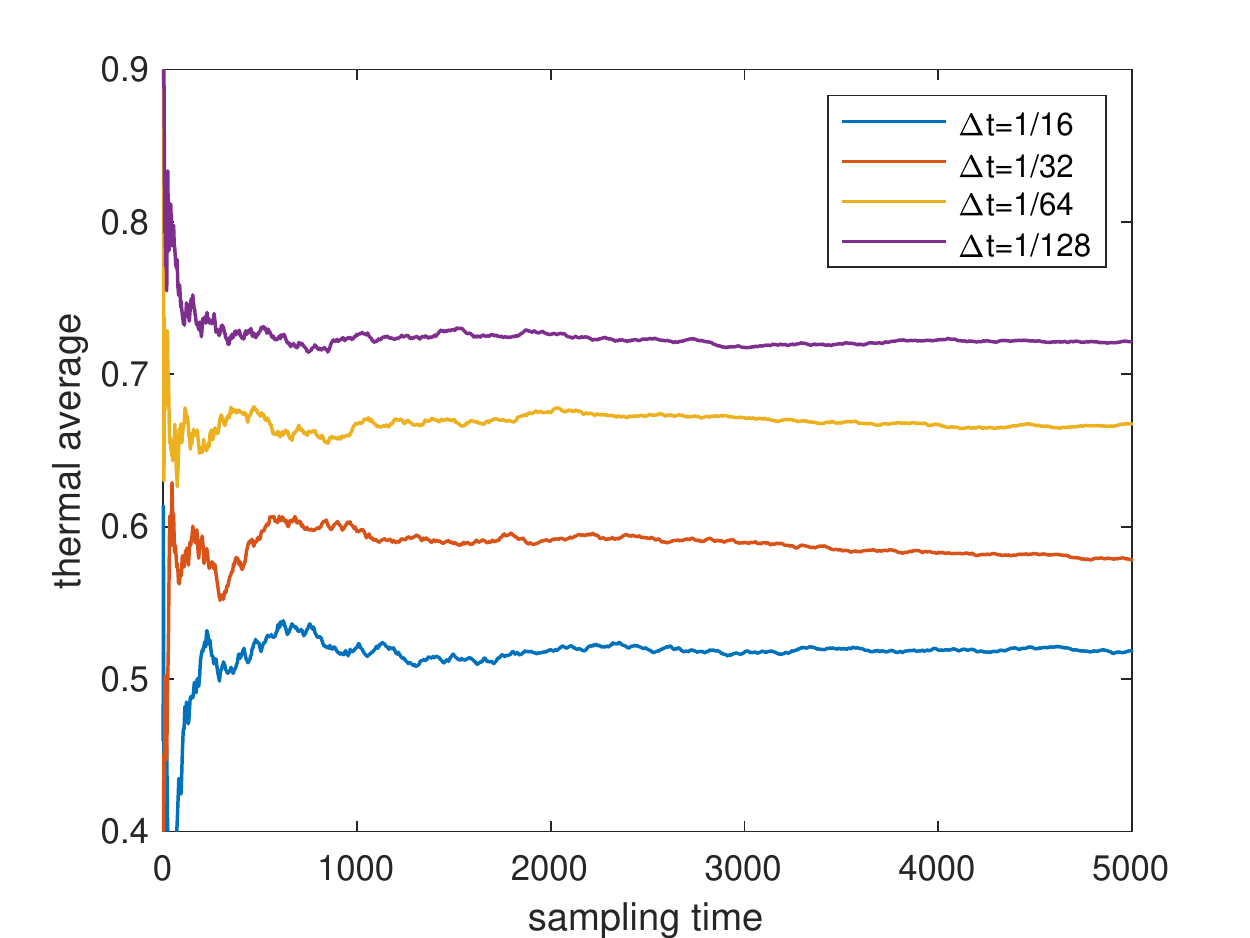}
	\includegraphics[width=0.4\textwidth]{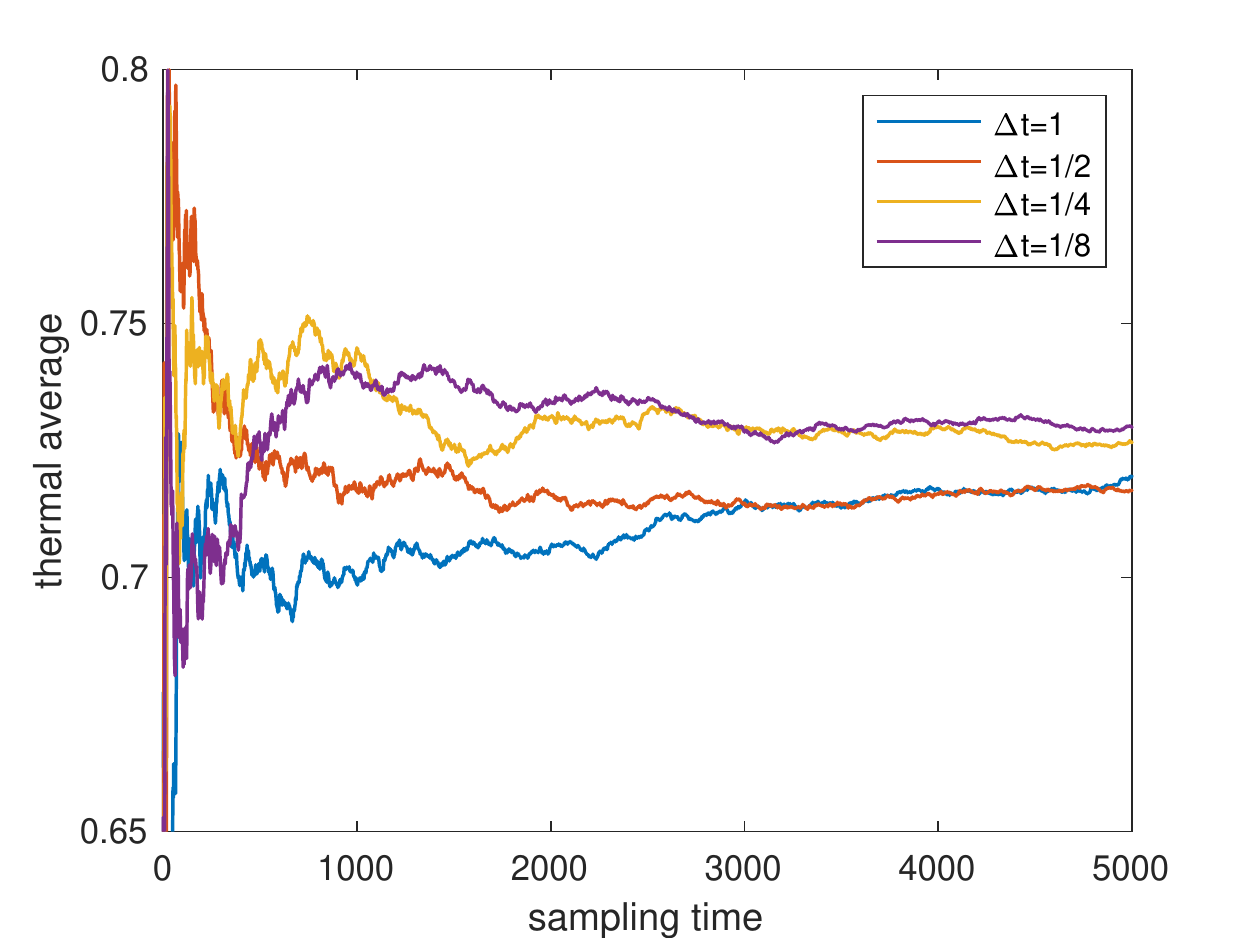} \\
	\includegraphics[width=0.4\textwidth]{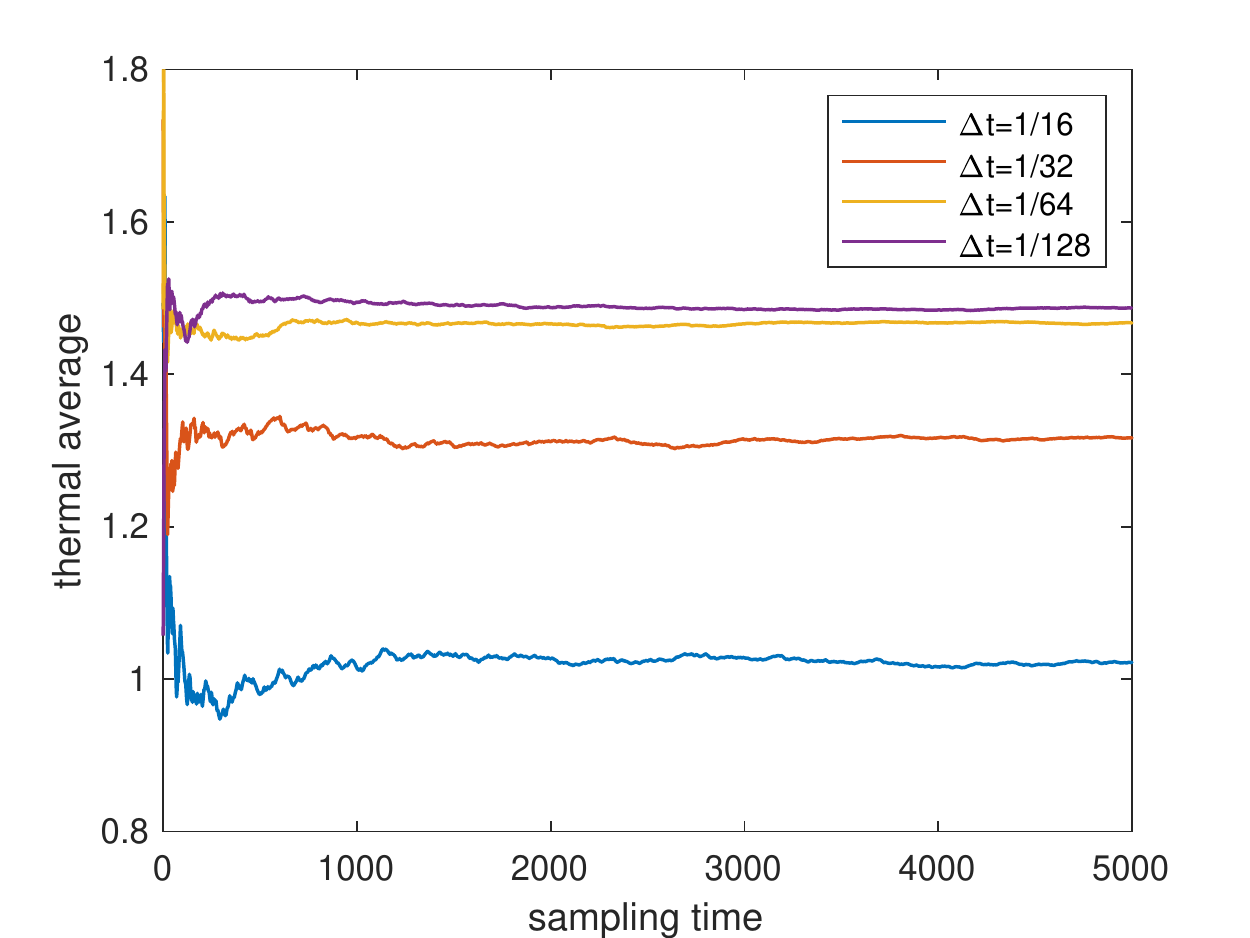}
	\includegraphics[width=0.4\textwidth]{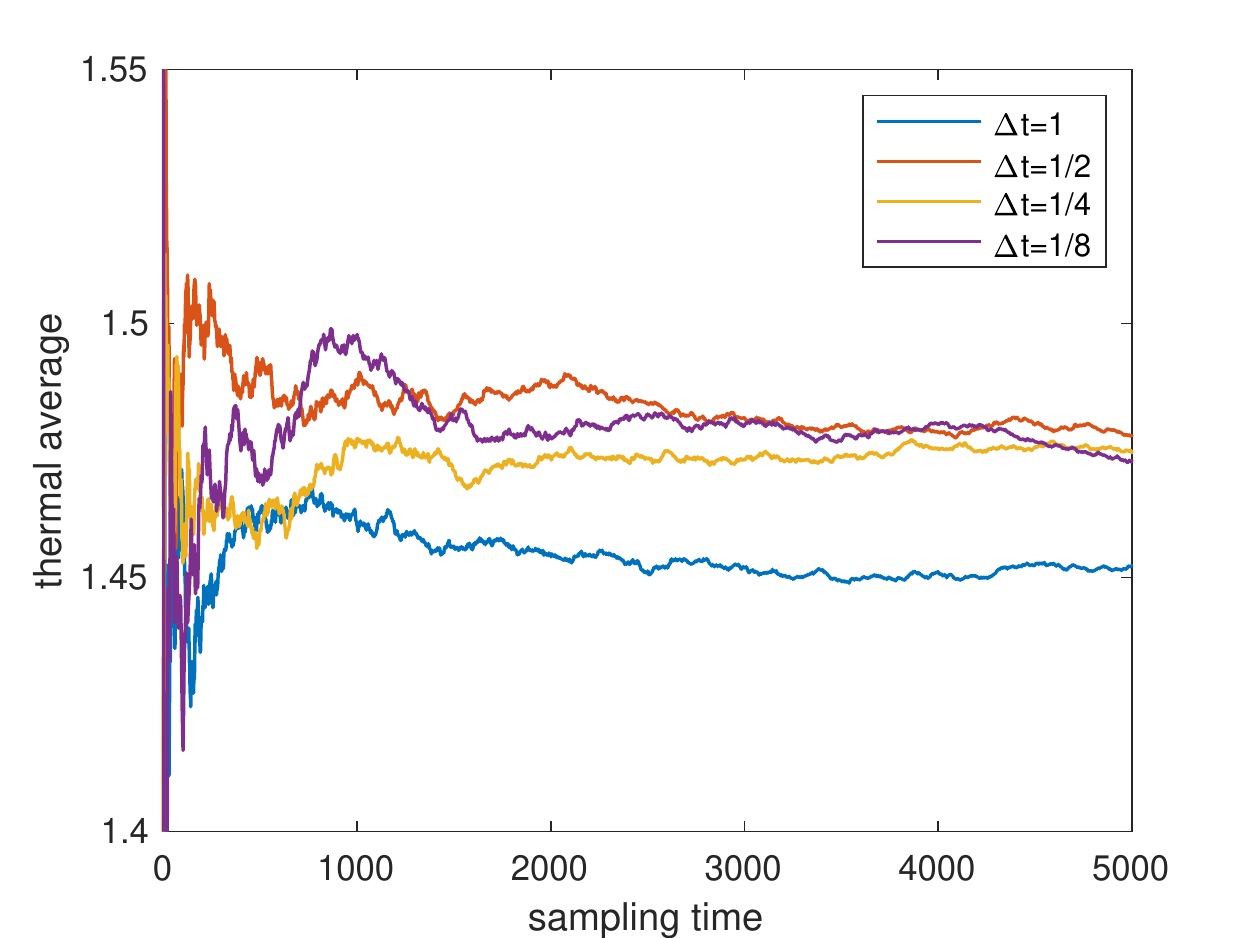}
	\caption{The time averages computed of the pmmLang and and the pmmLang+split in the mixed Coulomb-Lennard-Jones system, \blue{where the observable operator is defined in (\ref{mixed:observable})}.
	The left and right panels are for the pmmLang and the pmmLang+split, and the top and bottom figures are associated with the inverse temperature $\beta = 1,4$.
	The mass $m=1$, the number of particles $P=8$, the number of beads $N=8$, the total sampling time $T=6000$, the friction constant $\gamma=2$ and the batch size $p=2$.}
	\label{mixed:time average}
\end{figure*}
In the large interacting particle system, the computational cost of Algorithm \ref{dyn:pmmLang split} within a single timestep originates from evaluating the interaction forces $\nabla V_1^{(c)}(q_k^i-q_k^j)$ and the potential difference $U_2(\q^*) - U_2(\q)$. Note that $U_2(\q)$ defined in (\ref{U1})(\ref{U2}) is the sum of all short-ranged interacting potentials $V_2^{(c)}(q_k^i-q_k^j)$, thus can be efficiently calculated by cutoff or using data structures such as the cell list \cite{celllist1}. In this paper we employ the cutoff method to compute $U_2(\q)$, i.e., the particle pair $(q_k^i,q_k^j)$ is counted in the summation (\ref{U1})(\ref{U2}) only when their distance $|q_k^i-q_k^j|$ is less than the cutoff distance. Therefore, for large interacting particle systems, the majority of the computational cost is still the calculation of the interaction forces $\nabla V_1^{(c)}(q_k^i-q_k^j)$.\par
Now we employ the RBM to reduce the complexity due to the interaction forces $\nabla V_1^{(c)}(q_k^i-q_k^j)$. At each iteration, we randomly divide the $P$ particles to $n=P/p$ batches, where each batch $\mathcal C$ is of size $p$. The pmmLang driven by the smooth potential $U_1(\q)$ within the batch $\mathcal C$ is then\par
\begin{equation}
	\begin{aligned}
		\d\q^i & = \v^i\d t, \\
		\d\v^i & = -\q^i\d t - \frac{P-1}{p-1}(L^\alpha)^{-1}\sum_{j\in\mathcal{C},j\neq i}\nabla V_1^{(c)}(\q^i-\q^j)\d t
		\\
		& ~~~~ -\gamma\v^i \d t + \sqrt{\frac{2\gamma(L^\alpha)^{-1}}{\beta_N}}\d\B, ~~~~(i\in\mathcal C)
	\end{aligned}
	\label{dyn:pmmLang U1C}
\end{equation}
\par
By coupling the random-batch approximated pmmLang (\ref{dyn:pmmLang U1C}) driven by $U_1(\q)$ and the Metropolis-Hastings algorithm for $U_2(\q)$, we obtain the RBM with splitting Monte Carlo for the pmmLang (\blue{denoted by} the pmmLang+RBM+split, \blue{presented in} Algorithm \ref{dyn:pmmLang RBM split}).\par
\blue{To show how the splitting Monte Carlo method accelerates the sampling efficiency, we compare the performance of the pmmLang and the pmmLang+split in the numerical example below. In the mixed Coulomb-Lennard-Jones system (the potential function and the splitting scheme are defined in (\ref{mixed})(\ref{mixed:Vc1})(\ref{mixed:Vc2}) in Section \ref{sec:tests}), we plot in Figure \ref{mixed:time average} the time averages computed by the pmmLang and the pmmLang+split with various timesteps. Different inverse temperatures $\beta = 1,4$ are tested respectively.\par
We observe from Figure \ref{mixed:time average} that the time average of the pmmLang is very sensitive to the timestep $\Delta t$, while in the pmmLang+split we can adopt relatively large timesteps to obtain the correct thermal average, which greatly improves the efficiency of the simulation.\par
Despite of the satisfactory performance of the pmmLang+split in this example, the numerical efficiency of this Monte Carlo-type method can be influenced by the rejection rate along the sampling process. When the number of beads $N$ or the number of particles $P$ is large, it is likely that the high-dimensional nature of the dynamics will slow down the simulation since the the rejection rate is high. A detailed numerical investigation of the rejection rate with different parameters is presented in Section \ref{sec:tests}.}
\section{Numerical tests}
\label{sec:tests}
\subsection{Examples of the interacting particle system}
In Section \ref{sec:RBM} we have proposed and analyzed the efficient sampling method for the quantum interacting particle systems, the pmmLang+RBM. To further explore this method, we present in this section more numerical results, where different parameters and interaction potentials are tested. The efficiency and the error in the computing thermal average $\avg{\hat A}$ are primarily used to quantify the numerical performance of the pmmLang+RBM.\par
\begin{figure}
	\includegraphics[width=0.48\textwidth]{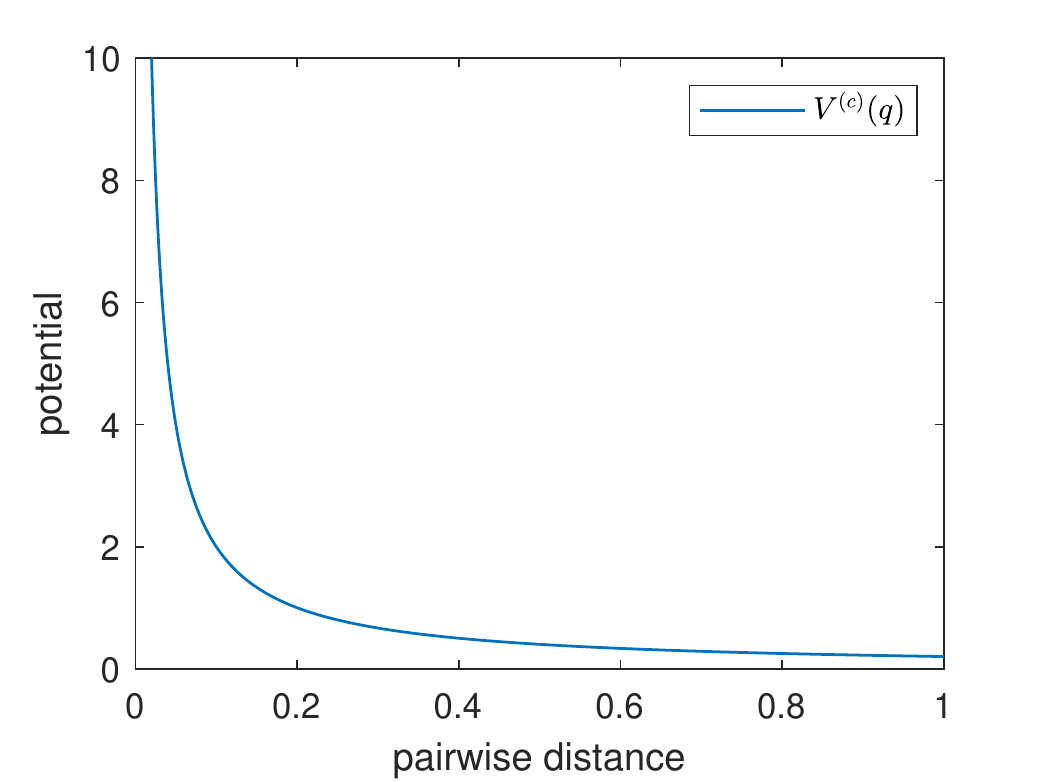}
	\includegraphics[width=0.48\textwidth]{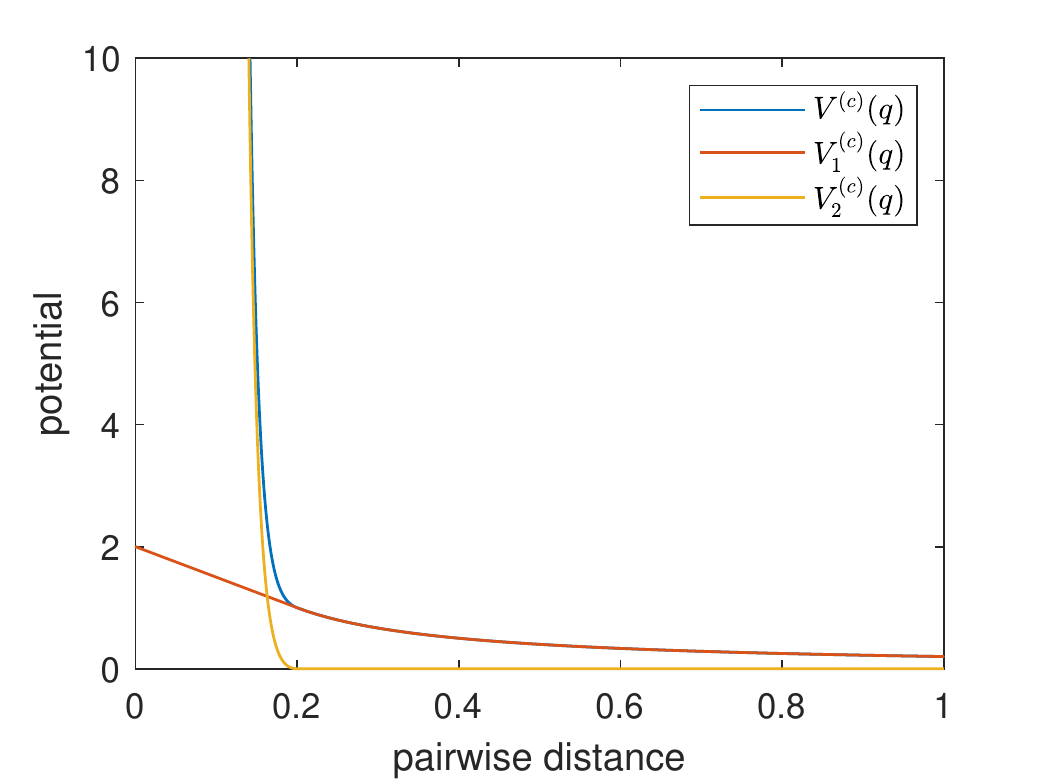}
	\caption{The interacting potential $V^{(c)}(q)$ with its splitting scheme. Top: the Coulomb potential (\ref{Coulomb}). Bottom: shows the mixed Coulomb-Lennard-Jones potential (\ref{mixed}) and the splitting potentials $V_1^{(c)}(q),V_2^{(c)}(q)$.}
	\label{fig:potential}
\end{figure}

In the quantum system (\ref{eq:V}), we choose the interacting potential $V^{(c)}(q)$ as either the Coulomb potential
\begin{equation}
	V^{(c)}(q) = \frac{\kappa}{r},~~~
	q\in\mathbb R^3
	\label{Coulomb}
\end{equation}
or the mixed Coulomb-Lennard-Jones potential
\begin{equation}
	V^{(c)}(q) =
	\left\{
	\begin{aligned}
		& \frac16\bigg(
			\Big(\frac{\sigma}r\Big)^{12} \hspace{-4pt} -
			\Big(\frac{\sigma}r\Big)^6 
		\bigg) + 1, && r<\sigma \\
		& \frac{\sigma}r, && r\Ge\sigma
	\end{aligned}
	\right.
	\label{mixed}
\end{equation}
where $\kappa = 1$, $\sigma = 0.3$ and $r = |q|$. The kernel of (\ref{mixed}) is provided by the Lennard-Jones potential, and is much more singular than the Coulomb potential. Hence we employ the splitting Monte Carlo method introduced in Section \ref{sec:split} to and split the mixed potential (\ref{mixed}) into the sum of
\begin{equation}
	V_1^{(c)}(q) = 
	\left\{
	\begin{aligned}
		& 2 - \frac{r}{\sigma}, && r<\sigma \\
		& \frac{\sigma}r, && r\Ge\sigma
	\end{aligned}
	\right.
	\label{mixed:Vc1}
\end{equation}
and
\begin{equation}
	V_2^{(c)}(q) =
	\left\{
	\begin{aligned}
		& \frac16\bigg(
		\Big(\frac{\sigma}r\Big)^{12} \hspace{-4pt} -
		\Big(\frac{\sigma}r\Big)^6 
		\bigg) + 1, && r<\sigma \\
		& 0, && r\Ge\sigma
	\end{aligned}
	\right.
	\label{mixed:Vc2}
\end{equation}
where $V_1^{(c)}(q)$ is smooth and $V_2^{(c)}$ is short-ranged.
The graphs of the Coulomb potential (\ref{Coulomb}) and the mixed potential (\ref{mixed}) with its splitting scheme (\ref{mixed:Vc1})(\ref{mixed:Vc2}) are shown in Figure \ref{fig:potential}.
\par
In the quantum system (\ref{eq:V}), the external potential is chosen \blue{to be} harmonic (\ref{eq:V^o}), where the parameter $\alpha$ is chosen \blue{as}
\begin{equation}
	\alpha = P^{-\frac23},
	\label{eq:alpha P23}
\end{equation}
where $P$ is the number of particles. Here the choice (\ref{eq:alpha P23}) is to assure that all the particles are evenly distributed in $\mathbb R^3$. While placing $P$ particles in the harmonic potential (\ref{eq:V^o}), the size of the potential well should be $O(P^{-\frac13})$ to ensure the particles have $O(1)$ pairwise distance, thus $\alpha$ in (\ref{eq:V^o}) is chosen as $O(P^{-\frac23})$.\par
The position-dependent observable $\hat A = A(\hat q)$ is chosen in the form of (\ref{eq:A RBM}).
For the Coulomb system (\ref{Coulomb}), choose $a(q) = V^{(c)}(q)$, then
\begin{equation}
	A(q) = \frac1P
	\sum_{1\Le i<j\Le P}
	\frac{\kappa}{|q^i-q^j|},~~~~
	q\in\mathbb R^{3P}
	\label{Coulomb:observable}
\end{equation}
is the average \blue{interacting potential} of the system. For the mixed system (\ref{mixed}), choose $a(q) = e^{-\theta |q|^2}$ with $\theta = 0.1$, then
\begin{equation}
	A(q) = \frac1P 
	\sum_{1\Le i<j\Le P}
	e^{-\theta|q^i-q^j|^2},~~~~
	q\in\mathbb R^{3P}
	\label{mixed:observable}
\end{equation}
\subsection{Outline of the numerical simulation method}
We outline the procedure to compute the thermal average $\avg{\hat A}$ defined in (\ref{avg:pmmLang RBM}) in this paper. By choosing sufficiently large $N$, $\avg{\hat A}$ is approximated by the ensemble average $\avg{W_N(\q)}_\pi$. Then the pmmLang and the pmmLang+RBM estimates $\avg{W_N(\q)}_\pi$ by the time averages (\ref{avg:pmmLang})(\ref{avg:pmmLang RBM}).\par
We stress that the calculation of the time averages only requires a single sampling path, thus (\ref{avg:pmmLang})(\ref{avg:pmmLang RBM}) are both random variables. That is to say, the values of (\ref{avg:pmmLang})(\ref{avg:pmmLang RBM}) depend on the Browian motion and the choices of random divisions along the samlping process. In the numerical experiments below, only one typical sampling path is used to compute the time avrages (\ref{avg:pmmLang})(\ref{avg:pmmLang RBM}).\par
We use the BAOAB scheme \cite{BAOAB1,BAOAB2} to integrate the pmmLang dynamics numerically. By combining the velocity Verlet method \cite{BAOAB2} with the Langevin thermostat, we obtain the BAOAB scheme for the pmmLang (\ref{dyn:pmmLang velocity}) in a timestep $\Delta t$,
\begin{align*}
	\v^{j*} & = \v^j - (\q^j + (L^\alpha)^{-1}\nabla U^\alpha(\q^j)) \frac{\Delta t}2 \\
	\q^{j+\frac12} & = \q^j + \v^{j*} \frac{\Delta t}2 \\
	\v^{j**} & = e^{-\gamma\Delta t}\v^{j*} + \sqrt{\frac{1-e^{-2\gamma\Delta t}}{\beta_N}}\bta \\
	\q^{j+1} & = \q^{j+\frac12} + \v^{j**} \frac{\Delta t}2 \\
	\v^{j+1} & = \v^{j**} - (\q^{j+1} + (L^\alpha)^{-1}\nabla U^\alpha(\q^{j+1})) \frac{\Delta t}2
\end{align*}
where $\bta\in\mathbb R^{N\times 3P}$ is a random variable and each column of $\bta$ obeys the Gaussian distribution $\mathsf N(0,(L^\alpha)^{-1})$. Similar BAOAB schemes can be derived for the pmmLang within the batch (\ref{dyn:pmmLang batch}) or the pmmLang in the splitting Monte Carlo method (\ref{dyn:pmmLang U1})(\ref{dyn:pmmLang U1C}).
\par
\blue{Since the spectrum of $L^\alpha$ is known, we cam sample from the Gaussian distribution $\mathsf N(0,(L^\alpha)^{-1})$ using the fast Fourier transform (FFT), whose complexity is only $O(N\log N)$ (see Appendix \ref{sec:precondition} for a detailed mathematical formulation). Therefore,} the computational cost due to algebraic operations is $O(N\log NP)$. In the interacting particle system with $N$ \blue{not relatively large}, the complexity from the interaction forces is still the \blue{dominant} difficulty in the numerical simulation.\par
\blue{Finally, we point out that the performance of the BAOAB scheme for the pmmLang may depend on $N$, although the continuous-time pmmLang dynamics has been shown to have a dimension-independent convergence rate\cite{precondition}. Designing a dimension-independent integrator for the pmmLang is beyond the scope of this article.}
\subsection{Tests of the pmmLang+RBM}
\subsubsection{Autocorrelation}
First, we aim to investigate the impacts of the RBM on the variance of the estimator, by computing the autocorrelation (abbreviated by AC) of the time averages (\ref{avg:pmmLang})(\ref{avg:pmmLang RBM}) numerically.\par
Recall that in Section \ref{sec:RBM}, we propose the RBM approach to reduce the complexity due to the weight function $W_N(\q)$, if the observable operator is in specific forms. Thus in practice, there are two ways to implement the pmmLang+RBM: one is to use the RBM in the dynamics (Algorithm \ref{dyn:pmmLang}) but compute $W_N(\q)$ as it is, and another is to use the RBM in both the dynamics and calculation of the weight function. To analyze the effects of the RBM, we will test both versions of the pmmLang+RBM.\par
\begin{figure}
	\centering
	\includegraphics[width=0.48\textwidth]{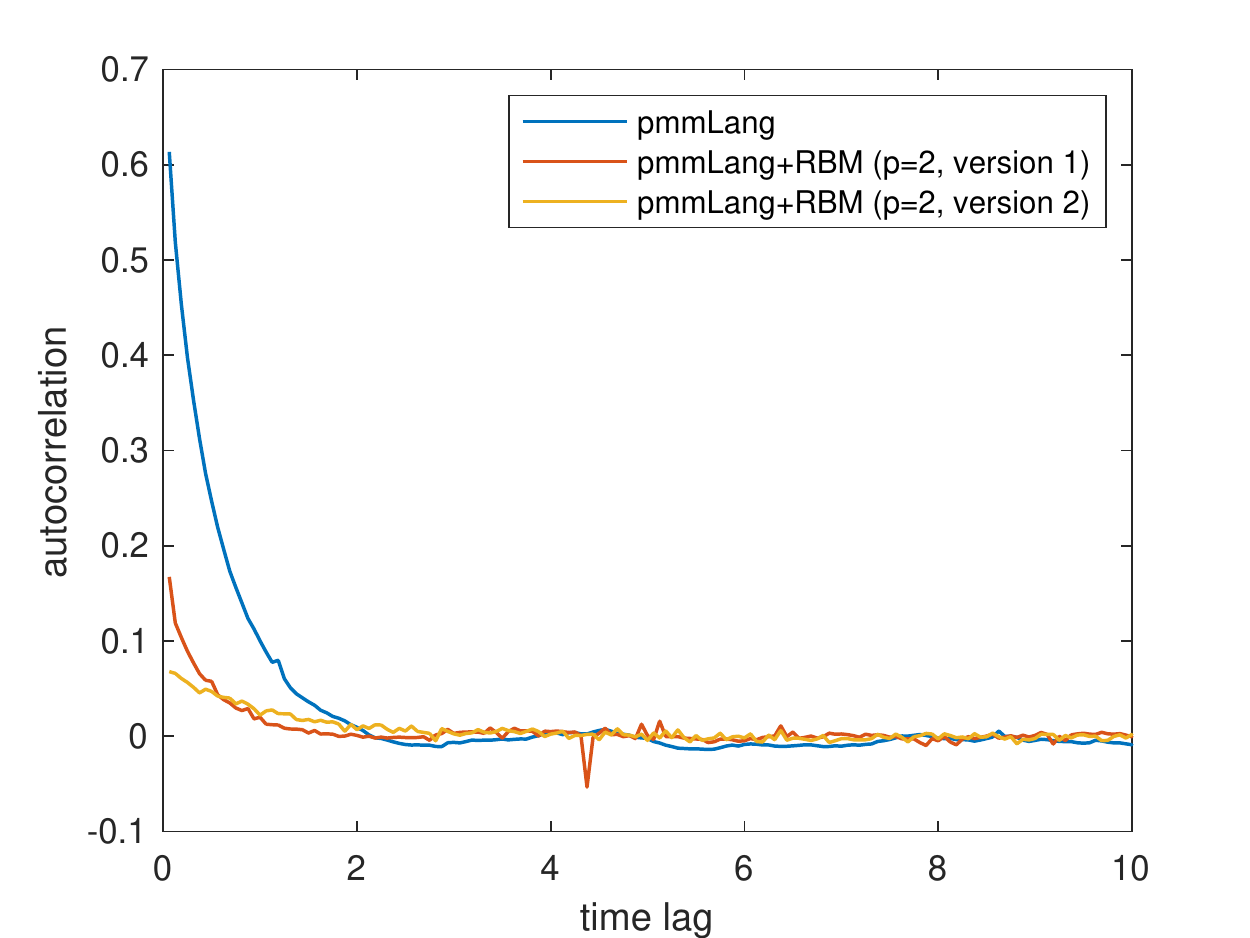}
	\caption{Autocorrelation of the time averages (\ref{avg:pmmLang})(\ref{avg:pmmLang RBM}) in the Coulomb interacting system.
	The blue, red and yellow curves are the autocorrelations of the pmmLang, the pmmLang+RBM (version 1) and the pmmLang+RBM (version 2) respectively.
	The mass $m=1$, the inverse temperature $\beta = 4$, the number of particles $P=16$, the number of beads $N=16$, the timestep $\Delta t = 1/16$, the total sampling time $T = 6000$, the friction constant $\gamma=2$ and the batch size $p=2$.}
	\label{Coulomb:autocorrelation}
\end{figure}

In the Coulomb system, we compute the pmmLang and the pmmLang+RBM time averages (\ref{avg:pmmLang})(\ref{avg:pmmLang RBM}), where the observable operator of interest is the kinetic energy and both versions of the pmmLang+RBM are considered. The autocorrelation of the time averages is shown in Figure \ref{Coulomb:autocorrelation}.
\begin{table}
	\begin{ruledtabular}
	\begin{tabular}{c|ccc}
		methods & pmmLang & w/RBM (ver.1) & w/RBM (ver.2) \\
		\hline
		AC time & 1.10 & 0.38 & 0.40 \\
		variance & \n{2.18}{-5} & \n{5.94}{-5} & \n{2.40}{-4} \\
		MSE & \n{1.10}{-3} & \n{5.37}{-3} & \n{5.55}{-3} \\
	\end{tabular}
	\end{ruledtabular}
	\caption{The autocorrelation time, the effective variance and the mean squared error of the time averages (\ref{avg:pmmLang})(\ref{avg:pmmLang RBM}).
	The first column is for the pmmLang, and
	the second and the third columns are for the two versions of the pmmLang+RBM respectively. The reference value is calculated by the pmmLang with $\Delta t = 1/64$.}
	\label{Coulomb:variance}
\end{table}
In Figure \ref{Coulomb:autocorrelation} and Table \ref{Coulomb:variance}, the autocorrelation of the pmmLang+RBM decays faster than the pmmLang.
Since the random divisions at different timesteps are independent, it is likely that the weak correlation of the interaction forces in different timesteps leads to a small autocorrelation in the pmmLang+RBM. Still, the randomness in the batch force approximation (\ref{eq:batch force}) enlarged the effective variance of the estimator.
Compared to the pmmLang of the version 1, the version 2 computes the weight function $W_N(\q)$ approximately, thus has additional variance in the estimator.\par
Since the random approximation to the weight function $W_N(\q)$ is necessary to reduce the computational cost per timestep, in the following tests we always employ the pmmLang+RBM of the version 2. That is to say, we use the RBM in both the dynamics and the calculation of the weight function.
\subsubsection{Convergence with the number of beads}
We study the convergence of pmmLang+RBM with the number of beads $N$. 
In the PIMD representation, the ensemble average $\avg{W_N(\q)}_\pi$ converges to the exact thermal average $\avg{\hat A}$ as the number of beads $N\rightarrow\infty$, hence it's necessary to check if the pmmLang+RBM time averages (\ref{avg:pmmLang RBM}) also possess this convergence property.\par
\begin{figure}
\centering
\includegraphics[width=0.23\textwidth]{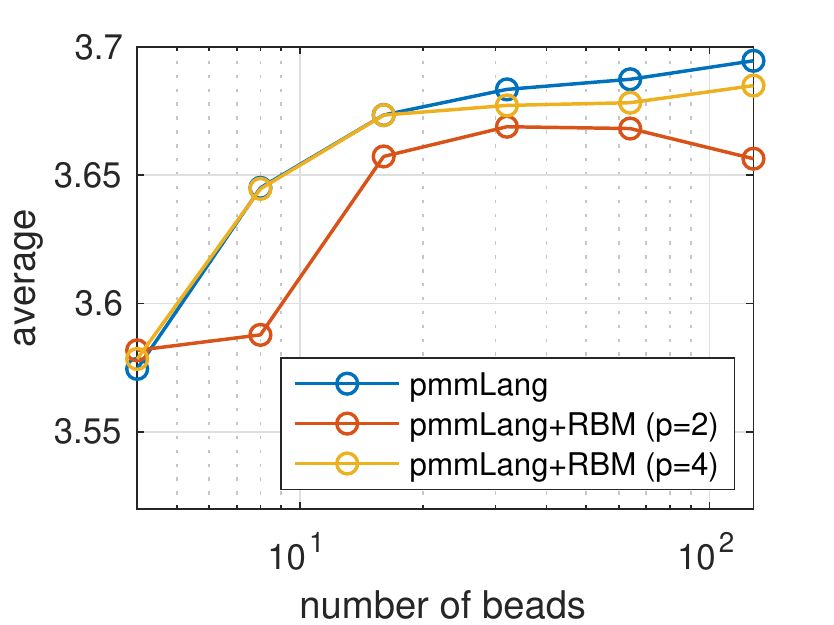}
\includegraphics[width=0.23\textwidth]{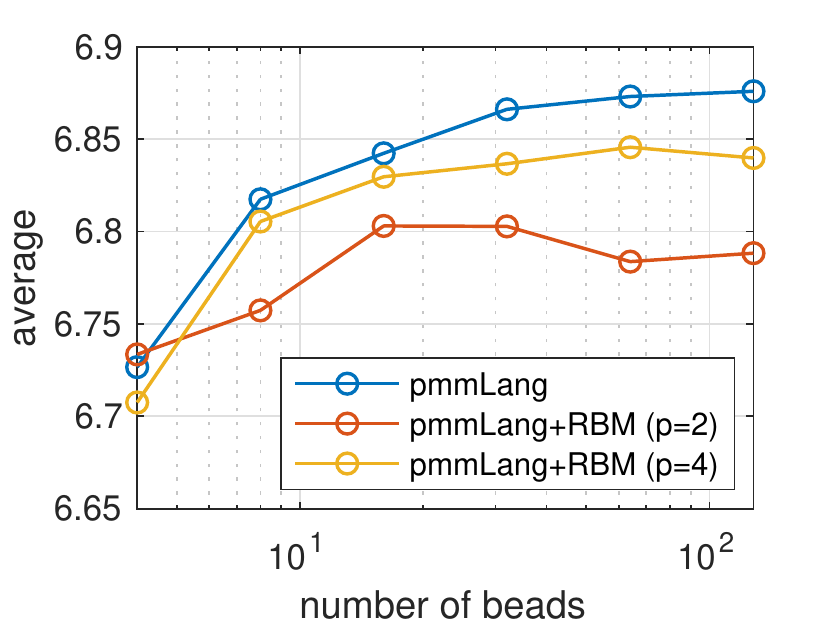}
\includegraphics[width=0.23\textwidth]{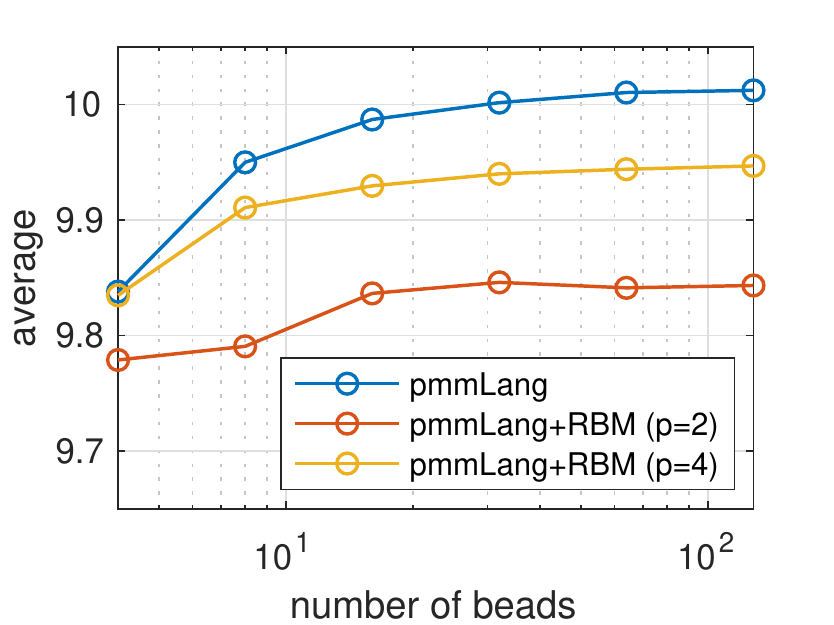}
\includegraphics[width=0.23\textwidth]{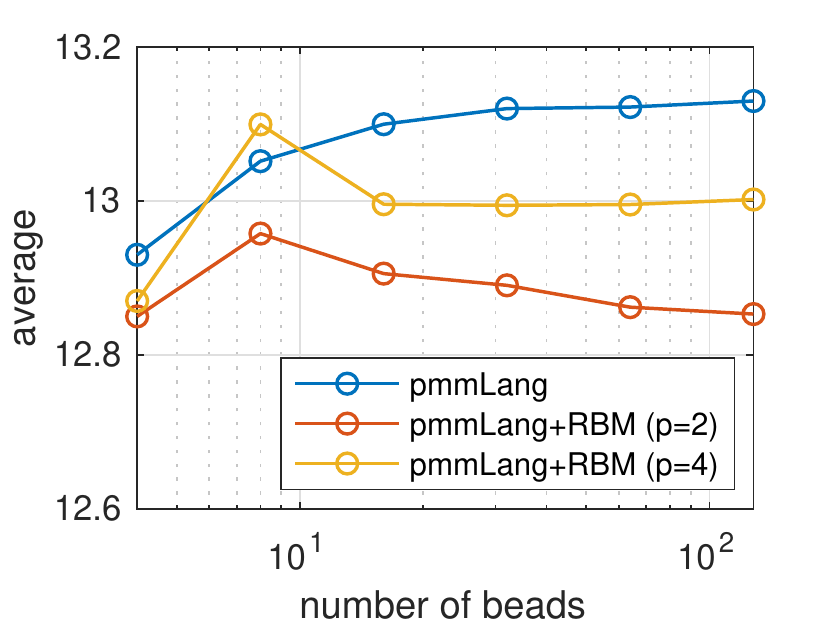}
\caption{Time averages (\ref{avg:pmmLang})(\ref{avg:pmmLang RBM}) of the pmmLang and the pmmLang+RBM in the Coulomb interacting system.
Figures at the top left, top right, bottom left and bottom right are associated with the number of particles $P=8,16,24,32$ respectively.
The blue curve is for the pmmLang, and the red and yellow curves are for the pmmLang+RBM with the batch size $p=2,4$.
The mass $m=1$, the inverse temperature $\beta = 4$, the timestep $\Delta t = 1/16$ and the total sampling time $T=6000$.
In the $x$-axis, the number of beads $N$ varies in $4,8,\cdots,128$.}
\label{Coulomb:convergence N}
\end{figure}

In the Coulomb interacting system, we plot the in Figure \ref{Coulomb:convergence N} the time averages (\ref{avg:pmmLang})(\ref{avg:pmmLang RBM}) of the pmmLang and the pmmLang+RBM, where the observable operator of interest is the kinetic energy.
Different numbers of beads are tested to study the convergence of the pmmLang and the pmmLang+RBM.
When the number of particles $P=16$, we record the effective variance of (\ref{avg:pmmLang})(\ref{avg:pmmLang RBM}) in Table \ref{Coulomb:convergence variance}.
\begin{table}
\begin{ruledtabular}
\begin{tabular}{c|ccc}
\#beads & pmmLang & w/RBM, $p=2$ & w/RBM, $p=4$ \\
\hline
4 & \n{2.50}{-5} & \n{9.83}{-4} & \n{1.09}{-4} \\
8 & \n{2.76}{-5} & \n{1.13}{-3} & \n{1.85}{-4} \\
16 & \n{2.42}{-5} & \n{2.87}{-4} & \n{8.34}{-5} \\
32 & \n{2.18}{-5} & \n{2.40}{-4} & \n{5.03}{-5} \\
64 & \n{2.22}{-5} & \n{2.42}{-4} & \n{4.82}{-5} \\
128 & \n{2.03}{-5} & \n{1.47}{-4} & \n{4.18}{-5} \\ 
\end{tabular}
\end{ruledtabular}
\caption{The effective variance 
of the time averages (\ref{avg:pmmLang})(\ref{avg:pmmLang RBM}) in the pmmLang and the pmmLang+RBM when the number of particles $P=16$.}
\label{Coulomb:convergence variance}
\end{table}
In Figure \ref{Coulomb:convergence N}, we observe  that pmmLang+RBM time averages (\ref{avg:pmmLang RBM}) converge as the number of beads $N$ enlarges. Besides, Table \ref{Coulomb:convergence variance} shows that the variance of the estimator is not sensitive to the number of beads $N$, thus it's safe increase $N$ in the pmmLang+RBM to obtain a more accurate approximation of the thermal average.\par
We also note that, even if $N$ is sufficiently large, there is a bias of the pmmLang+RBM time average (\ref{avg:pmmLang RBM}) from $\avg{\hat A}$. Moreover, the bias becomes significant when the number of particles $P$ grows large or the batch size $p$ is small as 2. This bias though can be reduced with fixed batch size $p$ by decreasing the timestep $\Delta t$, which are to be confirmed in the tests of the next subsection. Overall, the pmmLang+RBM time average (\ref{avg:pmmLang RBM}) is still an accurate approximation of $\avg{\hat A}$. When $N=128$, the relative error of (\ref{avg:pmmLang RBM}) from $\avg{\hat A}$ is no more than $2.5\%$. This bias though can be reduced with fixed batch size $p$ by decreasing the timestep $\Delta t$, which are to be confirmed in the \blue{follow-up tests. For simplicity, in the following we will take the number of beads $N=16$ in both the pmmLang and the pmmLang+RBM.}\par
\subsubsection{Error in the calculation of the thermal average}
\begin{figure*}
	\centering
	\includegraphics[width=0.4\textwidth]{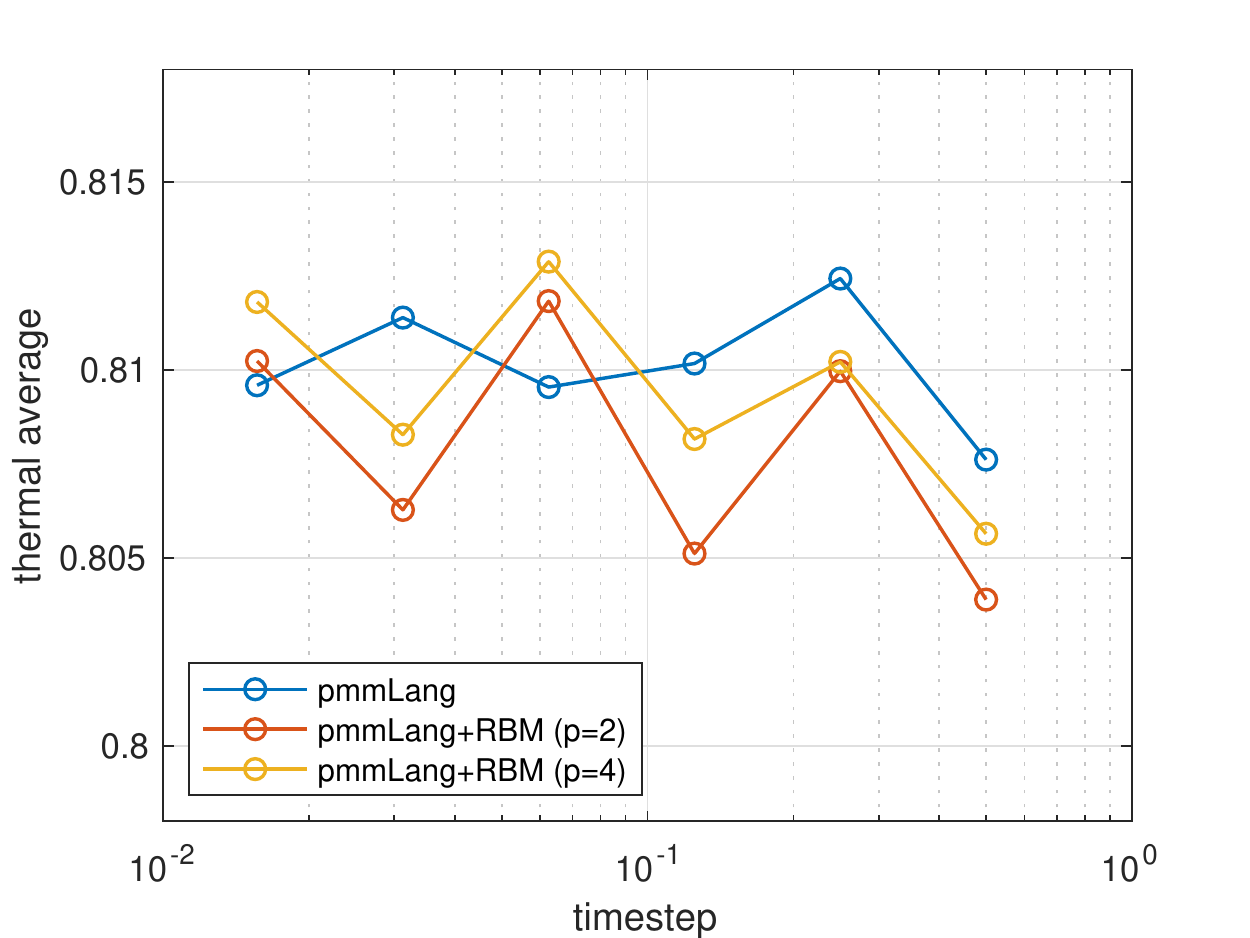}
	\includegraphics[width=0.4\textwidth]{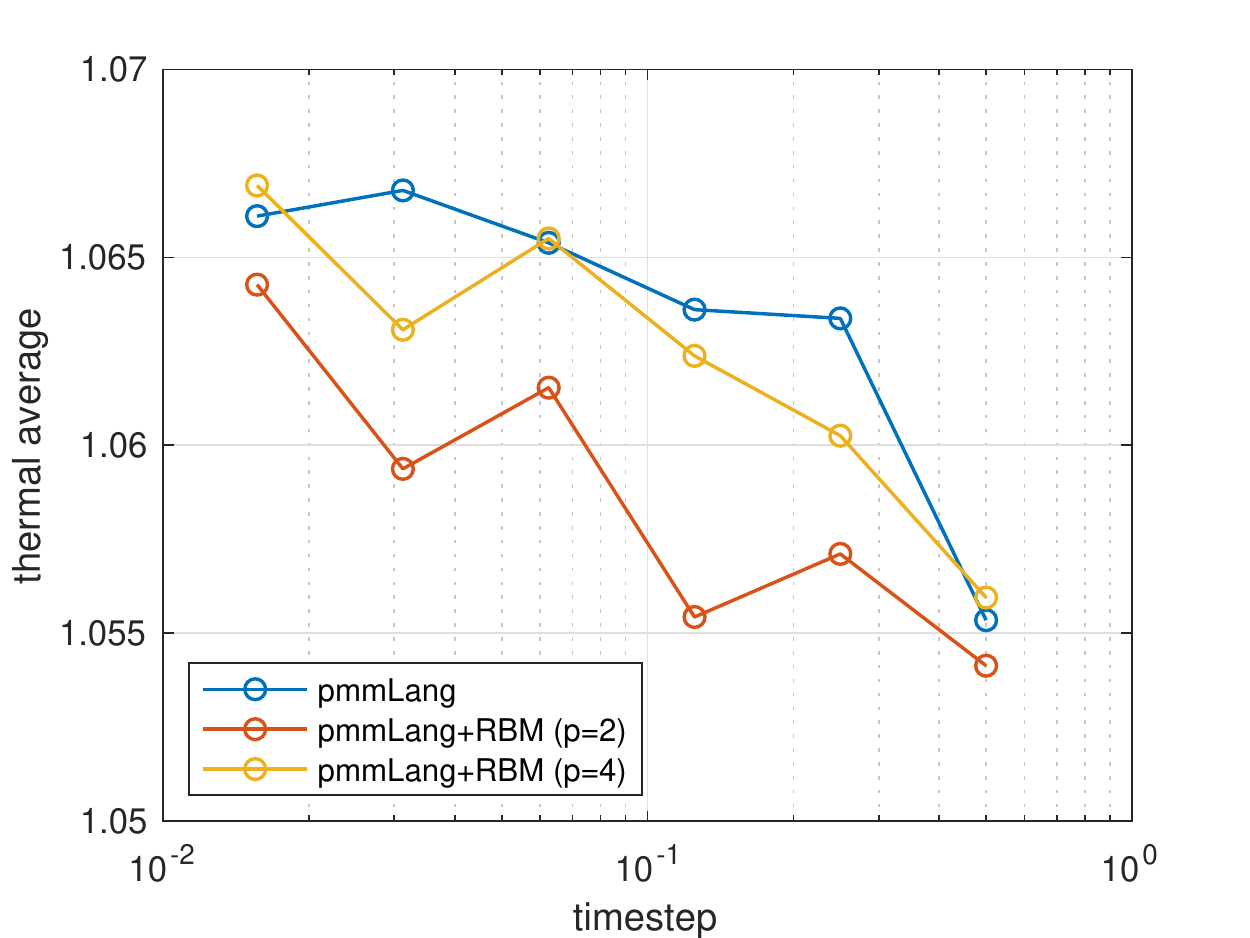} \\
	\includegraphics[width=0.4\textwidth]{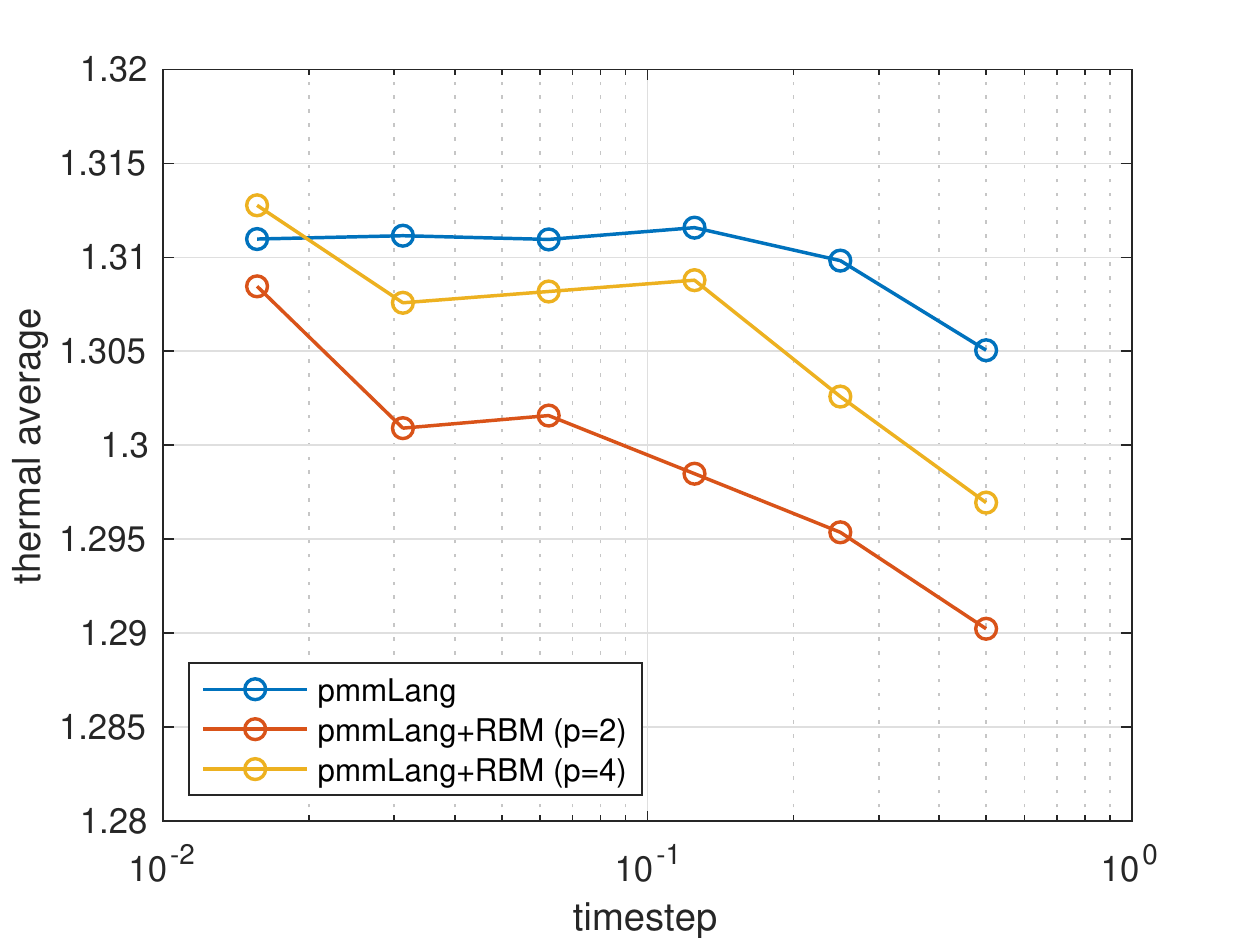}
	\includegraphics[width=0.4\textwidth]{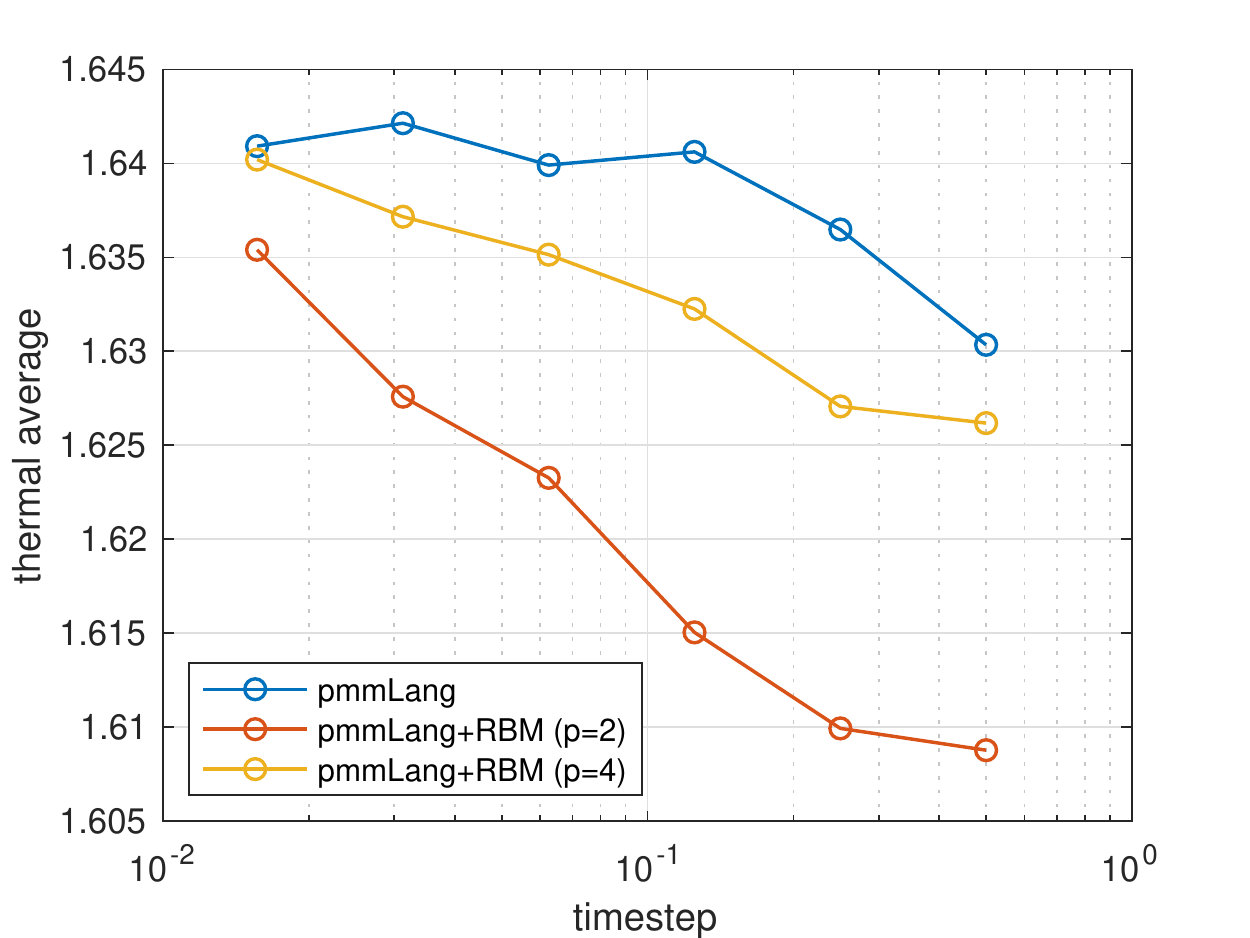} \\
	\includegraphics[width=0.4\textwidth]{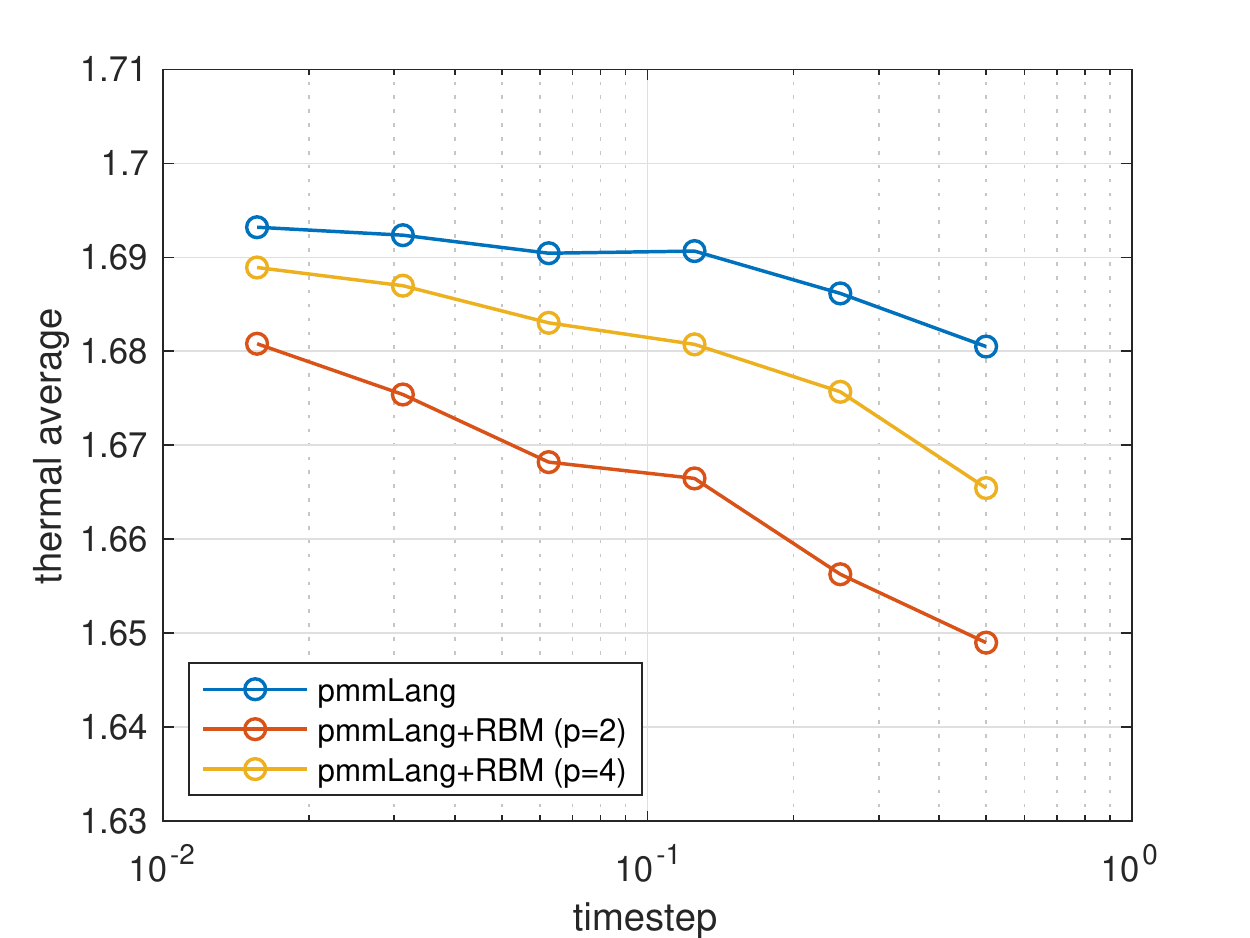}
	\includegraphics[width=0.4\textwidth]{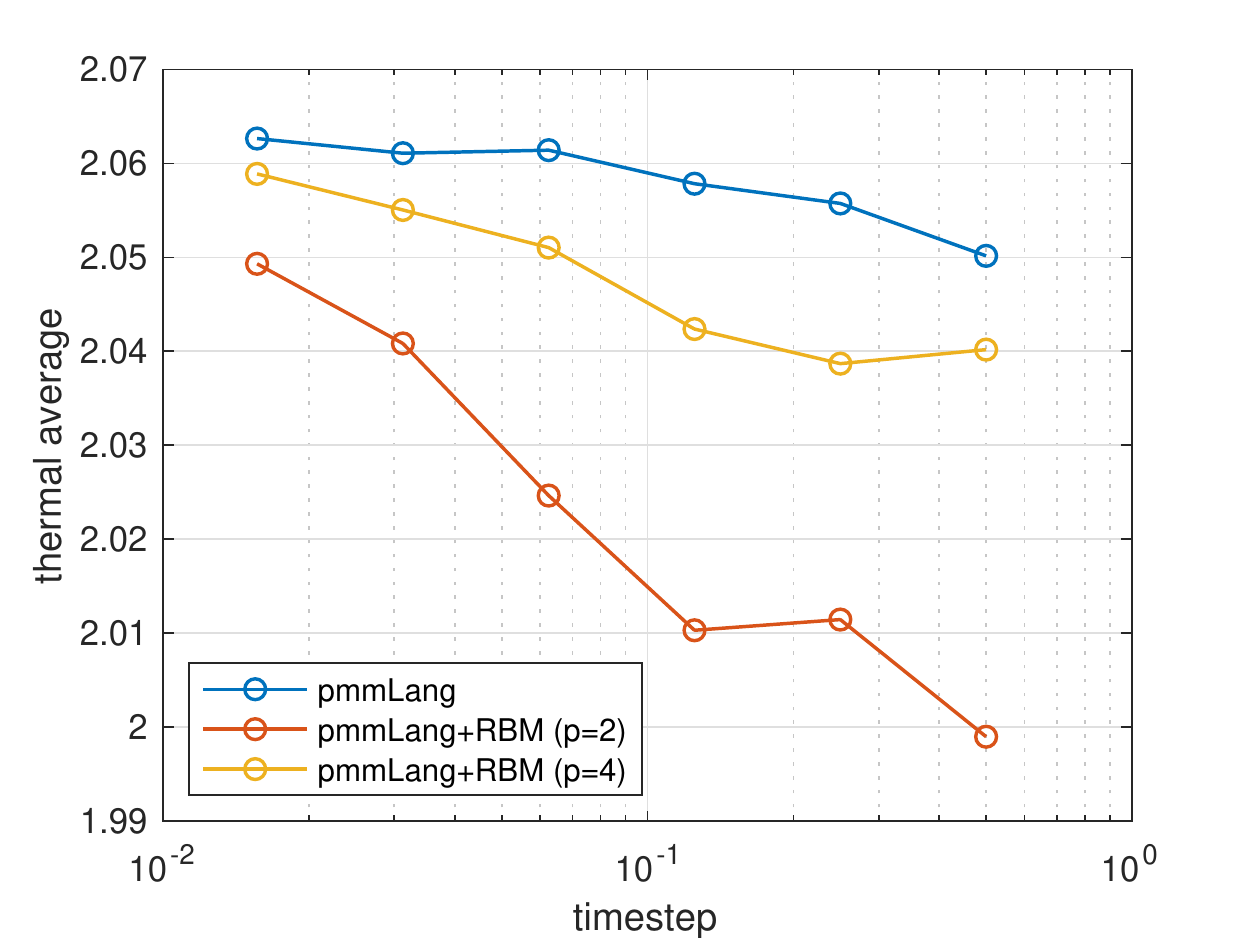} \\
	\includegraphics[width=0.4\textwidth]{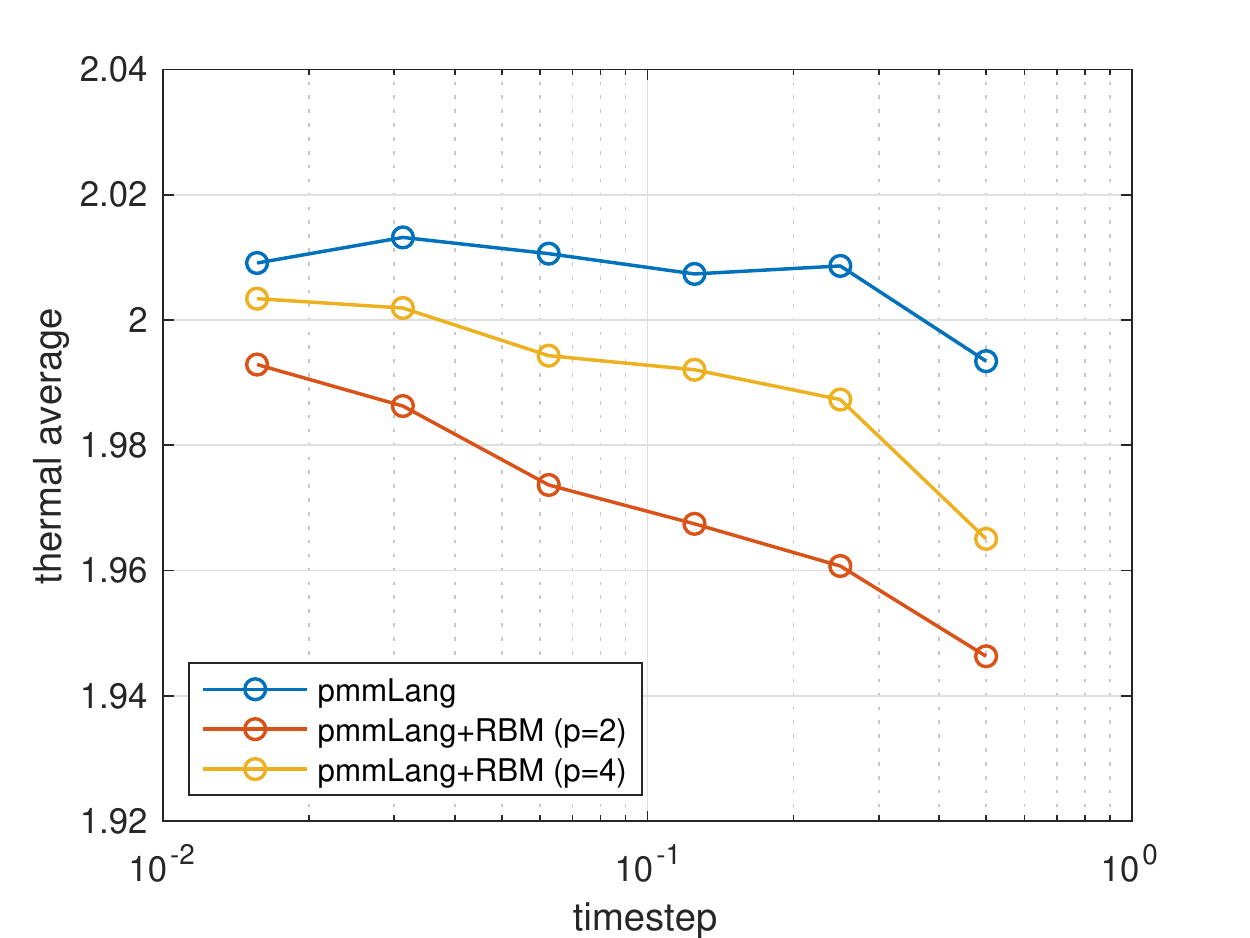}
	\includegraphics[width=0.4\textwidth]{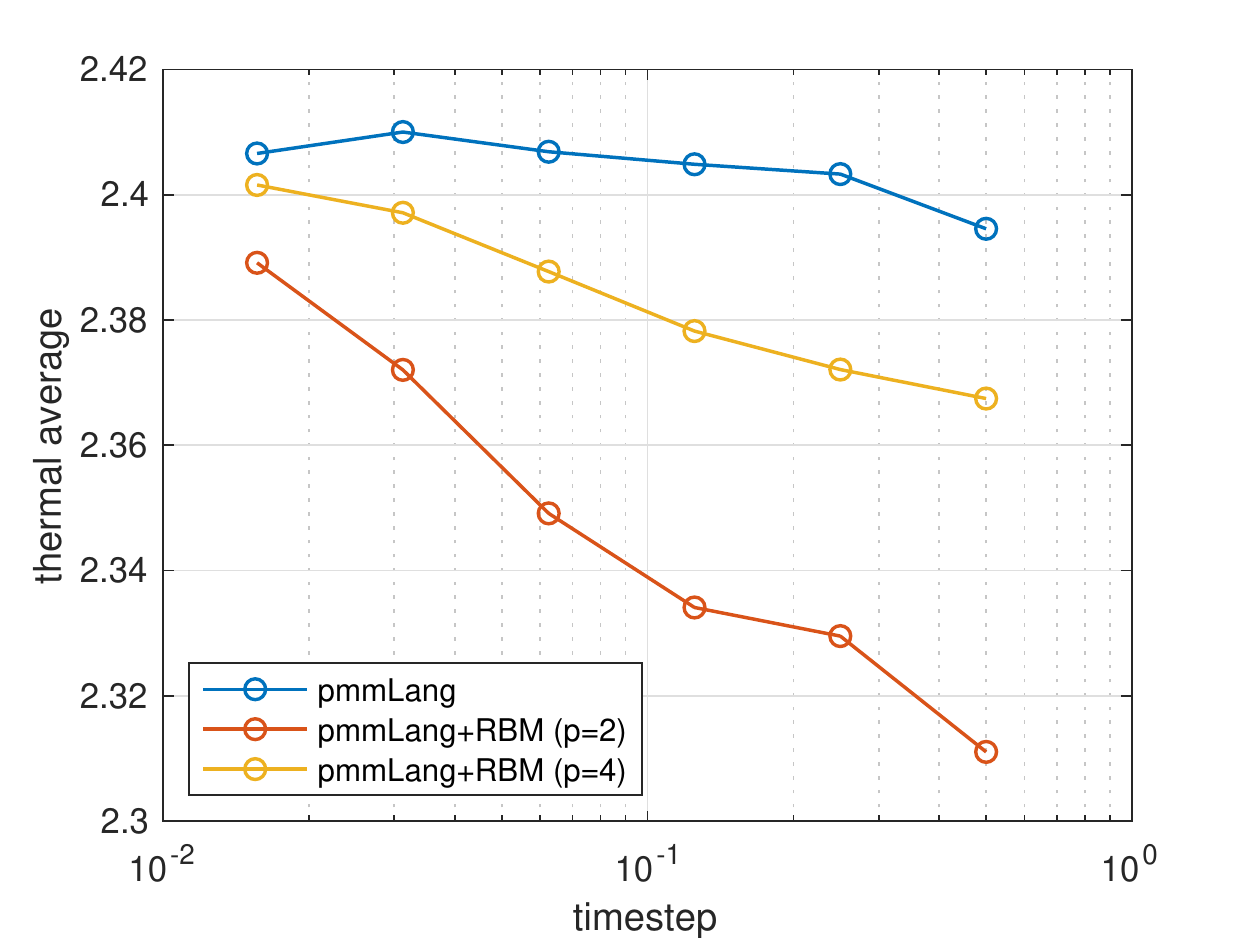} \\
\caption{
	The time averages (\ref{avg:pmmLang})(\ref{avg:pmmLang RBM}) of the pmmLang and the pmmLang+RBM in the Coulomb interacting system.
	The left and right panels represent the inverse temperature $\beta = 1$ and $4$. The figures from top to bottom are associated with the number of particles $P=8,16,24,32$ respectively, and the corresponding spread in the $y$-axis is $0.02, 0.04, 0.08, 0.12$.
	The mass $m=1$, the number of beads $N=16$, the timestep $\Delta t = 1/16$, the total sampling time $T=10000$, the friction constant $\gamma=2$ and the batch size $p=2$. In the $x$-axis, the timestep $\Delta t$ varies in $1/2,1/4,\cdots,1/64$.
}
\label{Coulomb:compare pmmLang RBM}
\end{figure*}

In this part we test the error of time average (\ref{avg:pmmLang RBM}) of the pmmLang+RBM with respect to varying timesteps $\Delta t$. In the Coulomb interacting system, we employ the pmmLang and the pmmLang+RBM to compute the time averages (\ref{avg:pmmLang})(\ref{avg:pmmLang RBM}), where the observable of interest is the position-dependent one with $A(q)$ given in (\ref{Coulomb:observable}).
In Figure \ref{Coulomb:compare pmmLang RBM}, we plot the time averages (\ref{avg:pmmLang})(\ref{avg:pmmLang RBM}) with different timesteps. Different numbers of particles are used to test the sampling methods.
The relative error of the time averages for $\beta=4$ is shown in Table \ref{Coulomb:compare pmmLang RBM error}.
\begin{table}
\begin{ruledtabular}
\begin{tabular}{c|ccc}
\#particles & pmmLang & w/RBM, $p=2$ & w/RBM, $p=4$ \\
\hline
8 & 0.26\% & 0.84\% & 0.55\% \\
16 & 0.27\% & 1.89\% & 0.84\% \\
24 & 0.33\% & 2.48\% & 1.16\% \\
32 & 0.14\% & 3.20\% & 1.43\% \\
\hline\hline
\#particles & pmmLang & w/RBM, $p=2$ & w/RBM, $p=4$ \\
\hline
8 & 0.07\% & 0.43\% & 0.06\% \\
16 & 0.06\% & 1.07\% & 0.35\% \\
24 & 0.06\% & 1.84\% & 0.56\% \\
32 & 0.01\% & 2.39\% & 0.78\% \\
\end{tabular}
\end{ruledtabular}
\caption{The relative error of the time averages (\ref{avg:pmmLang})(\ref{avg:pmmLang RBM}) of the pmmLang and the pmmLang+RBM when $\beta = 4$. The top and bottom figures are for the timestep $\Delta t = 1/4,1/16$ respectively. The reference values are computed by the pmmLang with $\Delta t = 1/64$.}
\label{Coulomb:compare pmmLang RBM error}
\end{table}
In Figure \ref{Coulomb:compare pmmLang RBM}, we observe that the bias of the pmmLang+RBM time average (\ref{avg:pmmLang RBM}) from $\avg{\hat A}$ diminishes as the timestep $\Delta t$ approaches 0. Also, the bias associated with the batch size $p=4$ is much less than $p=2$. The numerical results above confirm our arguments on the constant $C_{p,\Delta t}$ in the weak error analysis (\ref{eq:limit RBM}).\par
Finally, we report the time compexity of the pmmLang and the pmmLang+RBM in Table \ref{Coulomb:CPU time}. When the number of particles $P$ is large, the pmmLang+RBM is much more efficient than the pmmLang, while the relative error always still keeps small.
\begin{table}
\begin{ruledtabular}
\begin{tabular}{c|ccc}
\#particles & pmmLang & w/RBM, $p=2$ & w/RBM, $p=4$ \\
\hline
8 & \n{7.6}{-3} & \n{1.3}{-3} & \n{2.7}{-3} \\
16 & \n{2.2}{-2} & \n{2.4}{-3} & \n{5.3}{-3} \\
24 & \n{5.1}{-2} & \n{3.3}{-3} & \n{7.5}{-3} \\
32 & \n{1.0}{-1} & \n{4.5}{-3} & \n{9.6}{-3} \\
\end{tabular}
\end{ruledtabular}
\caption{The cost of CPU time \blue{(second)} in one single timestep of the pmmLang and the pmmLang+RBM.}
\label{Coulomb:CPU time}
\end{table}
\subsection{Tests of the pmmLang+RBM+split}
\subsubsection{Rejection rates}
We focus on the mixed Coulomb-Lennard-Jones system, where the pairwise interacting potential $V^{(c)}(q)$ is very singular.
As we have demonstrated in Section \ref{sec:split}, we shall use the splitting Monte Carlo method to avoid using too small timesteps. \blue{In this case, the rejection rate is an important index to assess the numerical efficiency of the method. Below we compute the rejection rates of the pmmLang+split and the pmmLang+RBM+split with different parameters including $N,P,p$ and $\Delta t$.\par 
In Table \ref{mixed:rejection rate P} and \ref{mixed:rejection rate N}, we fix the number of beads $N=16$ and the number of particles $P=16$ respectively and compute the rejection rate of the pmmLang+split and the pmmLang+RBM+split. The inverse temperature $\beta = 4$, the mass $m=1$, and different batch sizes $p$ and timesteps $\Delta t$ are used.\par
Under the proper scaling $\alpha = P^{-\frac23}$ specified in (\ref{eq:alpha P23}), Table \ref{mixed:rejection rate P} shows that the rejection rate gently grows with the number of particles $P$, and can be significantly reduced by shrinking the timestep. For example, when $N=16$ and $P=32$, the timestep $\Delta t = 1/16$ makes sure the rejection rate not larger than $25\%$. Additionally, the application of the RBM does not change the rejection rate too much.\par
However, when we fix the number of particles $P$, Table \ref{mixed:rejection rate N} shows that the rejection rate rapidly deteriorates with the number of beads $N$, even in the pmmLang+split without the use of random batches. This is due to the growing $N$ makes a particle $\q^i\in\mathbb R^{N\times3}$ more easily to collide with other particles. 
How to properly design the splitting Monte Carlo method to overcome the effects of the singular potential within the PIMD framework remains a topic to study in the future.}
\begin{table}
\begin{ruledtabular}
\begin{tabular}{c|ccc}
\#particles & pmmLang+split & w/RBM, $p=2$ & w/RBM, $p=4$ \\
\hline
8 & 9.73\% & 9.40\% & 9.53\% \\
16 & 15.37\% & 17.92\% & 15.25\% \\
24 & 19.43\% & 23.60\% & 21.75\% \\
32 & 26.40\% & 29.93\% & 27.72\% \\
\hline\hline
\#particles & pmmLang+split & w/RBM, $p=2$ & w/RBM, $p=4$ \\
\hline
8 & 6.05\% & 5.70\% & 4.79\% \\
16 & 10.85\% & 11.54\% & 10.27\% \\
24 & 13.05\% & 18.71\% & 14.22\% \\
32 & 17.50\% & 23.54\% & 20.59\% \\
\hline\hline
\#particles & pmmLang+split & w/RBM, $p=2$ & w/RBM, $p=4$ \\
\hline
8 & 2.95\% & 2.58\% & 3.60\% \\
16 & 6.00\% & 5.33\% & 5.45\% \\
24 & 8.26\% & 9.31\% & 8.93\% \\
32 & 10.50\% & 13.69\% & 11.69\% \\
\end{tabular}
\end{ruledtabular}
\caption{\blue{The rejection rates of the pmmLang+split and the pmmLang+RBM+split with numbers of particles $P$ along the sampling process. The tables from top to bottom correspond to the timestep $\Delta t = 1/8,1/16,1/32$ respectively. 
The number of beads $N=16$, the inverse temperature $\beta = 4$ and the mass $m=1$. The rejection rates are computed with the total sampling time $T=500$.}}
\label{mixed:rejection rate P}
\end{table}
\begin{table}
\begin{tabular}{c|ccc}
\hline
\hline
\#beads & pmmLang+split & w/RBM, $p=2$ & w/RBM, $p=4$ \\
\hline
16 & 15.54\% & 18.29\% & 18.21\% \\
32 & 26.19\% & 30.83\% & 28.63\% \\
64 & 47.38\% & 47.00\% & 47.75\% \\
128 & 74.62\% & 75.62\% & 71.25\% \\
\hline\hline
\#beads & pmmLang+split & w/RBM, $p=2$ & w/RBM, $p=4$ \\
\hline
16 & 9.94\% & 10.87\% & 11.10\% \\
32 & 19.10\% & 16.13\% & 18.94\% \\
64 & 34.31\% & 37.81\% & 38.84\% \\
128 & 66.44\% & 58.28\% & 67.52\% \\
\hline\hline
\end{tabular}
\caption{\blue{The rejection rates of the pmmLang and the pmmLang+RBM+split with different numbers of beads $N$ along the sampling process. The tables from top to bottom correspond to the timestep $\Delta t = 1/8,1/16$ respectively. 
The number of particles $P=16$, the inverse temperature $\beta = 4$ and the mass $m=1$. The rejection rates are computed with the total sampling time $T=500$.}}
\label{mixed:rejection rate N}
\end{table}
\subsubsection{Error in the calculation of the thermal average}
\begin{figure*}
	\centering
	\includegraphics[width=0.4\textwidth]{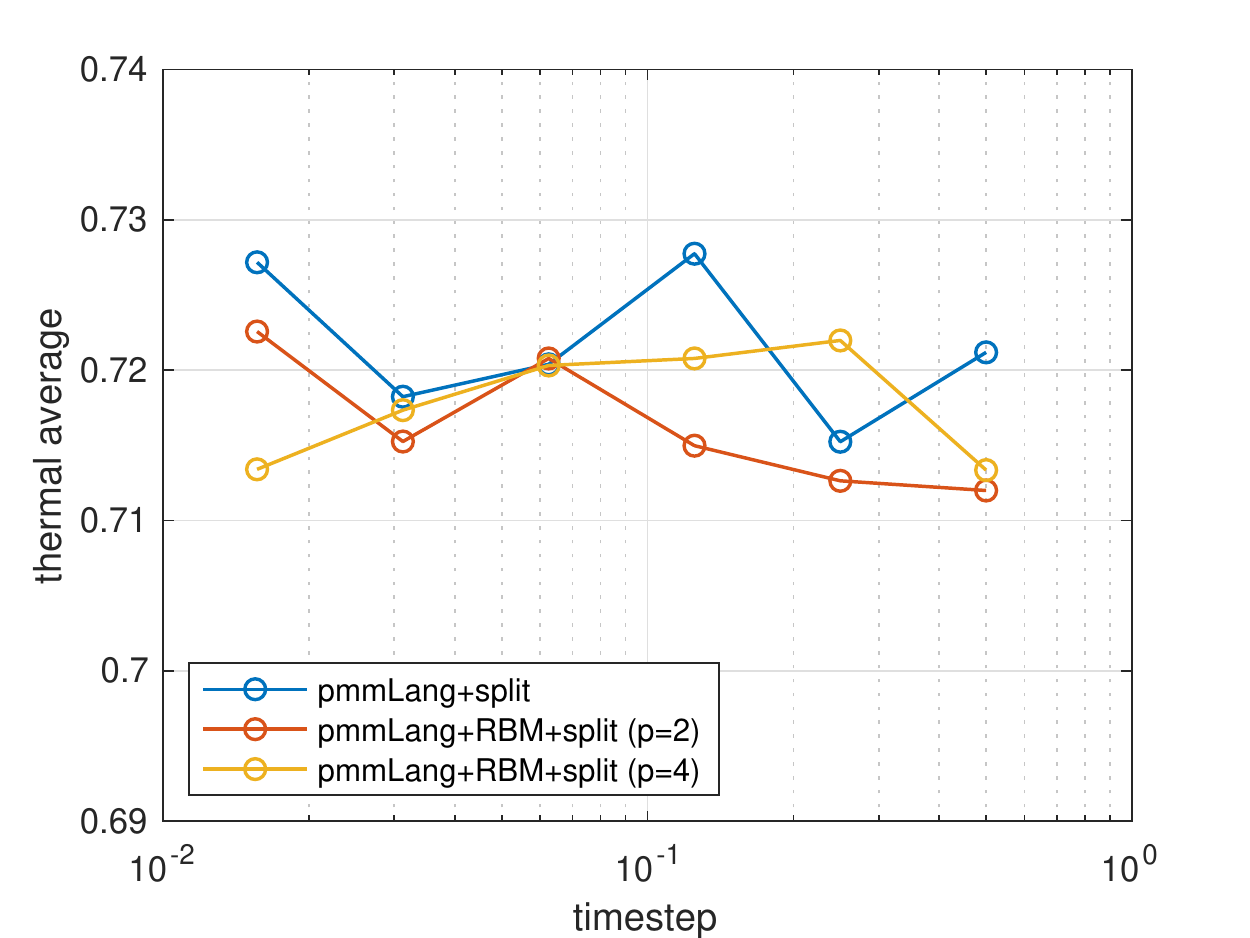}
	\includegraphics[width=0.4\textwidth]{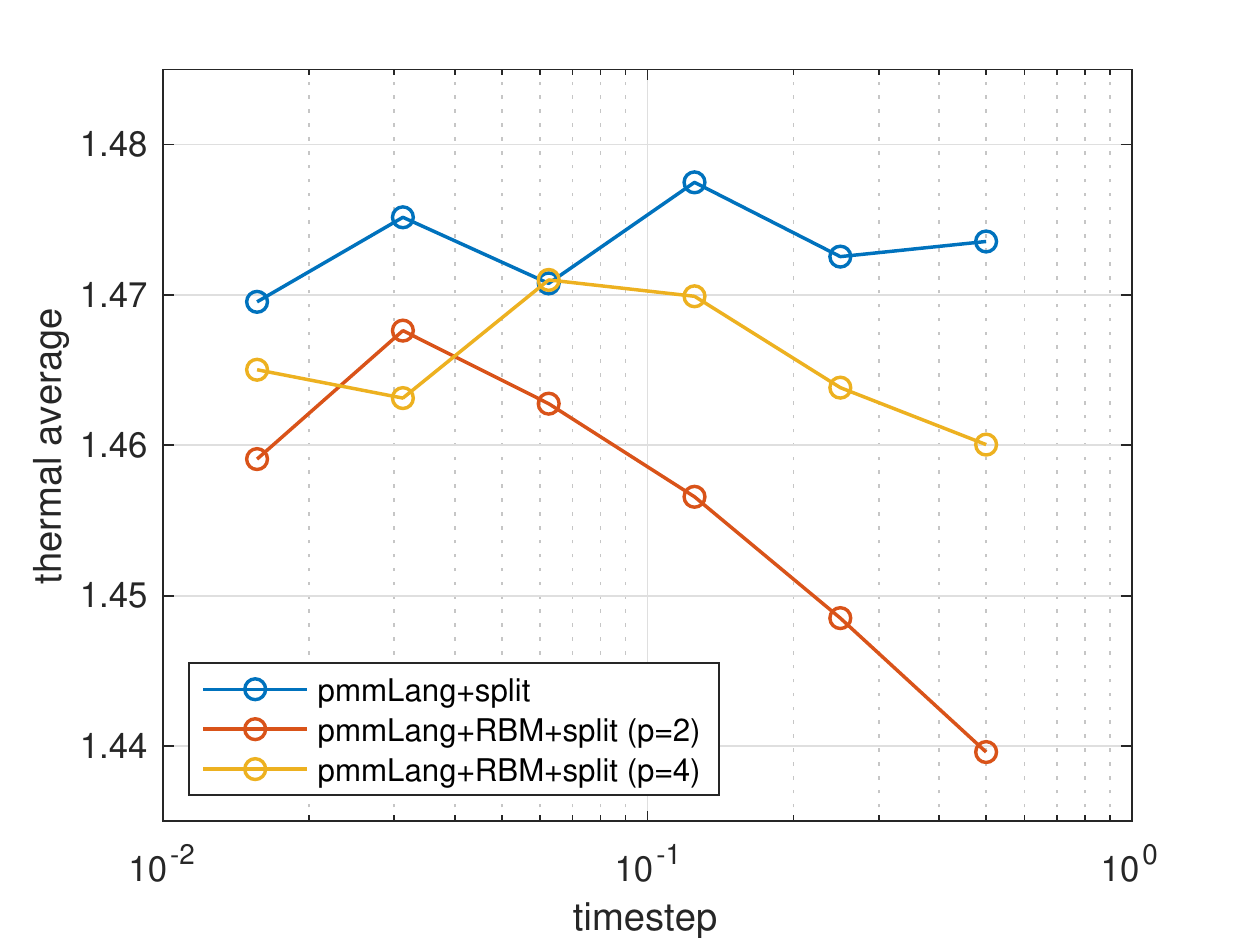} \\
	\includegraphics[width=0.4\textwidth]{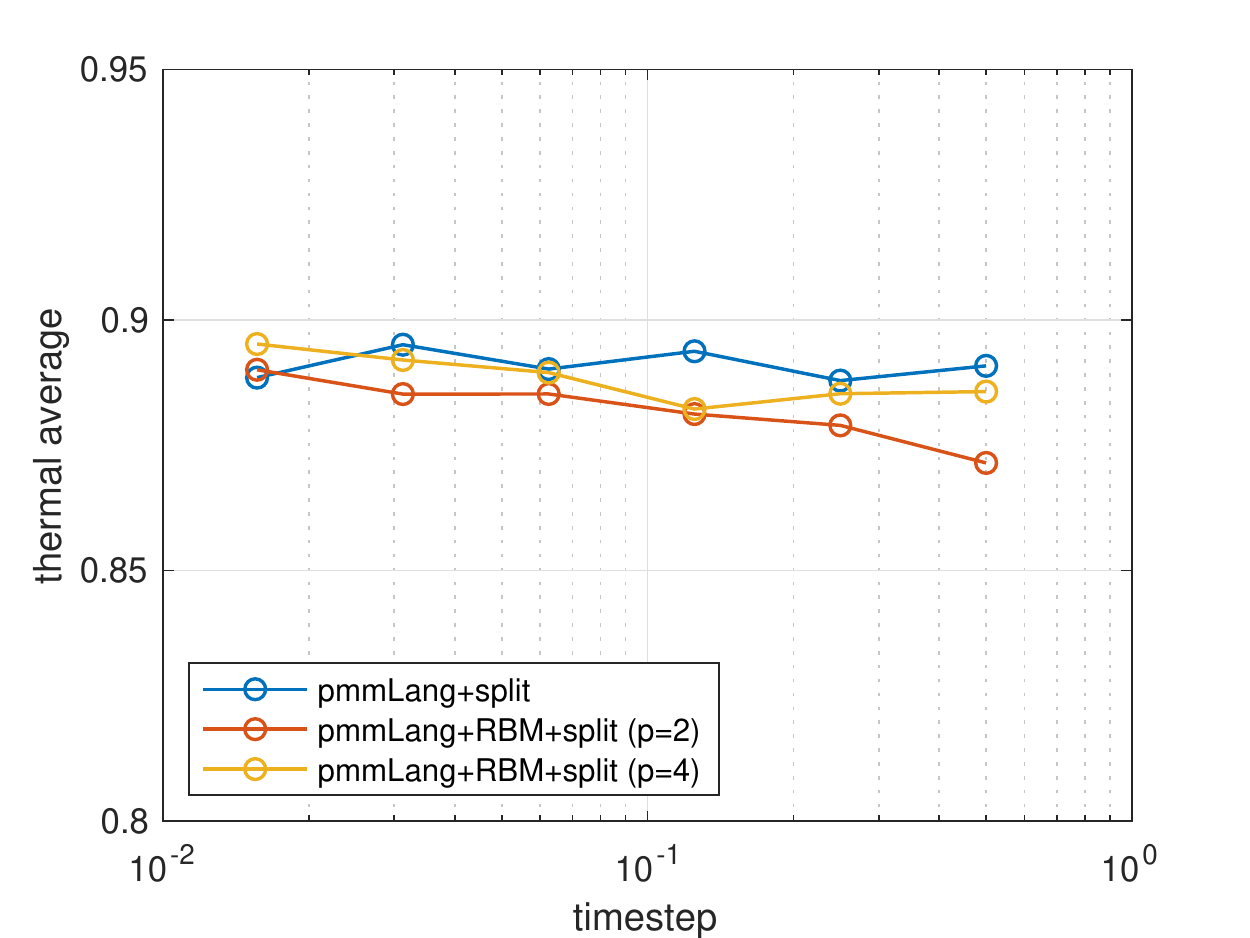}
	\includegraphics[width=0.4\textwidth]{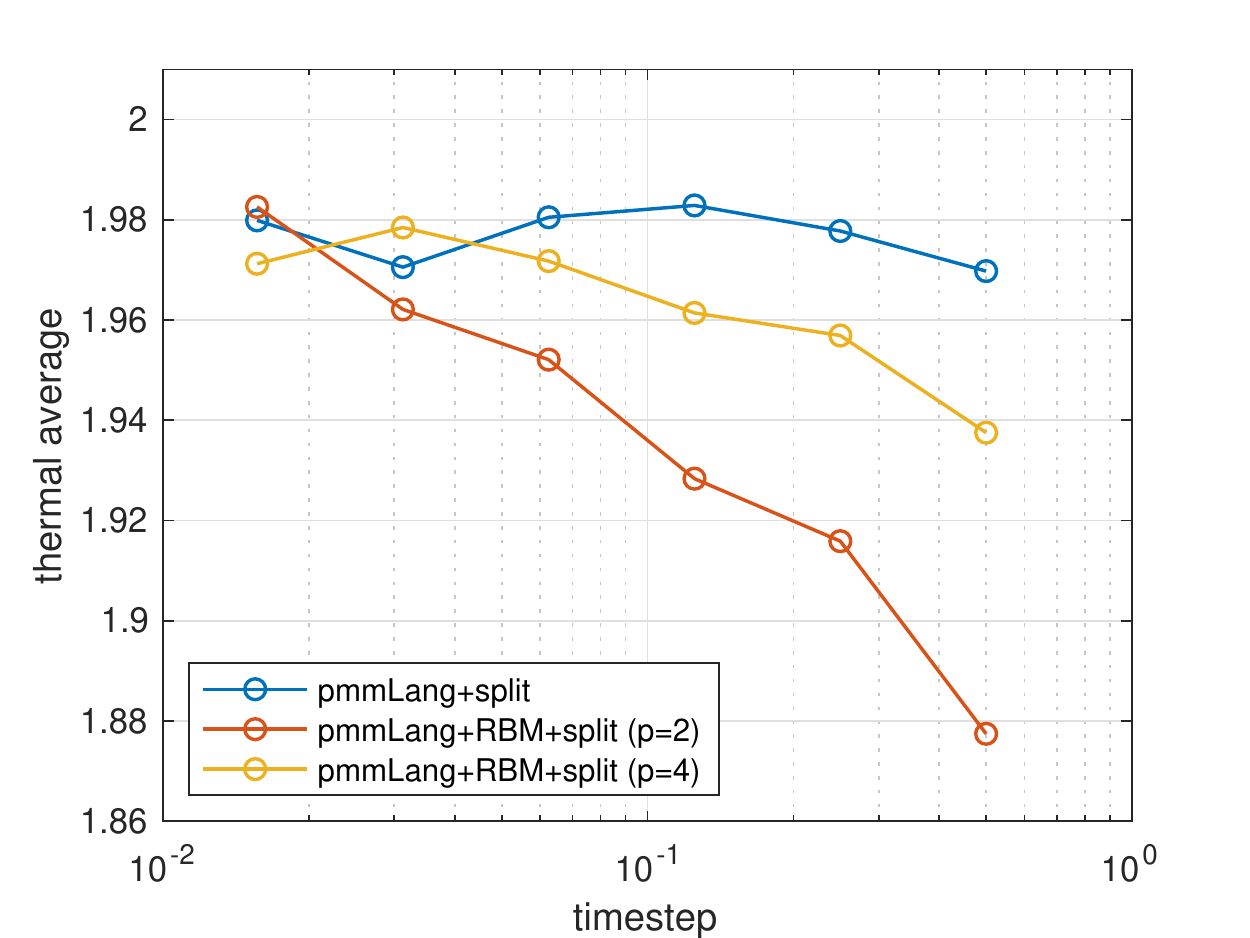} \\
	\includegraphics[width=0.4\textwidth]{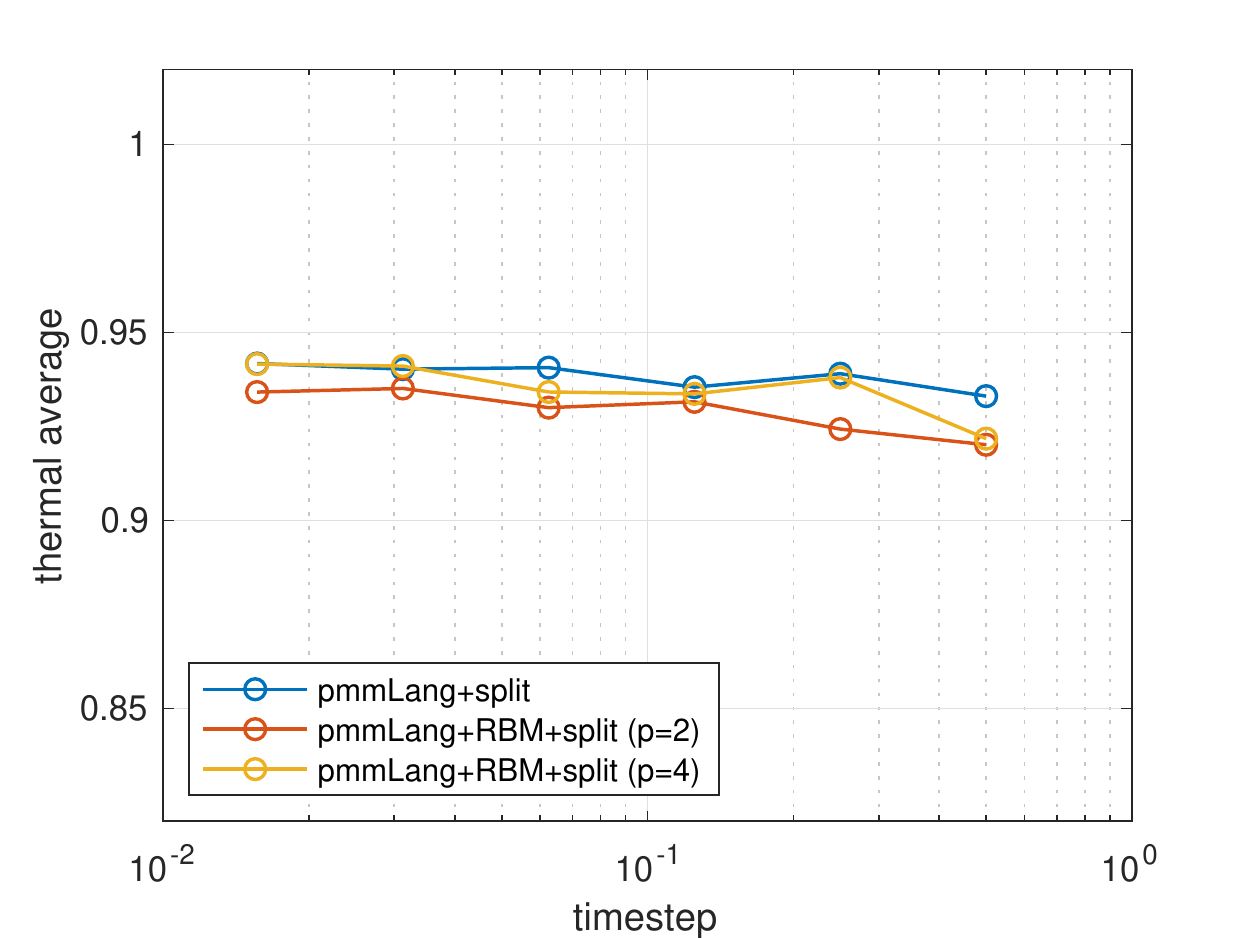}
	\includegraphics[width=0.4\textwidth]{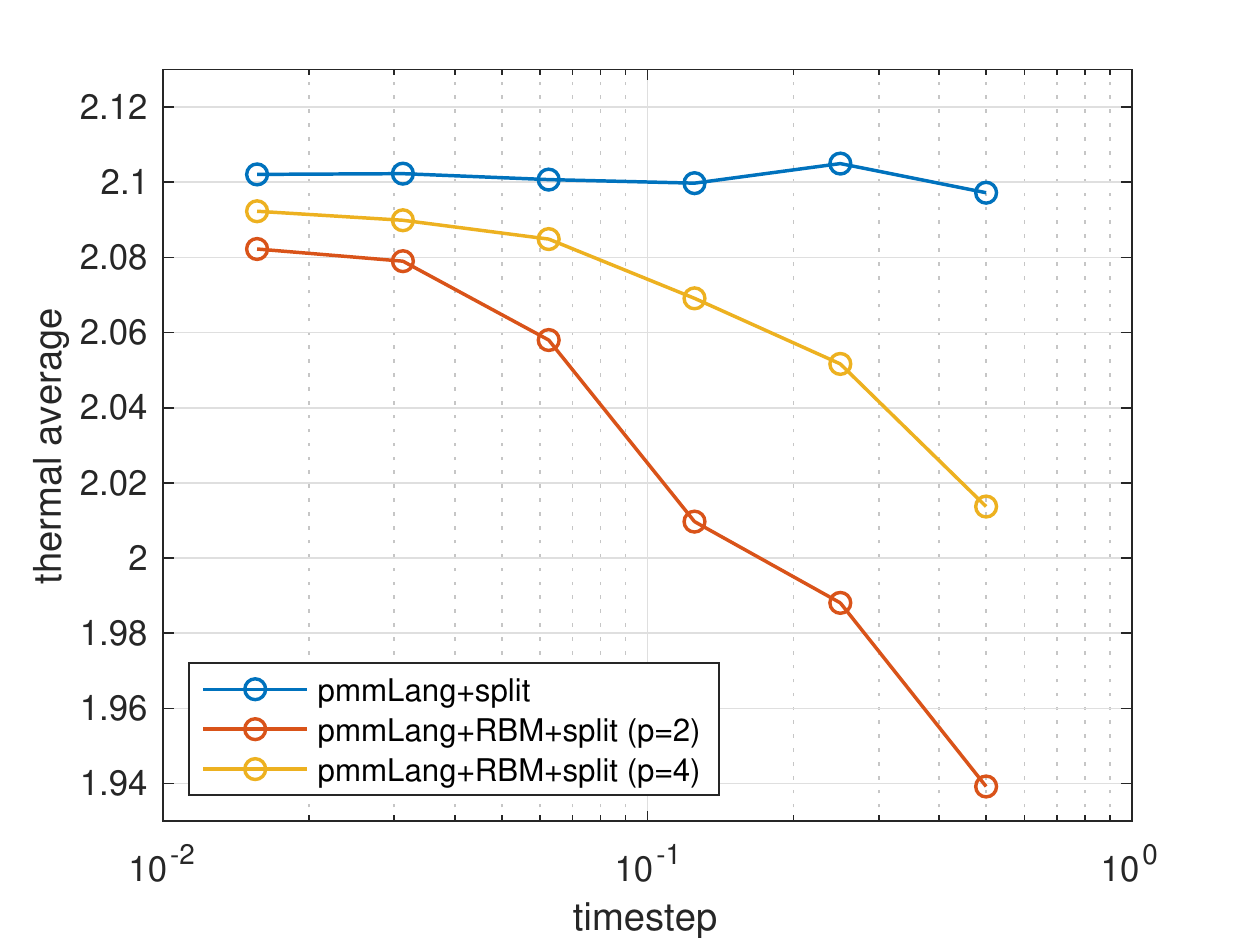} \\
	\includegraphics[width=0.4\textwidth]{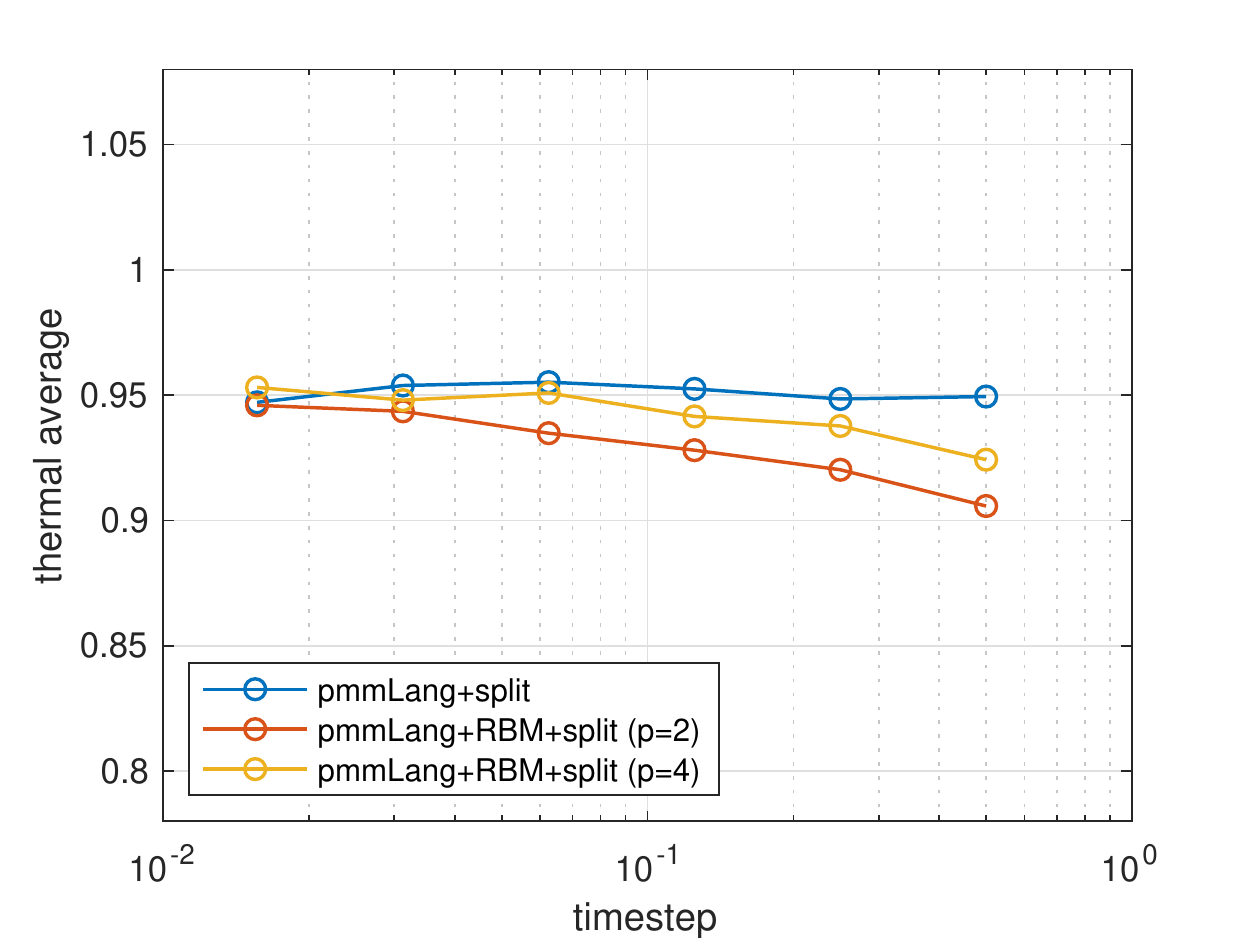}
	\includegraphics[width=0.4\textwidth]{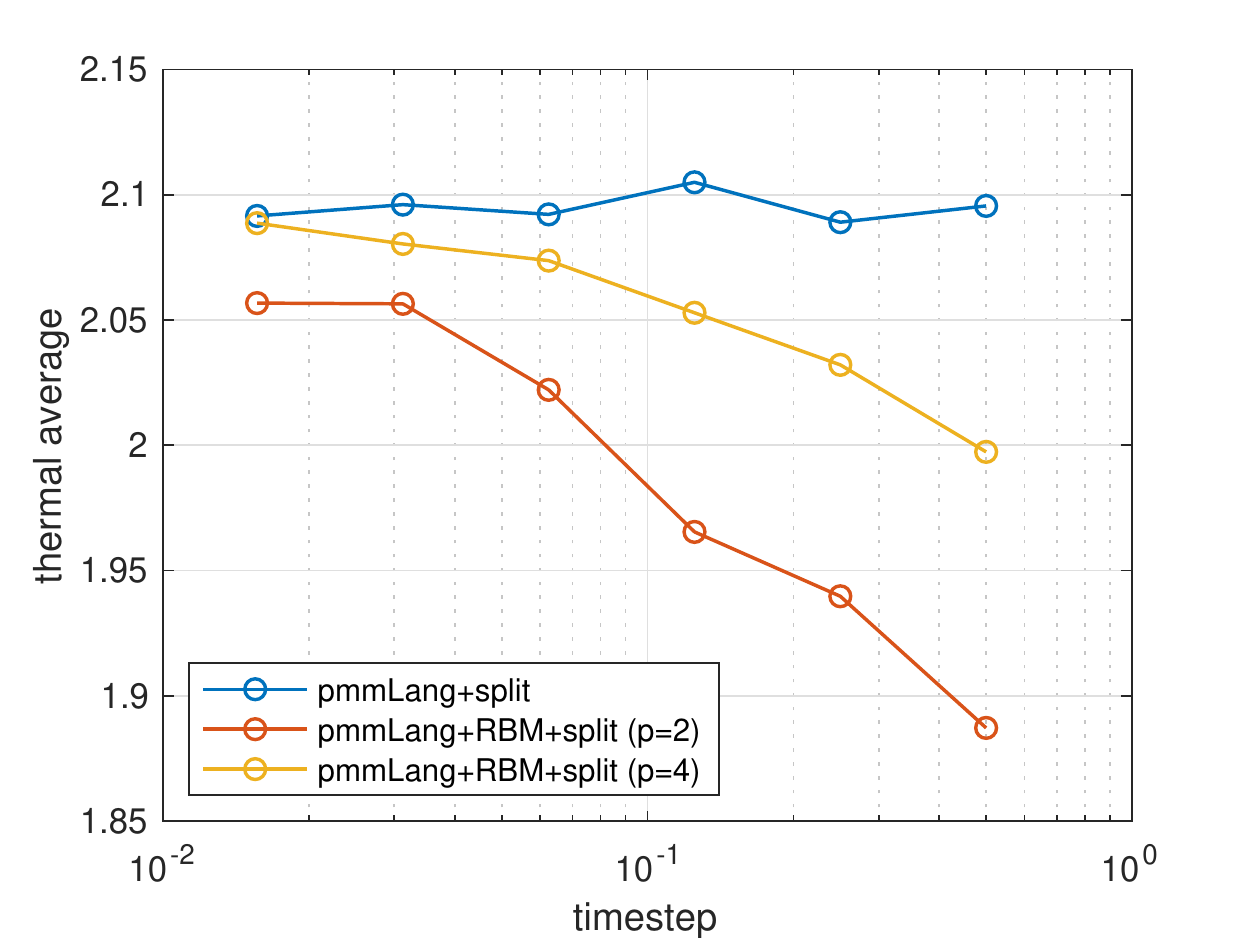} \\
\caption{The time averages of the pmmLang+split and the pmmLang+RBM+split in the mixed Coulomb-Lennard-Jones system.
The left and right panels represent the inverse temperature $\beta = 1$ and $4$.
The figures from top to bottom are associated with the number of particles $P=8,16,24,32$ respectively, and the corresponding spread in the $y$-axis is $0.05, 0.15, 0.2, 0.3$.
The mass $m=1$, the number of beads $N=16$, the timestep $\Delta t = 1/16$, the total sampling time $T=10000$, the friction constant $\gamma=2$ and the batch size $p=2$. In the $x$-axis, the timestep $\Delta t$ varies in $1/2,1/4,\cdots,1/64$.}
\label{mixed:compare pmmLang RBM}
\end{figure*}

Next, we test the error of the time average computed by the pmmLang+RBM+split with respect to varying timesteps $\Delta t$. In the mixed Coulomb-Lennard-Jones system, we employ the pmmLang+split and the pmmLang+RBM+split to compute the time averages, where the observable of interest is the position-dependent one with $A(q)$ given in (\ref{mixed:observable}).
In Figure \ref{mixed:compare pmmLang RBM}, we plot the time averages computed by the pmLang+split and the pmmLang+RBM+split the with different timesteps.
Different numbers of particles are used to test the sampling methods.
The relative error of the time averages for $\beta = 4$ is in Table \ref{mixed:compare pmmLang RBM error}.
\begin{table}
\begin{ruledtabular}
\begin{tabular}{c|ccc}
\#particles & pmmLang+split & w/RBM, $p=2$ & w/RBM, $p=4$ \\
\hline
8 & 0.20\% & 1.43\% & 0.39\% \\
16 & 0.11\% & 3.23\% & 1.16\% \\
24 & 0.14\% & 5.42\% & 2.40\% \\
32 & 0.12\% & 7.27\% & 2.84\% \\
\hline\hline
\#particles & pmmLang+split & w/RBM, $p=2$ & w/RBM, $p=4$ \\
\hline
8 & 0.08\% & 0.46\% & 0.10\% \\
16 & 0.03\% & 1.40\% & 0.41\% \\
24 & 0.06\% & 2.10\% & 0.82\% \\
32 & 0.03\% & 3.32\% & 0.85\% \\
\end{tabular}
\end{ruledtabular}
\caption{The relative error of the time averages of the pmmLang+split and the pmmLang+RBM+split when $\beta = 4$. The top and bottom figures are for the timestep $\Delta t = 1/4,1/16$ respectively. The reference values are computed by the pmmLang+split with $\Delta t = 1/64$.}
\label{mixed:compare pmmLang RBM error}
\end{table}
It can be seen from Figure \ref{mixed:compare pmmLang RBM} and Table \ref{mixed:compare pmmLang RBM error} that the error estimation (\ref{eq:limit RBM}) also holds for the pmmLang+RBM+split. That is to say, the bias of the time average computed by the pmmLang+RBM+split diminishes as the timestep $\Delta t\rightarrow0$ or the batch size $p$ increases.\par
Finally, we report the time complexity of the pmmLang and the pmmLang+RBM in Table \ref{mixed:CPU time}. Comparing the results in Table \ref{Coulomb:CPU time} and Table \ref{mixed:CPU time}, we observe that the pmmLang+RBM+split is slightly slower than the pmmLang+RBM, but is still much more efficient than the pmmLang+split.
Therefore, the additional calculation of the potential different $U_2(\q^*) -U_2(\q)$ in the splitting Monte Carlo method does not increases the total computational cost too much.
\begin{table}
\begin{ruledtabular}
\begin{tabular}{c|ccc}
\#particles & pmmLang+split & w/RBM, $p=2$ & w/RBM, $p=4$ \\
\hline
8 & \n{5.7}{-3} & \n{1.6}{-3} & \n{3.2}{-3} \\
16 & \n{2.5}{-2} & \n{2.9}{-3} & \n{6.1}{-3} \\
24 & \n{5.8}{-2} & \n{4.0}{-3} & \n{8.6}{-3} \\
32 & \n{1.1}{-1} & \n{5.9}{-3} & \n{1.1}{-2} \\
\end{tabular}
\end{ruledtabular}
\caption{The cost of CPU time \blue{(second)} in one single timestep of the pmmLang+split and the pmmLang+RBM+split.}
\label{mixed:CPU time}
\end{table}
\section{Conclusion}
\blue{We have proposed the pmmLang+RBM, an efficient sampling method of the quantum interacting particle system in the PIMD framework. The pmmLang+RBM properly combines the preconditioned mass-modified Langevin dynamics (pmmLang) and the random batch method (RBM) to resolve the stiffness of the ring polymer and reduce the complexity due to interaction forces. In the pmmLang+RBM, the computational cost due to interaction forces in a timestep is reduced from $O(NP^2)$ to $O(NP)$, where $N$ is the number of beads in the ring polymer and $P$ is the number of particles.\par
In the extensive numerical tests, the pmmLang+RBM shows fine performance in the calculation of the thermal average. The pmmLang+RBM shares a similar convergence mechanism with the original pmmLang, and results in a small bias from the target distribution even for large $N$ and $P$.
Nevertheless, a rigorous error estimation of the pmmLang+RBM remains to be studied further.\par
Under the circumstances of singular interacting potentials (e.g., the Lennard-Jones potential), we introduce the pmmLang+split and the pmmLang+RBM+split to avoid using extremely small timesteps. When the pmmLang+split has a low rejection rate, the pmmLang+RBM+split greatly reduces the computational cost per timestep, keeps small error in computing the thermal average, and does not increase the rejection rate too much.
It will be valuable to explore how to control the rejection rate when $N$ or $P$ is large.}
\appendix
\section{Discussion on the preconditioning methods}
\label{sec:precondition}
\subsection{Design of the preconditioning methods}
In this section we briefly introduce the theory of the preconditioning methods in the PIMD, which aim to resolve the stiffness in the ring polymer. As we have shown in Section \ref{sec:PIMD}, the stiffness originates from the stiffness matrix $L\in\mathbb R^{N\times N}$, which is defined by
$$
L = \frac m{\beta_N^2}
\begin{bmatrix}
	2 & -1 & & \cdots & & -1  \\
	-1 & 2 & -1& \cdots \\
	& -1 & 2 & \cdots \\
	\vdots & \vdots  &\vdots & \ddots & -1\\
	& & & -1 & 2 & -1\\
	-1 &&&&-1&2
\end{bmatrix}.
$$
Now let's focus on the Hamiltonian dynamics of the ring polymer system,
\begin{equation}
\begin{aligned}
	\d\q & = -M^{-1}\p\d t, \\
	\d\p & = -L\q\d t,
\end{aligned}
\label{app:dyn:Hamiltonian}
\end{equation}
where $M\in\mathbb R^{N\times N}$ is the positive definite mass matrix.\par
The first step of preconditioning is to use a coordinate transformation $\q = D\tilde\q$ to decompose (\ref{app:dyn:Hamiltonian}) into different modes. To be specific, using the transformation
\begin{equation}
	\tilde\q = D^{-1}\q , ~~~~
	\tilde\p = D^\T\p,
\end{equation}
we can rewrite (\ref{app:dyn:Hamiltonian}) as
\begin{equation}
\begin{aligned}
	\d\tilde\q & = -(D^\T MD)^{-1}\tilde\p\d t, \\
	\d\tilde\p & = -(D^\T LD)\tilde\q\d t,
\end{aligned}
\label{app:dyn:Hamiltonian transformed}
\end{equation}
where we require $D^\T MD$ and $D^\T LD$ to be both diagonal. In this, a preconditioning method in the PIMD is designed via the following steps:
\begin{enumerate}
\item Find a transformation matrix $D\in\mathbb R^{N\times N}$ such that 
\begin{equation}
	D^\T LD = \mathrm{diag}\{\lambda_1,\cdots,\lambda_N\}.
	\label{eq:diag L}
\end{equation}
\item For some suitable constants $\mu_1,\cdots,\mu_N>0$, choose the mass matrix
\begin{equation}
	M = D^{-\T}
	\mathrm{diag}\{\mu_1,\cdots,\mu_N\} D^{-1}
\end{equation}
\end{enumerate}
The transformed Hamiltonian dynamics (\ref{app:dyn:Hamiltonian transformed}) is then decomposed into different modes, where the frequency of the $k$-th mode is
\begin{equation}
	\omega_k = \sqrt{\frac{\lambda_k}{\mu_k}},
	~~~~
	k=1,\cdots,N
\end{equation}
and the stiffness is resolved if the frequencies $\omega_k$ are uniform for $k=1,\cdots,N$.\par
In the following we introduce two specific preconditioning methods, the staging coordinates and the preconditioned mass-modified Langevin dynamics (pmmLang).
\subsection{Staging coordinates}
The staging coordinates transformation is given by
\begin{equation}
	\tilde q_1 = q_1,~~
	\tilde q_k = q_k - \frac{(k-1)q_{k+1}+q_1}k,~~
	k=2,\cdots,N
\end{equation}
which directly yields the transformation matrix $D$. Using this transformation, the ring polymer potential becomes
\begin{equation}
	\sum_{k=1}^N |q_k-q_{k+1}|^2 = \sum_{k=1}^N m_k |\tilde q_k|^2
\end{equation}
where $m_1=0$ and $m_k = \frac{k}{k-1}$ for $k=2,\cdots,N$. Therefore $L\in\mathbb R^{N\times N}$ is diagonalized as in (\ref{eq:diag L}), where
\begin{equation}
	\lambda_1 = 0,~~\lambda_k = \frac{k}{k-1}
	\cdot\frac{m}{\beta_N^2},~~
	k = 2,\cdots,N
\end{equation}
In the staging coordinates method, the constants $\mu_1,\cdots,\mu_N$ are chosen to be
\begin{equation}
	\mu_1 = m,~~\mu_k = \frac{k}{k-1}m,~~k=1,\cdots,N
\end{equation}
thus the mass matrix $M\in\mathbb R^{N\times N}$ is given by
\begin{equation}
	M = \beta_N^2 L + m D^{-\T} E_{11} D 
	\label{eq:M staging}
\end{equation}
where $E_{11}\in\mathbb R^{N\times N}$ is the matrix with only $(1,1)$ entry equal to 1. If one attempts to use the staging coordinates in the physical coordinates $\q,\p$, then it will be a tough task to deal with the algebra of $M$ in (\ref{eq:M staging}).
\subsection{Preconditioned mass-modified Langevin dynamics}
The pmmLang is based on the spectral decomposition of $L$, and the corresponding transformation matrix $D$ is orthogonal. For simplicity, assume the number of beads $N$ is even. Then $L\in\mathbb R^{N\times N}$ is orthogonally diagonalized as in (\ref{eq:diag L}), where
\begin{equation*}
\begin{aligned}
	\lambda_1 = 0: &~
	D_{j,1} = \frac1{\sqrt{N}} \\
	\lambda_N = \frac{4m}{\beta_N^2}: &~
	D_{j,N} = \frac{(-1)^j}{\sqrt{N}} \\
	\lambda_{2k} = \frac{4m}{\beta_N^2}\sin^2 \frac{\pi k}N: &~
	D_{j,2k} = \sqrt{\frac2N} \cos \frac{2\pi kj}N \\
	\lambda_{2k+1} = \frac{4m}{\beta_N^2} 
	\sin^2 \frac{\pi k}N: &~
	D_{j,2k+1} = \sqrt{\frac2N}\sin \frac{2\pi kj}N \\
\end{aligned}
\end{equation*}
\begin{equation*}
	\Big(j=1,\cdots,N;k=1,\cdots,\frac N2-1\Big)
\end{equation*}
The constants $\mu_1,\cdots,\mu_N$ are simply chosen to be
\begin{equation}
	\mu_k = \lambda_k + \alpha,~~~~k=1,\cdots,N
\end{equation}
where the regularization parameter $\alpha>0$ is a fixed constant, and the corresponding mass matrix is $M = L^\alpha := L + \alpha I$.\par
The advantage of the pmmLang is that it is simple can be directly applied in the spatial coordinates $\q$. On the contrary, the mass matrix of the staging coordinates is complicated in the spatial coordinates $\q$, and is inconvenient to combine with the RBM, hence we choose the pmmLang as the preconditioning method in the interacting particle system.\par
Finally, we discuss the complexity due to algebraic operations in the pmmLang. In the BAOAB scheme in a timestep, one needs to solve the linear system
\begin{equation}
	(L^\alpha)^{-1} \nabla U^\alpha(\q)
\end{equation}
and compute the matrix-vector multiplication
\begin{equation}
	(L^\alpha)^{-\frac12}\bxi,~~~\bxi\sim\mathsf N(0,1)^{N\times 3P}
\end{equation}
to obtain the Gaussian random variable $\bta\sim\mathsf N(0,(L^\alpha)^{-1})$.
Since $L^\alpha$ is tridiagonal, the complexity of the linear system is $O(NP)$. With the use of the spectrum of $L$ given above, the matrix-vector multiplication $(L^\alpha)^{-\frac12}\bxi$ can be calculated by the fast Fourier transform, and the complexity is $O(N\log NP)$. In conclusion, the complexity due to algebraic operations in the pmmLang in a timestep is $O(N\log NP)$.
\section{Splitting Monte Carlo method}
In this section we establish the detailed balance for the splitting Monte Carlo method, and prove that the corresponding second-order Langevin dynamics preserves the desired Boltzmann distribution. To simplify our arguments, consider the target Boltzmann distribution
\begin{equation}
	\pi(q,v) = \exp\bigg(\hspace{-2pt}
	-\beta\Big(
	 \frac12\avg{v,Mv} + U(q)
	\Big)
	\bigg),
\end{equation}
where $q,v\in\mathbb R^d$ is the position and velocity, $M\in\mathbb R^{d\times d}$ is the positive definite mass matrix and $\beta>0$ is the inverse temperature.\par
The Langevin dynamics which preserving the distribution $\pi(q,v)$ is given by
\begin{equation}
	\begin{aligned}
		\d q & = v \d t, \\
		\d v & = -M^{-1}\nabla U(q)\d t - \gamma v \d t + \sqrt{\frac{2\gamma M^{-1}}{\beta}}\d B,
	\end{aligned}
	\label{app:Lang}
\end{equation}
where $B$ is the standard Brownian motion in $\mathbb R^d$, and $\gamma>0$ is the friction constant. The detailed balance for (\ref{app:Lang}) reads \cite{db}:
\begin{theorem}
	Let $T((q,v),(q',v'))$ be the transition probability density of the Langevin dynamics (\ref{app:Lang}) in time $t$, then the detailed balance holds:
	\begin{equation}
		\pi(q,v) T((q,v),(q',v')) = \pi(q',-v')
		T((q',-v'),(q,-v)),
	\end{equation}
which implies $\pi(q,v)$ is the invariant distribution of (\ref{app:Lang}):
\begin{equation}
	\int_{\mathbb R^{2d}}
	\pi(q,v) T((q,v),(q',v'))\d q\d v = 
	\pi(q',v').
\end{equation}
\end{theorem}
When the potential function $U(q)$ is singular, we split it into
\begin{equation}
	U(q) = U_1(q) + U_2(q),
	\label{U}
\end{equation}
where $U_1(q)$ is smooth and $U_2(q)$ is short-ranged and singular.
Then we obtain the splitting Monte Carlo method method (Algorithm \ref{app:dyn:pmmLang split}), which preserves the Boltzmann distribution $\pi(q,v)$ and avoids the gradient of the singular part of $U_1(q)$. The detailed balance of this method is given in the following theorem.
\begin{algorithm*}
	\caption{Splitting Monte Carlo method for the Langevin dynamics in a timestep $\Delta t$}
	Evolve the Langevin dynamics in a timestep $\Delta t$:
	$$
		\begin{aligned}
			\d q & = v\d t \\
			\d v & = -M^{-1}\nabla U_1(q)\d t - \gamma v\d t + \sqrt{\frac{2\gamma}{\beta}}M^{-\frac12}\d B
		\end{aligned}
	$$
	Let $(q^*,v^*)$ be the proposal calculated above. Set $(q,v) = (q^*,v^*)$ with probability 
	$$
		a(q,q^*) = \min\{1,e^{-\beta(U_2(q^*)-U_2(q))}\}
	$$
	otherwise set $(q,v) = (q,-v)$.
	\label{app:dyn:pmmLang split}
\end{algorithm*}
\begin{theorem}
	Let $T((q,v),(q',v'))$ be the transition probability density of Algorithm \ref{app:dyn:pmmLang split}, then the detailed balance holds:
	\begin{equation}
	\pi(q,v) T((q,v),(q',v'))
	=
	\pi(q',-v') T((q',-v'),(q,-v))
	\label{app:detailed balance split}
	\end{equation}
	which implies
	$\pi(q,v)$ is the invariant distribution of the splitting Monte Carlo method (Algorithm \ref{app:dyn:pmmLang split}):
	\begin{equation}
	\int_{\mathbb R^{2d}}
	\pi(q,v) T((q,v),(q',v'))\d q\d v = 
	\pi(q',v')
	\end{equation}
	\label{app:thm:split MC}
\end{theorem}
\textbf{Proof}
Define the distributions
\begin{align}
	\pi_1(q,v) & = \exp\bigg(-\beta\bigg(\frac12\avg{v,Mv} + U_1(q)\bigg)\bigg), \\
	\pi_2(q) & = \exp\Big(-\beta U_2(q)\Big),
\end{align}
then the target distribution $\pi(q,v) = \pi_1(q,v) \pi_2(q)$.
Define the acceptance probability
\begin{equation}
	a(q,q^*) = \min\{1,e^{-\beta(U_2(q^*)-U_2(q))}\},
\end{equation}
then $\pi_2(q)$ satisfies
\begin{equation}
	\pi_2(q) a(q,q') = \pi_2(q') a(q',q)
	\label{app:eq1}
\end{equation}
with the acceptance probability $a(\cdot,\cdot)$ defined above.
Let $T_1((q,v),(q',v'))$ be the transition probability density of the Langevin dynamics in a timestep $\Delta t$:
\begin{equation}
\begin{aligned}
	\d q & = v\d t \\
	\d v & = -M^{-1}\nabla U_1(q)\d t - \gamma v\d t + \sqrt{\frac{2\gamma}{\beta}}M^{-\frac12}\d B
\end{aligned}
\end{equation}
then from the detailed balance (\ref{eq:detailed balance}) we obtain
\begin{align}
	\pi_1((q,v),(q',v')) T_1((q,v),(q',v')) \hspace{2cm} \notag \\
	 = \pi_1(q',-v') T_1((q',-v'),(q,-v)) 
\end{align}
Note that the transition probability density of Algorithm \ref{app:dyn:pmmLang split} is
\begin{align}
	T((q,v),(q',v')) & = 
	T_1((q,v),(q',v'))a(q,q') \notag \\ 
	& + \delta(q'-q) \delta(v'+v)(1-A(q,v))
	\label{app:eq2}
\end{align}
where $A(q,v)$ is the average acceptance probability at $(q,v)$:
\begin{equation}
	A(q,v) = 
	\int_{\mathbb R^{2d}}
	T_1((q,v),(q',v'))
\end{equation}
To prove the detailed balance (\ref{app:detailed balance split}), we just need to verify 
\begin{align}
	\pi(q,v) T_1((q,v),(q',v')) a(q,q') \hspace{2cm} \notag \\
	=\pi(q',-v') T_1((q',-v'),(q,-v)) a(q',q)
	\label{app:n1} 
\end{align}
and
\begin{align}
	\pi(q,v) \delta(q'-q)\delta(v'+v)
	(1-A(q,v)) \hspace{2cm}\notag\\ 
	=\pi(q',-v') \delta(q'-q)\delta(v'+v)
	(1-A(q',-v')).
	\label{app:n2}
\end{align}
In fact, (\ref{app:n1}) is the product of (\ref{app:eq1}) and (\ref{app:eq2}), and (\ref{app:n2}) holds for $q = q',v = -v'$. Hence Theorem \ref{app:thm:split MC} is proved.\par
Theorem \ref{app:thm:split MC} directly applies to the pmmLang (\ref{dyn:pmmLang}) with the splitting scheme $U^\alpha(\q) = U_1(\q) + U_2(\q)$, yielding Theorem \ref{thm:split MC} within the PIMD framework.

\section*{Acknowledgement}
Z. Zhou is supported by the National Key R\&D Program of China, Project Number 2020YFA0712000 and NSFC grant No. 11801016, No. 12031013. Z. Zhou is also partially supported by Beijing Academy of Artificial Intelligence (BAAI). The authors thank Prof. Shi Jin and Prof. Jian Liu for helpful discussions.

\section*{Data Availability}
The cosdes and data that support the findings of this study are available from the corresponding author upon reasonable request.

\section*{References}
\bibliography{reference}% Produces the bibliography via BibTeX.

\end{document}